\documentclass[journal]{IEEEtran}

\usepackage{comment}
\usepackage{relsize}

\usepackage{cite}
\ifCLASSINFOpdf
   \usepackage[pdftex]{graphicx}
   \graphicspath{{pdf/}{jpg/}}
   \DeclareGraphicsExtensions{.pdf,.jpg,.png}
\else
\fi
\usepackage[cmex10]{amsmath}
\usepackage{algorithmic}
\usepackage[boxed,Algorithm]{algorithm}%
 \usepackage{url}
 \usepackage{breakurl}
\usepackage{enumitem}
\usepackage{amsfonts}
\usepackage{color}
\usepackage{epstopdf}
\usepackage{dsfont}
\usepackage{amsmath,amssymb,amsthm,mathrsfs,amsfonts,dsfont}
\usepackage{soul}
\soulregister\cite7
\soulregister\ref7
\soulregister\pageref7

\newtheorem{proposition}{Proposition}

\newcommand*{\figref}[1]{\figurename~\ref{#1}}
\renewcommand*{\figurename}{Fig.}

\newcommand*{\eqn}[1]{(\ref{#1})}

\newcommand{\fx}{f}
\newcommand{\fk}{\hat{f}}
\newcommand{\tfx}{g}
\newcommand{\tfk}{\hat{g}}
\newcommand{\dtild}{\mathbf{\tilde{d}}}
\newcommand{\oversamplingfactor}{\sigma}

\newcommand{\SNs}{S^*}
\newcommand{\SN}{S}
\newcommand{\SNb}{\mathcal{S}^\bot}
\newcommand{\mSN}{\mathcal{S}}

\newcommand{\supp}{\operatorname{supp}}

\begin{document}
\title{The SPURS Algorithm for Resampling an Irregularly Sampled Signal onto a Cartesian Grid }

\author{Amir~Kiperwas,
        Daniel~Rosenfeld,~\IEEEmembership{Member,~IEEE,}
        and~Yonina~C.~Eldar,~\IEEEmembership{Fellow,~IEEE}%
\thanks{A. Kiperwas and Y. C. Eldar are with the Department
of Electrical Engineering, Technion, Israel Institute of Technology}%
\thanks{D. Rosenfeld is with RAFAEL Advanced Defense Systems Ltd., Israel.}%
\thanks{This research was supported by a grant from RAFAEL Advanced Defense Systems Ltd., Israel.}
\thanks{Manuscript received Month XX, 2016; revised Month XX, 2016.}}

\markboth{IEEE TRANSACTIONS ON MEDICAL IMAGING,~Vol.~X, No.~X, MARCH~2016}%
{Kiperwas \MakeLowercase{\textit{et al.}}: The SPURS Algorithm for Resampling an Irregularly Sampled Signal onto a Cartesian Grid}
\maketitle

\begin{abstract}
We present an algorithm for resampling a function from its values on a non-Cartesian grid onto a Cartesian grid.
This problem arises in many applications such as MRI, CT, radio astronomy and geophysics.
Our algorithm, termed SParse Uniform ReSampling (SPURS), employs methods from modern sampling theory to achieve a small approximation error while maintaining low computational cost.
The given non-Cartesian samples are projected onto a selected intermediate subspace, spanned by integer translations of a compactly supported kernel function. This produces a sparse system of equations describing the relation between the nonuniformly spaced samples and a vector of coefficients representing the projection of the signal onto the chosen subspace.
This sparse system of equations can be solved efficiently using available sparse equation solvers. The result is then projected onto the subspace in which the sampled signal is known to reside. The second projection is implemented efficiently using a digital linear shift invariant (LSI) filter and produces uniformly spaced values of the signal on a Cartesian grid.
The method can be iterated to improve the reconstruction results.

We then apply SPURS to reconstruction of MRI data from nonuniformly spaced k-space samples.
Simulations demonstrate that SPURS outperforms other reconstruction methods while maintaining a similar computational complexity over a range of sampling densities and trajectories as well as various input SNR levels.
\end{abstract}


\IEEEpeerreviewmaketitle

\section{Introduction}
\IEEEPARstart{R}{econstruction} of a signal from a given set of nonuniformly spaced samples of its representation in the frequency domain is a problem encountered in a vast range of scientific fields: radio astronomy, seismic and geophysical imaging such as geophysical diffraction tomography (GDT) and ground penetrating radar (GPR)\cite{song2006high,song2006two},  SAR imaging \cite{subiza2003approach} and medical imaging systems including magnetic resonance imaging (MRI), computerized tomography (CT) and diffraction ultrasound tomography \cite{bronstein2002reconstruction}.

In the last decades nonuniform sampling patterns have become increasingly popular amongst MRI practitioners. In particular, radial \cite{rasche1995continuous,ferreira2013cardiovascular} and spiral \cite{ahn1986high,delattre2010spiral,pipe2014spiral} trajectories allow faster and more efficient coverage of k-space, thereby reducing scan time and giving rise to other desirable properties such as lower motion sensitivity \cite{gmitro1993use,van1998two}. Other notable non-Cartesian sampling patterns in MRI are stochastic \cite{scheffler1996frequency} and rosette \cite{noll1997multishot,noll1998simultaneous} trajectories which benefit from less systematic shifting or blurring artifacts.
A popular approach for recovering the original image is to resample the signal on a Cartesian grid in k-space and then use the inverse fast Fourier transform (IFFT) in order to transform back into the image domain. It has been shown \cite{schomberg1995gridding} that this approach is advantageous in terms of computational complexity.

In MRI, the most widely used resampling algorithm is convolutional gridding \cite{jackson1991selection,sedarat2000optimality}, which consists of four steps: 1) precompensation
for varying sampling density; 2) convolution with a Kaiser-Bessel window onto a Cartesian grid; 3) IFFT; and
4) postcompensation by dividing the image by the transform of the window.

Two other notable classes of resampling methods employed in medical imaging are the least-squares (LS) and the nonuniform-FFT (NUFFT) algorithms.
LS techniques, in particular URS/BURS \cite{rosenfeld1998optimal,rosenfeld2002new} are methods for calculating the LS solution for the equation describing the relationship between the acquired nonuniformly spaced k-space samples and their uniformly spaced counterparts, as given by the standard sinc-function interpolation of the sampling theorem. These methods invert this relationship using the regularized pseudoinverse by means of a singular value decomposition. Finding a solution to problems of common sizes using URS is computationally intractable. BURS offers an approximate tractable solution to the LS problem.

The NUFFT \cite{dutt1993fast,song2009least,fessler2003nonuniform} is a computationally efficient family of algorithms for approximating the Fourier transform, its inverse and its transpose of a function sampled on a Cartesian grid in one domain onto non-Cartesian locations in the other domain.
A nonuniform Fourier matrix $A$ \cite{nguyen1999regular} is approximated efficiently by performing the following three operations consecutively: 1) Pre-compensation/weighting of the samples taken on the Cartesian grid; 2) FFT/IFFT onto an oversampled Cartesian grid; 3) interpolation from this uniform grid to the nonuniform sample locations using a compactly supported interpolation kernel. The Hermitian adjoint of $A$, denoted $A^*$, which is approximated by performing the Hermitian conjugate of operations 1--3 in reverse order\footnote{It can be shown that the operation performed by convolutional gridding is equivalent to $A^*$.}, is used along with $A$ to solve the inverse problem --- transforming from the non-Cartesian onto the Cartesian grid. This is usually performed using variants of the conjugate gradient method which operates with $A$ and $A^*$ alternately until convergence.

In recent years the concepts of sampling and reconstruction have been generalized within the mathematical framework of function spaces \cite{unser1994general,eldar2009beyond,eldar2014sampling}. Methods were developed for reconstructing a desired signal, or an approximation of this signal, beyond the restrictions of the classic Shannon-Nyquist sampling theorem.

In this paper we apply these concepts to the reconstruction of a function from non-uniformly spaced samples in the spatial frequency domain. Resampling is performed onto a Cartesian grid in a computationally efficient manner while maintaining a low reconstruction error.
First, the given non-Cartesian samples are projected onto an intermediate subspace, spanned by integer translations of a compactly supported kernel function.
A sparse system of equations is produced which describes the relation between the nonuniformly spaced samples and a vector of coefficients representing the projection of the signal onto the auxiliary subspace.
This sparse system of equations is then solved efficiently using available sparse equation solvers.
The result is next projected onto the subspace in which the sampled signal is known to reside.
The second projection is implemented efficiently using a digital linear shift invariant (LSI) filter to produce uniformly spaced values of the signal on a Cartesian grid in k-space. Finally, the uniform samples are inverse Fourier transformed to obtain the reconstructed image.

Our algorithm, termed SParse Uniform ReSampling (SPURS) allows handling large scale problems while maintaining a small approximation error at a low computational cost.
We demonstrate that the reconstruction error can be traded off for computational complexity by controlling the kernel function spanning the auxiliary subspace and by oversampling the reconstruction grid.

These methods are applied to the problem of MR image reconstruction from nonuniformly spaced measurements in k-space.
The SPURS algorithm is compared using numerical simulations with other prevalent reconstruction methods in terms of its accuracy, its computational burden and its behavior in the presence of noise. %
The results demonstrate that a single iteration of SPURS outperforms other reconstruction methods while maintaining a similar computational complexity over a range of sampling densities and trajectories as well as various input SNR levels.
Iterating SPURS yields further improvement of the reconstruction result and allows for faster trajectories employing less sampling points.

We provide a freely available package\cite{SPURS2016code}, which contains Matlab (The MathWorks, Inc., Natick, MA, USA) code implementing the SPURS algorithm along with examples reproducing some of the results presented herein.

This paper is organized as follows.
Section \ref{sec:GS_methods} introduces generalized sampling methods which are employed throughout the paper.
Section \ref{sec:MRI_problem} formulates the non-Cartesian MRI resampling problem.
In Section \ref{sec:GSURS_alg} the basic SPURS algorithm is presented and then extended in Section \ref{sec:Extensions}.
Numerical simulations and their results are provided in Section \ref{sec:Simulations} and further discussed in Section \ref{sec:Discussion}.
We conclude in Section \ref{sec:Conclusion}.

\section{Generalized sampling methods}\label{sec:GS_methods}

This section reviews some concepts and methods which generalize the classic approach to sampling and reconstruction of signals and are used throughout the paper.

Unless noted, the notations in this paper are given for a 1D problem; the extension to higher dimensions is straightforward using separable functions.

In the classic approach to signal sampling a signal $\fk$ is represented by measurements which are its values at given sampling points. In recent years %
\cite{unser1994general,unser2000sampling,thevenaz2000interpolation,eldar2014sampling}
 this idea was extended and generalized within a function-space framework.
The processes of sampling and reconstruction can be viewed as an expansion of a signal onto a set of vectors
that span a signal subspace $\mathcal{A}$ of a Hilbert space $\mathcal{H}$:
  \begin{equation}\label{eq:set}
 \fk = \sum\limits_{n} {d\left[ n \right]{\mathbf{a}_n}}  = A\mathbf{d},
 \end{equation}
 where $\mathbf{d} \in \ell_2$, and $A:\ell_2 \rightarrow \mathcal{H}$ is a set transform corresponding to a set of vectors $\left\{ {\mathbf{a}_n} \right\}$ which span the subspace $\mathcal{A}$ and constitute a Riesz basis or a frame.
 Thus, applying $A$ is equivalent to taking linear combinations of the set of vectors $\left\{ {\mathbf{a}_n} \right\}$.
 Measurements are expressed as inner products of the function $\fk$ with a set of vectors $\left\{ {\mathbf{s}_m} \right\}$ that span the sampling subspace $\mSN  \subseteq \mathcal{H}$. Using this notation, the vector of samples $\mathbf{b}$ is given by $\mathbf{b} = {S}^*\fk$ where ${b}\left[ m \right] = {}\langle {{\mathbf{s}_m},\fk} {}\rangle$ and $\SN ^*$ is the adjoint of $\SN $.
Note that knowing the samples ${b}\left[ m \right]$ is equivalent
to knowing the orthogonal projection of ${\fk}$ onto $\mathcal{S}$, denoted by ${\fk}_{\mathcal{S}}$:
  \begin{equation*}\label{eq:sampling_ort_proj}
{\fk}_{\mathcal{S}} = P_{\mathcal{S}}{\fk} = S(S^*S)^{-1}S^*\fk = S(S^*S)^{-1}\mathbf{b},
 \end{equation*}
where
\begin{equation*}\label{eq:S_ort_proj}
P_{\mathcal{S}} = S(S^*S)^{-1}S^*,
 \end{equation*}
is the orthogonal projection operator defined by its range space $\mathcal{R}\left( {P_{\mathcal{S}}}\right) = \mathcal{S}$ and its null space $\mathcal{N}\left( {P_{\mathcal{S}}} \right) = {\mathcal{R}\left( {P_{\mathcal{S}}}\right)}^\bot $.

\begin{figure}
\centering
\begin{tabular}{cc}
\includegraphics[trim={0cm 0cm 0cm 0cm},clip,width=0.225\textwidth]{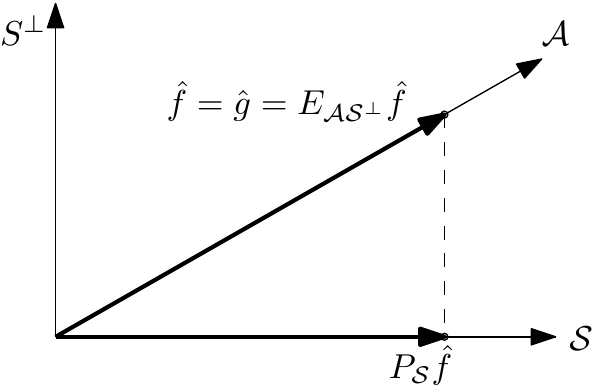} & \includegraphics[trim={0cm 0cm 0cm 0cm},clip,width=0.225\textwidth]{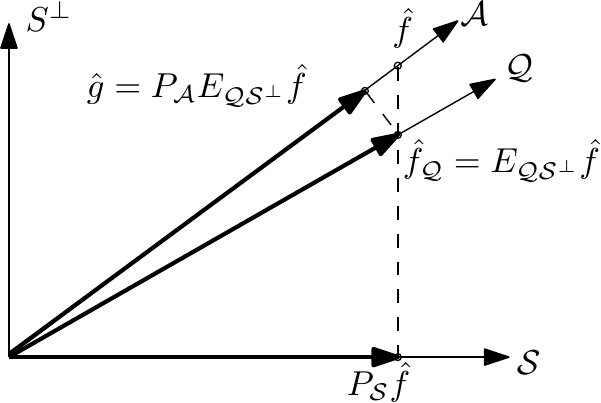} \\
(a)  & (b)\\
\end{tabular}
\caption{Geometrical interpretation. (a) An oblique projection in a perfect reconstruction scenario; (b) The SPURS scheme.}
\label{fig:GEO}
\end{figure}

A standard sampling problem is to reconstruct a signal ${\fk} \in \mathcal{A}$
from its vector of samples $\mathbf{b} = {S}^*\fk$. Geometrically, this amounts to finding a signal in $\mathcal{A}$ with the projection ${\fk}_{\mathcal{S}}$ onto $\mathcal{S}$ (see \figref{fig:GEO}(a)).
In order to be able to reconstruct any ${\fk}$ in $\mathcal{A}$ from samples in $\mathcal{S}$ it is required that $\mathcal{A}$ and $\mathcal{S}^{\bot}$ intersect only at zero. Otherwise, any non-zero signal in the intersection of $\mathcal{A}$ and $\mathcal{S}^{\bot}$ will yield zero samples and cannot be recovered.
For a unique solution we also need $\mathcal{A}$ and $\mathcal{S}$ to have the same numbers of degrees of freedom. These two requirements are fulfilled by the direct-sum condition
  \begin{equation}\label{eq:direct_sum}
\mathcal{H} = \mathcal{A} \oplus \mathcal{S}^\bot,
 \end{equation}
which implies that $\mathcal{A}$ and $\mathcal{S}^\bot$ are disjoint, and together span the space $\mathcal{H}$.

The reconstructed signal $\tfk$ is constructed to lie in the signal subspace $\mathcal{A}$.
Any signal $\tfk \in \mathcal{A}$ can be represented by $\tfk = A {\dtild}$, where ${\dtild} \in \ell_2$.
Restricting attention to linear recovery methods, we can write ${\dtild} = H \mathbf{b}$ for some transformation $H:\ell_2 \rightarrow \ell_2$, such that
\begin{equation*}\label{eq:f_reconstruction1}
\tfk  =  A{\dtild} = AH\mathbf{b} = AH{\SN ^ * }\fk = AH{\SN ^ * }A\mathbf{d},
\end{equation*}
where we use \eqn{eq:set}.
Perfect reconstruction means that $\tfk = \fk$.
Our problem then reduces to finding $H$ which satisfies
\begin{equation*}\label{eq:f_reconstruction2}
\tfk = AH{\SN ^ * }A\mathbf{d}  =  A\mathbf{d} = \fk
\end{equation*}
for any $\fk \in \mathcal{A}$, i.e., for any choice of $\mathbf{d}$. It is easily seen that choosing $H = {{{\left( {{\SN ^*}A} \right)}^{-1}}}$\cite{eldar2005general} satisfies this equation, where \eqn{eq:direct_sum} ensures that the inverse exists. In this case,
  \begin{equation}\label{eq:S_sol}
\tfk = {A{{\left( {{\SN ^*}A} \right)}^{-1}}{\SN ^*}}\fk.
 \end{equation}

The operator in \eqn{eq:S_sol} is the oblique projection\cite{tang2000oblique} onto $\mathcal{A}$ along ${\mSN ^ \bot }$:
\begin{equation}\label{eq:EAS}
{E_{\mathcal{A}{\mSN ^ \bot }}} = {A{{\left( {{\SN ^*}A} \right)}^{-1}}{\SN ^*}}.
\end{equation}
An operator $E$ is a projection if it satisfies $E^2=E$.
The oblique projection operator \eqn{eq:EAS} is a projection operator that is
not necessarily Hermitian. The notation ${E_{\mathcal{A}{\mSN ^ \bot }}}$ denotes an oblique projection
with range space $\mathcal{R}\left( {E_{\mathcal{A}{\mSN ^ \bot }}}\right) = \mathcal{A}$ and null space $\mathcal{N}\left( {E_{\mathcal{A}{\mSN ^ \bot }}} \right) = \SNb $. If $\mathcal{A} = \mathcal{S}$, then ${E_{\mathcal{A}{\mSN ^ \bot }}} = {P_{\mathcal{A}}}$.
A geometric interpretation of the perfect reconstruction scheme of \eqn{eq:S_sol} is illustrated in \figref{fig:GEO}(a).

When \eqn{eq:direct_sum} is not satisfied, we use the Moore-Penrose pseudoinverse\cite{ben2003generalized}, denoted by ${{{\left( {{\SN ^*}A} \right)}^{\dag}}}$. When ${{{\left( {{\SN ^*}A} \right)}^{\dag}}}$ is invertible, we have ${{{\left( {{\SN ^*}A} \right)}^{\dag}}} = {{{\left( {{\SN ^*}A} \right)}^{-1}}} $. For the sake of generality, we shall use the pseudoinverse henceforth.

A desired property of a reconstructed signal $\tfk$ is that it obeys the consistency condition which requires that injecting $\tfk$ back into the system must result in the same measurements as the original system itself, i.e., $S^* \tfk = \mathbf{b}$. %
Even when the input does not lie entirely in $\mathcal{A}$, for example, due to mismodeling or noise and regardless of $\mathcal{S}$, the property
$S^*{E_{\mathcal{A}{\mSN ^ \bot }}} = S^*$ %
ensures that $\tfk$ is consistent.

The consistency principle may be employed to perform reconstruction into a subspace, say $\mathcal{Q}$, which differs from $\mathcal{A}$. In this case, perfect reconstruction can no longer be achieved. Instead, we may seek a signal ${\fk_{\mathcal{Q}}} \in Q$ which satisfies the consistency condition: $S^* {\fk_{\mathcal{Q}}}  = \mathbf{b}$.
It is easily seen that the desired ${\fk_{\mathcal{Q}}}$ can be obtained using ${E_{\mathcal{Q}{\mSN ^ \bot }}}$, the oblique projection onto $\mathcal{Q}$ along $\mathcal{S}^\bot$,
 \begin{equation*}\label{eq:consistent_f_Q}
{\fk_{\mathcal{Q}}} = {Q{{\left( {{\SN ^*}Q} \right)}^{-1}}{\SN ^*}}\fk.
\end{equation*}
The consistency property of the oblique projection was introduced in \cite{unser1994general}. It was later extended in \cite{eldar2003sampling,eldar2004sampling,eldar2005general,eldar2014sampling} to a broader framework, alongside a geometric interpretation of the sampling and reconstruction schemes, and is employed later on.

The generalized approach to sampling and reconstruction allows for increased flexibility in the recovery process. This will be utilized in this paper to develop a computationally efficient implementation for the reconstruction of $\fk$ from a given set of non-uniformly spaced k-space samples $\mathbf{b}$, at the cost of a small amount of approximation error.

\section{The MRI problem}\label{sec:MRI_problem}
An MRI image is represented by a gray level function $\fx\left( x \right)$, where $x$ denotes the spatial coordinate in $2$ or $3$ spatial dimensions.
The Fourier transform of the image is denoted ${\fk}\left( k \right)$, where $k$ is the spatial frequency domain coordinate, termed ``k-space'':
 \begin{equation*}\label{eq:Ftransform}
{\fk}\left( k \right) = \int\limits_{ - \infty }^\infty  {\fx\left( x \right){e^{ - j2\pi kx}}dx}.
 \end{equation*}

The MRI tomograph collects a finite set of k-space raw data samples $\{{\fk}\left( {{\kappa _m}} \right)\}, m = 1, \cdots ,M$. The set of sampling points $\left\{ {{\kappa _m}} \right\}$ may be nonuniformly distributed in 2D or 3D k-space.
The vector of samples is denoted by $\mathbf{b}$, with, ${b}\left[ m \right] = {}\langle {{\mathbf{s}_m},\fk} {}\rangle = {{\fk}\left( {{\kappa _m}} \right)}$, where
 \begin{equation}\label{eq:s_m}
    {\mathbf{s}_m}(k) = \delta \left( {k - {\kappa _m}} \right).
 \end{equation}
The sampling subspace is denoted $\mSN  = {\rm{span}}\left\{ \mathbf{s}_m\right\}$.

The field of view (FOV) in the image domain is limited, which implies that the k-space function of the image, ${\fk}\left( k \right)$, is spanned by a set of shifted ${\rm{sinc}}$ functions
   \begin{equation}\label{eq:a_n}
\mathbf{a}_n \left( k \right) = {\rm{sinc}}\left( k / {{\Delta}} - n\right), n \in \mathbb{Z},
 \end{equation}
where ${\rm{FOV}} \buildrel \Delta \over =  1/{{\Delta}}$.
We denote the signal subspace by $\mathcal{A} = {\rm{span}}\left\{ \mathbf{a}_n\right\}$.
We seek a computationally efficient solution to the reconstruction problem: Given a set of nonuniformly spaced k-space samples of an unknown image and the corresponding sampling coordinates, find a good approximation of the function on a Cartesian grid in k-space from which we can subsequently reconstruct an approximation of the image, using the IFFT.

A straightforward approach to reconstruction is to employ \eqn{eq:S_sol} within the framework described above which results in perfect reconstruction of $\fk$. It is easily shown that this solution is equivalent to the URS scheme \cite{rosenfeld1998optimal,rosenfeld2002new} mentioned above. In practical MRI scenarios, this solution requires inverting a huge full matrix of ${\rm{sinc}}$ coefficients which represents ${{\SN ^*}A}$. Storing this matrix on the computer, not to mention calculating its inverse, is intractable due to the sheer size the matrices involved.
Instead, we suggest using an auxiliary subspace and a series of two projections: an oblique projection onto the auxiliary subspace followed by an orthogonal projection onto the signal subspace. The first projection is implemented by solving a sparse system of equations whereas the second is implemented using an LSI filter. The details are presented in the following section.

\section{SParse Uniform ReSampling algorithm}\label{sec:GSURS_alg}
In this section we present the main ideas underlying our reconstruction method as well as the detailed steps performed by the algorithm. We also discuss the resulting approximation error.
\subsection{SPURS reconstruction}
The straightforward reconstruction approach, performed by implementation of ${E_{\mathcal{A}{\mSN ^ \bot }}}$ is computationally prohibitive for typical MRI problems.
Our algorithm, termed SParse Uniform ReSampling (SPURS) trades off reconstruction error for computational complexity, i.e., perfect reconstruction is sacrificed for the sake of efficiency, by relying on the notion of consistency introduced in Section \ref{sec:GS_methods}.

The pivot of the new algorithm is an interim subspace $\mathcal{Q}$ which is designed to enable efficient reconstruction of $\fk$.
We choose $\mathcal{Q}$ as a shift invariant subspace spanned by a compactly supported kernel, designed to be close to the signal subspace $\mathcal{A}$.
The reconstruction process comprises of two projections. The first is an oblique projection onto $\mathcal{Q}$, which recovers a consistent approximation of $\fk$ in $\mathcal{Q}$, denoted $\fk_{\mathcal{Q}}$. Consistency in this context implies that sampling $\fk_{\mathcal{Q}}$ with $S^*$ yields the original samples $\mathbf{b}$. %
The second projection is an orthogonal projection onto $\mathcal{A}$, which recovers the closest signal in $\mathcal{A}$ to the signal $\fk_{\mathcal{Q}}$.
It will be shown that the introduction of the interim subspace $\mathcal{Q}$ is instrumental to achieving low computational complexity while keeping the approximation error at bay. This is achieved by ensuring that ${{\left( {{\SN ^*}Q} \right)}^{\dag}}$ is easy to compute.

We begin by introducing an intermediate subspace $\mathcal{Q} \in \mathcal{H}$ which is spanned by the set $\left\{ {{\mathbf{q}_n}} \right\}$, comprising integer translations of a compactly supported function ${q\left( k \right)}$, i.e.,
 \begin{equation}\label{eq:q_n}
{\mathbf{q}_n}\left( k \right) = {q}\left( k / {{\Delta}} - n\right), n \in \mathbb{Z}.%
 \end{equation}
 We seek a consistent reconstruction of $\fk$ in $\mathcal{Q}$, represented by $\fk_{\mathcal{Q}} = Q\mathbf{c}$ which is given by an oblique projection onto $\mathcal{Q}$ along ${\mSN ^ \bot }$, i.e.,
 \begin{equation*}\label{eq:f_Q}
  \fk_{\mathcal{Q}} = Q\mathbf{c} ={Q{{\left( {{\SN ^*}Q} \right)}^{\dag}}\mathbf{b}} = {Q{{\left( {{\SN ^*}Q} \right)}^{\dag}}{\SN ^*}}\fk = {{E_{\mathcal{Q}{\mSN ^ \bot }}}}{{\fk}}.
 \end{equation*}
Consistency in this context implies that sampling $\fk_{\mathcal{Q}}$ using ${\SN ^*}$ produces the original samples: ${\SN ^*}\fk_{\mathcal{Q}}=\mathbf{b}$.
As we show below, the compact support of ${q\left( k \right)}$ allows for efficient computation of $\fk_{\mathcal{Q}}$. Choosing $\mathcal{Q} = \mathcal{A}$ results in perfect reconstruction \eqn{eq:S_sol}, however, since $\mathcal{A}$ is spanned by a non-compact kernel \eqn{eq:a_n} the computational burden is prohibitive in practical scenarios.

We obtain $\mathbf{c}$ by formulating and solving the equation which relates the nonuniform samples $\mathbf{b}$ to the coefficient vector $\mathbf{c}$:
 \begin{equation}\label{eq:SNQc}
{\mathbf{b}} = {\SN ^*}Q{\mathbf{c}}.
 \end{equation}
Using $\mathbf{c}$, which defines $\fk_{\mathcal{Q}}$, and given the knowledge that $\fk \in {\mathcal{A}}$, we next project $\fk_{\mathcal{Q}}$ onto $\mathcal{A}$. The closest solution in the $L_2$ sense is an orthogonal projection of $\fk_{\mathcal{Q}}$ onto $\mathcal{A}$, denoted $P_{\mathcal{A}}$. %
Due to the fact that both $\mathcal{A}$ and $\mathcal{Q}$ are shift-invariant subspaces ${P_{\mathcal{A}}}\fk_{\mathcal{Q}}$ can be calculated efficiently by employing an LSI filter, as discussed below.

Summarizing , the SPURS reconstruction process comprises a sequence of two projections:
 \begin{equation}\label{eq:GSsystem2}
  {\tfk} = \underbrace {A{{\left( {{A^*}A} \right)}^{\dag}}{A^*}}_{{P_{\mathcal{A}}}}\underbrace {Q{{\left( {{\SN ^*}Q} \right)}^{\dag}}{\SN ^*}}_{{E_{\mathcal{Q}{\mSN ^ \bot }}}}{{\fk}}.
 \end{equation}
A geometrical interpretation of \eqn{eq:GSsystem2} is depicted in \figref{fig:GEO}(b).

Let us split the sequence of operators in \eqn{eq:GSsystem2} into two steps.
First, given the vector of samples $\mathbf{b}$, the vector of coefficients $\mathbf{c}$ is calculated by solving \eqn{eq:SNQc} in the least squares sense:
\begin{equation}\label{eq:cSQb}
\mathbf{c} = {{\left( {{\SN ^*}Q} \right)}^{\dag}}\mathbf{b}.
\end{equation}
Subsequently, the vector $\mathbf{c}$ is used to calculate the coefficients $\mathbf{d}$, given by
\begin{equation}\label{eq:dSASQc}
\mathbf{d} = {{\left( {{{{{A}}}^*}A} \right)}^{\dag}}{\left( {{{{{A}}}^*}Q} \right)}\mathbf{c}.
\end{equation}
Reconstruction is then given by $\tfk = A\mathbf{d}$.
Here $A$, $S$ and $Q$ are the set transforms \eqn{eq:set} corresponding to $\mathbf{a}_n \left( k \right) = {\rm{sinc}}\left( k / {{\Delta}} - n\right)$, ${\mathbf{s}_m}(k) = \delta \left( {k - {\kappa _m}} \right)$ and ${\mathbf{q}_n}\left( k \right) = {q}\left( k / {{\Delta}} - n\right)$.

We next address the practical implementation details of each step, and show how the steps are implemented efficiently.

\subsection{Projection onto the subspace $\mathcal{Q}$}

In order to calculate $\mathbf{c}$ let us first formulate \eqn{eq:SNQc} explicitly for ${\mathbf{s}_m}(k)$ defined in \eqn{eq:s_m}:
\begin{equation}\label{eq:SparseSysMat2}
\begin{array}{l}
{b\left[ m \right] = \sum\limits_{n} {c\left[ n \right]{q}\left( {{\kappa _m} - {k_n}} \right)} },
\end{array}
\end{equation}
where we are given the locations in k-space of the nonuniformly distributed sampling points $\left\{ {{\kappa _m}} \right\}$ as well as the Cartesian reconstruction locations $\left\{ {{k_n}} = \Delta n \right\}$.
Due to the compact support of the function $q$, only a small number of coefficients $c\left[ n \right]$ in \eqn{eq:SparseSysMat2} contribute to the calculation of each value $b\left[ m \right]$. Therefore, \eqn{eq:SparseSysMat2} represents a sparse relation between the coefficient vectors $\mathbf{b}$ and $\mathbf{c}$, which can be expressed by an $M \times N$ sparse matrix $\mathbf{{\Phi}}$, with elements
\begin{equation}\label{eq:SparseSysMat3}
 {\left\{ \mathbf{{\Phi}} \right\}_{m,n}} = {\left\{ {{\SN ^*}Q} \right\}_{m,n}} = {q}\left( {{\kappa _m} - {k_n}} \right),
\end{equation}
where $M$ and $N$ are the number of coefficients in the vectors $\mathbf{b}$ and $\mathbf{c}$, respectively.

In order to find the vector $\mathbf{c}$, we formulate a weighted regularized least squares problem
 \begin{equation}\label{eq:WLS}
\mathbf{c} = \text{arg}\mathop {\min }\limits_{\mathbf{c'}} {\left\| {{\bar \Gamma}}\left( {{\mathbf{b}}-{\mathbf{\Phi c'}}} \right) \right\|^2} + \rho{\left\|{\mathbf{c'}}\right\|^2},
\end{equation}
where ${\rho}>0$ is a Tikhonov regularization parameter\cite{tikhonov1963solution}, ${\mathbf{{\bar \Gamma }}} = \mathbf{{\Gamma}}^{\frac{1}{2}}$ and $\mathbf{{\Gamma}}$ is an $M \times M$ diagonal weighting matrix with weights ${w}_i > 0$.
The regularization is required in order to prevent overfitting and to cope with the possible ill-posedness of the problem which is common in real life situations where the samples are contaminated by measurement noise. The weights, ${w}_i$, may contribute in cases when the noise density varies in k-space and can improve the numerical stability when facing challenging sampling patterns. %

By taking the derivative of \eqn{eq:WLS} we obtain the well known normal equations,
 \begin{equation}\label{eq:WLSNE}
 \left( {{\mathbf{{\Phi}}^T}\mathbf{{\Gamma}}\mathbf{{\Phi}} + {\rho }\mathbf{I}} \right)\mathbf{c} = {\mathbf{{\Phi}}^T}\mathbf{{\Gamma}}\mathbf{b}.
\end{equation}
To solve \eqn{eq:WLSNE} we note that although $\mathbf{{\Phi}}$ is sparse, there is no guarantee regarding the sparsity of ${\mathbf{{\Phi}}^T}\mathbf{{\Gamma}}\mathbf{{\Phi}}$. In fact, it could easily become a full matrix.
A useful sparsity conserving formulation of the normal equations is given by the sparse tableau approach \cite{heath1984numerical} also referred to as the Hachtel augmented matrix method \cite{duff1976comparison}.
We extend this formulation to accommodate for the weights and the regularization. By defining a residual term $\mathbf{r} = \mathbf{{\bar \Gamma }}\left(\mathbf{b} - \mathbf{{\Phi}}\mathbf{c}\right)$ we reformulate \eqn{eq:WLSNE} as
\begin{equation}\label{eq:SparseTableau}
\begin{array}{l}
\left( \begin{array}{l}
\mathbf{{\bar \Gamma }}\mathbf{b}\\
\mathbf{0}
\end{array} \right) =
\left( {\begin{array}{*{20}{c}}
\mathbf{I}&{\mathbf{{\bar \Gamma }}\mathbf{{\Phi}}}\\
{{\mathbf{{\Phi}}^T}{\mathbf{{\bar \Gamma }}^T}}&{ - {\rho}\mathbf{I}}
\end{array}} \right)\left( \begin{array}{l}
\mathbf{r}\\
\mathbf{c}
\end{array} \right) =
\mathbf{\Psi}\left( \begin{array}{l}
\mathbf{r}\\
\mathbf{c}
\end{array} \right).
\end{array}
\end{equation}
In this formulation $\mathbf{\Psi}$ maintains the sparsity of $\mathbf{\Phi}$.

The solution of this system of equations by means of directly inverting $\bf{\Psi }$ is of complexity $O((M \times N)^3)$ and easily becomes computationally prohibitive; so is the amount of computer memory required to store the non-sparse matrix $\bf{\Psi }^{-1}$ which is of order $O((M \times N)^2)$. Moreover, even if $\bf{\Psi }^{-1}$ were known, it would still require $O(\left(M \times N\right)^2)$ operations to compute $\mathbf{c}$ from $\mathbf{b}$ in \eqn{eq:SparseTableau}. Instead, sparse equation solvers are employed to calculate the LU factors of $\bf{\Psi }$. This factorization reduces both the memory requirements and the computational effort employed for the solution to the order of $O\left({\rm{NNZ}}\left(\mathbf{\Psi}\right)\right)$\cite{tinney1967power,tinney1973solution}, where ${\rm{NNZ}}\left(\mathbf{\Psi}\right)$ is the number of non-zero elements in the sparse matrix $\mathbf{\Psi}$ and ${\rm{NNZ}}\left(\mathbf{\Psi}\right) \ll \left(M \times N\right)$ (see Section \ref{ComputationalComplexity}). This process enables a computationally efficient solution of \eqn{eq:SparseTableau}, which gives us $\mathbf{c}$.

In practice \eqn{eq:SparseTableau} is solved in two steps: In the first step, the sparse solver package UMFPACK\cite{davis1997unsymmetric,davis2004algorithm} is used to calculate the $\text{LU}$ factorization of $\mathbf{\Psi}$. In particular, the matrix $\mathbf{\Psi}$ is factored as:
 \begin{equation*}\label{eq:PRQLU}
\mathbf{P\left( {R^{-1} \Psi} \right)Q = LU},
 \end{equation*}
where $\mathbf{P}$ and $\mathbf{Q}$ are permutation matrices, $\mathbf{R}$ is a diagonal scaling matrix which helps achieving a sparser and more stable factorization, and $\mathbf{L}$, $\mathbf{U}$ are lower and upper triangular matrices respectively. For further details refer to \cite{davis2006direct}.
It is important to emphasize that the factorization process is performed offline only once for a given sampling pattern or trajectory defined by the set of sampling locations $\left\{ {{\kappa _m}} \right\}$. The $\mathbf{LU}$ factors maintain the sparsity of $\mathbf{\Psi}$ up to a small amount of zero fill-in, and can be stored for later use with a new sampling data set taken over the same trajectory.

In the second step, given a set of samples $\mathbf{b}$, calculation of $\mathbf{c}$ using $\mathbf{L}$ and $\mathbf{U}$ is done by means of forward substitution and backward elimination, operations which typically achieve a memory usage and computational complexity which is linear in the number of non zero (NNZ) elements of the sparse $\mathbf{L}$, $\mathbf{U}$ matrices. %

\subsection{Calculation of the values of $\fk$ on a Cartesian grid}\label{CalcOnGrid}
Once the vector of coefficients $\mathbf{c}$ is calculated, we proceed to compute the vector $\mathbf{d}$ in \eqn{eq:dSASQc}. Since both $Q$ and $A$ correspond to integer shifts of a kernel function, $\mathcal{Q}$ and $\mathcal{A}$ are SI subspaces and, therefore, \eqn{eq:dSASQc} can be implemented efficiently using an LSI %
filter\cite{eldar2014sampling}:%
\begin{equation}\label{eq:filter2}
{H_{{\rm{LSI}}}}\left( {{e^{j\omega}}} \right) = \frac{{{R_{\mathcal{AQ}}}\left( {{e^{j\omega}}} \right)}}{{{R_{\mathcal{AA}}}\left( {{e^{j\omega}}} \right)}},
 \end{equation}
where
 \begin{equation*}\label{eq:RAQ}
{R_{\mathcal{AQ}}}\left( {{e^{j\omega}}} \right) = \text{DTFT}\left\{ {r_{aq}}\left[ n \right] \right\}=\sum\limits_{n \in \mathbb{Z}} {{r_{aq}}\left[ n \right]{e^{ - j\omega n}}}
 \end{equation*}
is the discrete-time Fourier transform (DTFT) of the sampled correlation sequence
 \begin{equation*}\label{eq:R_aq}
{r_{aq}}\left[ n \right] = {}\langle {a\left(k \right),q\left( {k + n\Delta} \right)} {}\rangle  = \int_{ - \infty }^\infty  {a\left( k \right)q\left( {k + n\Delta} \right)}dk,
 \end{equation*}
resulting in
 \begin{equation}\label{eq:R_AQw}
{R_{\mathcal{AQ}}}\left( {{e^{j\omega }}} \right) = \frac{1}{\Delta }\sum\limits_{n \in \mathbb{Z}} {\overline {A\left( {\frac{\omega }{\Delta } - \frac{{2\pi n}}{\Delta }} \right)} } Q\left( {\frac{\omega }{\Delta } - \frac{{2\pi n}}{\Delta }} \right)\,\,\,.
 \end{equation}
${R_{\mathcal{AA}}}\left( {{e^{j\omega }}} \right)$ is similarly defined.
$Q\left(\omega\right)$ in \eqn{eq:R_AQw} is the continuous-time Fourier transform (CTFT) of ${q}\left(k/\Delta\right)$
\begin{equation*}
Q\left( \omega  \right) = {\text{CTFT}}\left\{ {{q}\left( k/\Delta \right)} \right\} = \int_{ - \infty }^\infty  {q\left( k/\Delta \right){e^{ - j\omega k}}dk},
 \end{equation*}
where $A\left(\omega\right)$ is similarly defined.
Since $\fk$ resides in the spatial frequency domain, the filter \eqn{eq:filter2} is defined in the spatial domain --- the image space,
using the change of variables
 \begin{equation*}
 {\omega  \to \frac{{2\pi x}}{{{\rm{FOV}} }} = {{2\pi \Delta }}{ }x}.
 \end{equation*}
$\mathbf{d}$ is given by
 \begin{equation*}
d\left[ n \right] = \sum\limits_{k \in Z} {c\left[ n \right]{h_{LSI}}\left[ {n - k} \right]},
 \end{equation*}
which, by the convolution property of the DTFT, is equivalent to
 \begin{equation*}
D\left( {{e^{j\omega }}} \right) = {H_{{\rm{LSI}}}}\left( {{e^{j\omega }}} \right)C\left( {{e^{j\omega }}} \right),
 \end{equation*}
where $C\left( {{e^{j\omega }}} \right)$ and $D\left( {{e^{j\omega }}} \right)$ are the DTFTs of $\mathbf{c}$ and $\mathbf{d}$, respectively.
Once $\mathbf{d}$ is calculated, it is used to reconstruct $\tfk$ by
 \begin{equation*}
\tfk \left( k \right) = A\mathbf{d} = \sum\limits_n {d\left[ n \right]{\rm{sinc}}\left( {{k/\Delta - n} } \right)}.
 \end{equation*}
For ${k_{n} = n\Delta}$, $d\left[n\right] = \tfk \left( {{k_n}} \right)$, which is a vector of the function values on a Cartesian grid in k-space.

\begin{figure}
 \centering
  \includegraphics[width=0.5\textwidth]{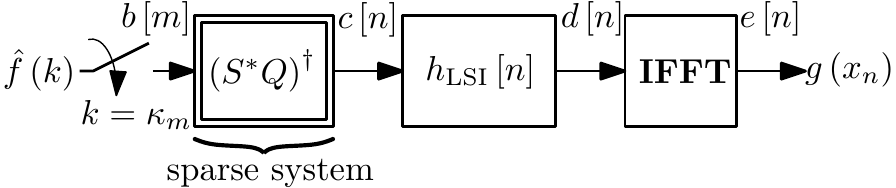} \\
    \caption{SPURS system block diagram with filtering implemented as convolution in k-space.}
    \label{fig:GSsystemPR}
\end{figure}

\begin{figure}
 \centering
  \includegraphics[width=0.5\textwidth]{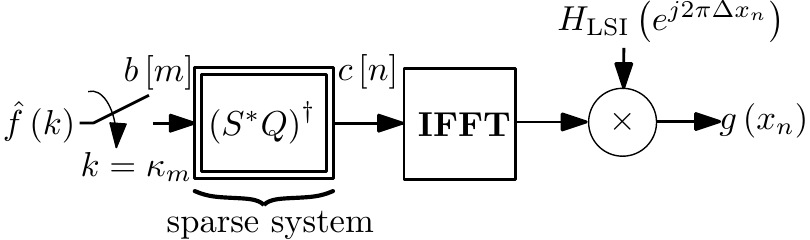} \\
    \caption{SPURS system block diagram with filtering implemented as point-wise multiplication in the image domain.}
    \label{fig:GSsystemPR_im}
\end{figure}

The reconstruction process is next completed by inverse Fourier transforming back into the image domain $\mathbf{e} = {\rm{IFFT}}\left\{ \mathbf{d} \right\} $. The estimate of the uniformly sampled image is then given by
${\tfx \left( {{x_n}} \right)} =  e{\left[ n \right]} $,
where
 \begin{equation}\label{eq:x_n}
{{x_n} = \frac{{\rm{FOV}}}{N}n = \frac{n}{{\Delta N}},\,\,n \in \left[ { - {N}/{2},{N}/{2}} \right) \cap \mathbb{Z}}.
 \end{equation}
The entire reconstruction process is depicted in \figref{fig:GSsystemPR},
where
\begin{equation*}
{h_{{\rm{LSI}}}}\left[ n \right] = {\rm{IDTFT}}\left\{ {{H_{{\rm{LSI}}}}} \right\}
= \frac{1}{{2\pi }}\int_{ - \pi }^\pi  {{H_{{\rm{LSI}}}}\left( {{e^{j\omega }}} \right){e^{j\omega n}}d\omega.}
\end{equation*}
We note that rather than performing the filtering operation of \eqn{eq:filter2} in k-space, we can employ the convolution property of the Fourier transform and implement it as a point-wise multiplication in the image domain following the IFFT, using the values of the filter ${H_{{\rm{LSI}}}}$ at the image grid coordinates $\{x_n\}$, i.e. $\left.{H_{{\rm{LSI}}}}\left( {{e^{j\omega }}} \right)\right| {_{\omega  = 2\pi \Delta {x_n}}} $ as depicted in \figref{fig:GSsystemPR_im}. %

\subsection{SPURS algorithm summary}
To summarize, the SPURS algorithm is divided into two stages; an offline stage which is performed only once for a given sampling trajectory, and an online stage which is repeated for each new set of samples.
\begin{description}
  \item [Phase 1 -- Offline preparation and factorization:] the sparse tableau system matrix $\mathbf{\Psi}$ of \eqn{eq:SparseTableau} is prepared and %
      its LU factorization computed. %
      See Algorithm \ref{algo:SPURSp1}.
  \item [Phase 2 -- Online solution:] The sparse system of equations \eqn{eq:SparseTableau} is solved for a given set of k-space samples $\mathbf{b}$. The result is subsequently filtered using the digital correction filter ${H_{\rm{LSI}}\left( {{e^{j \omega}}} \right)}$ of \eqn{eq:filter2} producing the vector of coefficients $\mathbf{d}$ which represent estimates of the function on the uniform reconstruction grid $\tfk \left(k_n\right)$.
      The result is transformed to the image domain using the IFFT giving $\tfx \left(x_n\right)$.
      See Algorithm \ref{algo:SPURSp3}.
  \item [Algorithm outputs:] $\mathbf{ d}$, ${\tfx \left( {{x_n}} \right)}$.
\end{description}

\begin{algorithm}[ht!]
  \caption{SPURS - Offline preparation and factorization.}
  \label{algo:SPURSp1}
  Input:
  \begin{itemize}
    \item $\left\{ {{\kappa _m},m = 1,...,M} \right\}$: nonuniform sampling grid.%
    \item $\left\{ {{k_n}=n \Delta ,n = 1,...,N} \right\}$: uniform reconstruction grid.%
    \item ${q\left(  \cdot  \right)} $: a compactly supported kernel (e.g. B-spline).
    \item $\mathbf{{\Gamma }}$: a $M \times M$ diagonal weighting matrix, ${\mathbf{{\bar \Gamma }}} = \mathbf{{\Gamma}}^{\frac{1}{2}}$.
    \item ${\rho}>0$: a regularization parameter.
  \end{itemize}

Algorithm:
  \begin{algorithmic}[1]
    \topsep=1.5ex
    \itemsep=1.5ex
    \STATE Construct the sparse $M \times N$ system matrix $ \mathbf{{\Phi}}$, with ${\left\{ \mathbf{{\Phi}} \right\}_{m,n}} = q \left( {{\kappa _m} - {k_n}} \right)$.

    \STATE Construct the $\left(M+N\right) \times \left(M+N\right)$ sparse tableau matrix:
    \[\mathbf{\Psi=\left( {\begin{array}{*{20}{c}}
    \mathbf{I}&{{\mathbf{\bar \Gamma }}{\mathbf{\Phi}}}\\
    {{{\mathbf{\Phi}}^T}{{\mathbf{\bar \Gamma }}^T}}&{ - {\rho}\mathbf{I}}
    \end{array}} \right)}.\]

    \STATE Factorize $\mathbf {\Psi}$ so that $\mathbf{P\left( {R^{-1} \Psi} \right)Q = LU}$.

    \STATE Calculate the LSI filter for use in the online stage:
    \[{H_{\rm{LSI}}\left( {{e^{j \omega}}} \right) = \frac{{{R_{AQ}}\left( {{e^{j \omega}}} \right)}}{{{R_{AA}}\left( {{e^{j \omega}}} \right)}} }.\]
  \end{algorithmic}
  Output: $H_{\rm{LSI}}\left( {{e^{j \omega}}} \right)$ and $\mathbf{L, U, P, Q, R}$.
\end{algorithm}

\begin{algorithm}[ht!]
  \caption{SPURS - Online solution.}
  \label{algo:SPURSp3}
  Input:
  \begin{itemize}
    \item $\mathbf{L}$, $\mathbf{U}$: sparse $\left(M+N\right) \times \left(M+N\right)$ lower and upper diagonal matrices.
    \item $\mathbf{P}$, $\mathbf{Q}$, $\mathbf{R}$: $\left(M+N\right) \times \left(M+N\right)$ permutation and scaling matrices.
    \item $\mathbf{{\Gamma }}$: an $M \times M$ diagonal weighting matrix, ${\mathbf{{\bar \Gamma }}} = \mathbf{{\Gamma}}^{\frac{1}{2}}$.
    \item $\mathbf{\mathbf{b} }$: an $M \times 1$ vector of nonuniformly spaced k-space sample values of ${{\fk}}$, where ${b\left[m\right]} = {\fk}\left( {{\kappa _m}} \right)$.
    \item ${H_{{\rm{LSI}}}\left( {{e^{j \omega}}} \right)}$.
  \end{itemize}
Algorithm:
  \begin{algorithmic}[1]
    \topsep=1.5ex
    \itemsep=1.5ex

    \STATE Construct the $\left(M+N\right) \times 1$ vector $\mathbf{\check{b}} = \left( \begin{array}{l}
    \mathbf{{\bar \Gamma }}\mathbf{b}\\
    \mathbf{0}
    \end{array} \right).$

    \STATE Scale $\mathbf{\check{b}}$ by $\mathbf{R}$ and permute the result using $\mathbf{P}$. \\Store it in the full length vector $\mathbf{y} = \mathbf{P}{\mathbf{R}^{-1}\mathbf{\check{b}}}$.
    \STATE Solve  $\mathbf{L}\mathbf{z} = \mathbf{y}$  and $\mathbf{U}\mathbf{w} = \mathbf{z}$ by forward substitution and backward elimination.
    \STATE Permute  $\mathbf{w}$ by $\mathbf{Q}$ and store the results in the $\left(M+N\right) \times 1$ vector $\mathbf{\check{c}}$, where $\mathbf{{\check{c}} = Qw} = \left( \begin{array}{l}
    \mathbf{r}\\
    \mathbf{c}
    \end{array} \right)$.
    \STATE Filter the $N \times 1$ vector $\mathbf{c}$ using ${H_{\rm{LSI}}\left( {{e^{j \omega}}} \right)}$ and store the results in $\mathbf{d}$.
    \STATE Compute $\mathbf{e} = \textnormal{IFFT}\left\{ \mathbf{d} \right\}$.
  \end{algorithmic}
  Output: $\mathbf{d}$ and the image %
  ${\tfx \left( {{x_n}} \right)} = e{\left[ n \right]}$.
\end{algorithm}

\subsection{SPURS approximation error}
We now analyze the error introduced by the SPURS method and provide geometrical insight for this error using the concept of angles between subspaces.

We begin by defining the reconstruction error introduced by SPURS as
 \begin{equation*}\label{eq:ort_err1}
{\mathlarger{\varepsilon}_{\mathsmaller{{\rm{SPURS}}}}}( \fk ) = \fk - \tfk = \fk - {P_{\cal A}}{E_{{\cal Q}{\cal S}^ \bot}}\fk. %
 \end{equation*}
Since $\fk$ is in $\mathcal{A}$, ${P_{\cal A}}\fk = \fk$ and
 \begin{equation}\label{eq:ort_err2}
{\mathlarger{\varepsilon}_{\mathsmaller{{\rm{SPURS}}}}}( \fk ) = {P_{\cal A}}\left( {I - {E_{{\cal Q}{\cal S}^ \bot}}} \right)\fk = {P_{\cal A}}{E_{{\cal S}^ \bot{\cal Q}}}\fk.
 \end{equation}
Note that when  $\mathcal{Q}=\mathcal{A}$, ${P_{\cal A}}{E_{{\cal A}{\cal S}^ \bot}} = {P_{\cal A}}$ and ${\mathlarger{\varepsilon}_{\mathsmaller{{\rm{SPURS}}}}}( \fk ) = 0$.
We employ the concept of angles between closed subspaces \cite{unser1994general}:
\begin{equation}\label{eq:def_cos}
{{{\cos }}\left( {{{\cal A}},{\cal S}} \right)} =  \mathop {\inf }\limits_{v \in {\cal A},\left\| v \right\| = 1} \left\| {{P_{\cal S}}v} \right\|,
 \end{equation}
\begin{equation}\label{eq:def_sin}
{\sin \left( {{\cal A},{\cal S}} \right) = \mathop {\sup }\limits_{v \in {\cal A},\left\| v \right\| = 1} \left\| {{P_{{{\cal S}^ \bot }}}v} \right\|
.}
 \end{equation}

\begin{proposition}
\label{prop:first}
Let $\mathcal{A}$, $\mathcal{S}$ and $\mathcal{Q}$ be closed subspaces of a Hilbert space $\mathcal{H}$.
Then
\begin{multline}\label{eq:bounded_err3}
{\| {{{\mathlarger{\varepsilon}_{\mathsmaller{{\rm{SPURS}}}}}( \fk )}} \|^2} \le \frac{{{{\sin }^2}\left( {{{\cal A}},{\cal S}} \right)}}{{{{\cos }^2}\left( {{{\cal Q}},{\cal S}} \right)}}{\| {{P_{{{\cal Q}^ \bot }}}\fk} \|^2}\\ \le \frac{{{{\sin }^2}\left( {{{\cal A}},{\cal S}} \right)}{{\sin }^2 \left( {{\cal A}} ,{{\cal Q} } \right)}}{{{{\cos }^2}\left( {{{\cal Q}},{\cal S}} \right)}}{\| {\fk} \|^2}.
\end{multline}
\end{proposition}
 \begin{IEEEproof}
The error in \eqn{eq:ort_err2} can be bounded by defining $v = {E_{{{\cal S}}^ \bot{\cal Q}}}\fk/\| {{E_{{{\cal S}}^ \bot{\cal Q}}}}\fk \|$ which is a normalized vector in ${\cal S}^ \bot$.
Orthogonally projecting $v$ onto $\mathcal{A}$ and using definition \eqn{eq:def_sin} we get
\begin{equation}\label{eq:bound}
{\| {{P_{\cal A}}v} \|^2} \le \sin^2 \left( {{\cal S}}^ \bot ,{{\cal A}^ \bot } \right).
\end{equation}
Plugging \eqn{eq:bound} into \eqn{eq:ort_err2} we obtain
\begin{equation}\label{eq:bounded_err}
{\| {{{\mathlarger{\varepsilon}_{\mathsmaller{{\rm{SPURS}}}}}}( \fk )} \|^2} \le
\sin^2 \left( {{\cal S}{}^ \bot ,{{\cal A}^ \bot }} \right){\| {{E_{{\cal S}{}^ \bot {\cal Q}}}\fk} \|^2}.
 \end{equation}
Using the definition in \eqn{eq:def_cos},
 \begin{equation*}\label{eq:cosy}
\cos \left( {{\cal S}^\bot,{\cal Q}^\bot} \right) \le \frac{{\| {{P_{{\cal Q}^\bot}}u} \|}}{{\| u \|}},\,\,\forall u \in {\cal S}^\bot \backslash \left\{ 0 \right\}.
 \end{equation*}
  Choosing $u = {{E_{{\cal S}{}^ \bot {\cal Q}}}\fk}$ leads to
   \begin{equation}\label{eq:ort_bound}
 \| {{E_{{\cal S}{}^ \bot {\cal Q}}}\fk} \| \le \frac{{\| {{P_{{\cal Q}^\bot}}{{E_{{\cal S}{}^ \bot {\cal Q}}}\fk}} \|}}{{\cos \left( {{\cal S}^\bot,{\cal Q}^\bot} \right)}}.
 \end{equation}
Substituting the algebraic expressions for ${P_{{\cal Q}^\bot}}$ and ${{E_{{\cal S}{}^ \bot {\cal Q}}}}$
we can immediately verify that ${P_{{{\cal Q}^\bot}}}{{E_{{\cal S}{}^ \bot {\cal Q}}}} = {P_{{\cal Q}^\bot}}$.
Using definition \eqn{eq:def_sin},
\begin{equation*}\label{eq:bound2}
\sin \left( {{\cal A},{\cal Q}} \right) \ge \frac{{\| {P_{{\cal Q}^\bot}\fk} \|}}{{\| \fk \|}},\,\,\forall \fk \in {\cal A} \backslash \left\{ 0 \right\}.
\end{equation*}
Substituting these results into \eqn{eq:ort_bound} we obtain,
 \begin{equation}\label{eq:bounded_obl1}
\| {{E_{{\cal S}{}^ \bot {\cal Q}}}\fk} \| \le \frac{{\| {{P_{{{\cal Q}^ \bot }}}\fk} \|}}{{\cos \left( {{\cal S}{}^ \bot ,{{\cal Q}^ \bot }} \right)}}
\le \frac{{{\sin } \left( {{\cal A}} ,{{\cal Q} } \right)}}{{\cos \left( {{\cal S}{}^ \bot ,{{\cal Q}^ \bot }} \right)}}{\| \fk \|}
.
\end{equation}
Plugging \eqn{eq:bounded_obl1} into \eqn{eq:bounded_err} and using the relations\cite{unser1994general}
\begin{equation*}\label{eq:def_id_cos}
\centering
\begin{array}{l}
\cos \left( {{\cal Q},{\cal S}} \right) = \cos \left( {{\cal S^{\bot}},{\cal Q^{\bot}}} \right)\\
\sin \left( {{\cal A},{\cal S}} \right) = \sin \left( {{\cal S^{\bot}},{\cal A^{\bot}}} \right),%
\end{array}
 \end{equation*}
completes the proof.
\end{IEEEproof}
Proposition \ref{prop:first} implies that the reconstruction error is controlled by the angle between the sampling subspace $\mathcal{S}$ and signal subspace $\mathcal{A}$, by the angle between the sampling subspace $\mathcal{S}$ and the auxiliary subspace $\mathcal{Q}$ as well as by ${\| {{P_{{{\cal Q}^ \bot }}}\fk} \|^2}$, which is the energy of the signal $\fk$ which does not reside in $\mathcal{Q}$.
For the MRI problem formulated, we only have control over $\mathcal{Q}$, whereas $\mathcal{A}$ (determined by the FOV) and $\mathcal{S}$ (determined by the sampling pattern) are predefined. Therefore, it would be preferable to maximize ${{{{\cos }^2}\left( {{{\cal Q}},{\cal S}} \right)}}$ and minimize ${\| {{P_{{{\cal Q}^ \bot }}}\fk} \|^2}$ over $\mathcal{Q}$ for a given computational budget\footnote{Otherwise, we would choose $\mathcal{Q} = \mathcal{A}$ and achieve perfect reconstruction at a high computational cost.}. This is done by an appropriate selection of the kernel function ${q}\left(k\right)$ which spans $\mathcal{Q}$ (Section \ref{par:KernelFunc}) and by employing dense grid interpolation as described in Section \ref{par:DenseGridInt}.

\section{Extensions of SPURS}\label{sec:Extensions}
\subsection{Selection of the kernel function spanning $\cal Q$}\label{par:KernelFunc}
The selection of the function ${q}\left(k\right)$ which spans $\mathcal{Q}$, both the function itself and its support, has a considerable effect on the quality of the reconstructed image. In this work we use basis splines (B-splines\cite{schonberg1946contributions}) which have gained popularity in signal processing applications\cite{unser1999splines}. They are commonly used in image processing because of their ability to represent efficiently smooth signals and the low computational complexity needed for their evaluation at arbitrary locations.
A B-spline of degree $p$ is a piecewise polynomial with the
pieces combined at knots, such that the function is continuously differentiable $p-1$ times. A B-spline of degree $p$, denoted $\beta ^p$, is the function obtained by the $(p+1)$-fold convolution of of a rectangular pulse ${\beta ^0}\left( k \right)$:
\begin{equation*}\label{eq:B-spline_p}
{\beta ^p}\left( k \right) = \underbrace{{\beta ^0}\left( k \right) *  \ldots  * {\beta ^0}\left( k \right)}_{(p+1)\, {\rm{times}}},
\end{equation*}
with a support of $p+1$, where
\begin{equation*}\label{eq:B-spline_0}
{\beta ^0}\left( k \right) = \left\{ \begin{array}{l}
1,\,\, - {\textstyle{\frac{1}{2}}} < k < {\textstyle{\frac{1}{2}}}\\
{\textstyle{\frac{1}{2}}},\,\,\left| k \right| = {\textstyle{\frac{1}{2}}}\\
0,\,\,{\rm{otherwise}}.
\end{array} \right.
\end{equation*}

Increasing the degree of the spline increases its order of approximation and improves the image quality at the expense of increased computational burden, as a result of the larger support.
It can be shown\cite{unser1999splines} that as the order of the spline increases, the subspace $\mathcal{Q}$ tends to $\mathcal{A}$, subsequently, decreasing ${\| {{P_{{{\cal Q}^ \bot }}}\fk} \|}$ in \eqn{eq:bounded_err3}.
Similarly, it is well known that the selection of the kernel function has a significant influence on the performance of both NUFFT \cite{fessler2003nonuniform} and convolutional gridding \cite{jackson1991selection,sedarat2000optimality}.

For ${q}\left(k\right) = {\beta ^p}\left( k \right)$ and ${a}\left(k\right) = {\rm{sinc}}\left( k \right)$ the LSI filter \eqn{eq:filter2} can be expressed explicitly by
evaluating %
\begin{equation*}\label{eq:Spline_freq}
Q\left( \omega  \right){\rm{ = CTFT}}\left\{ {{\beta ^p}\left( {\frac{k}{\Delta }} \right)} \right\} %
= \Delta {{\rm{sinc}}^{p + 1}}\left( {\Delta \frac{\omega }{{2\pi }}} \right),
\end{equation*}
\begin{equation*}
A\left( \omega  \right){\rm{ = CTFT}}\left\{ {{\rm{sinc}}\left( {\frac{k}{\Delta }} \right)} \right\} = \Delta {\rm{rect}}\left( {\Delta \frac{\omega }{{2\pi }}} \right),
\end{equation*}
where,
\begin{equation*}\label{eq:sinc_freq}
{{\rm{rect}}(\xi ) = \left\{ {\begin{array}{*{20}{l}}
{1,{\mkern 1mu} {\mkern 1mu} \left| \xi  \right| < \frac{1}{2}}\\
{0,{\mkern 1mu} {\mkern 1mu} \left| \xi  \right| \ge \frac{1}{2}}
\end{array}} \right.,\,\,\,{\rm{sinc}}(\xi ) = \left\{ {\begin{array}{*{20}{l}}
{\frac{{\sin \left( {\pi \xi } \right)}}{{\pi \xi }},{\mkern 1mu} {\mkern 1mu} \xi  \ne 0}\\
{1,{\mkern 1mu} \,\,\,\,\,\,\,\,\,\,\,\,\,\,{\mkern 1mu} \xi  = 0}
\end{array}} \right.}.
\end{equation*}
Plugging $Q\left( \omega  \right)$ and $A\left( \omega  \right)$ into \eqn{eq:R_AQw} results in
 \begin{equation}\label{eq:filterHort}
{H_{{\rm{LSI}}}}\left( {{e^{j\omega }}} \right) = \frac{{\sum\limits_{n \in \mathbb{Z}} {{\rm{rect}}\left( {\frac{{{} \omega }}{{2\pi }} - {} n} \right){{\rm{sinc}}^{p + 1}}\left( {\frac{{{} \omega }}{{2\pi }} - {} n} \right)} }}{{\sum\limits_{n \in \mathbb{Z}} {{\rm{rect}}\left( {\frac{{{} \omega }}{{2\pi }} - {} n} \right)} }}.
 \end{equation}

The reconstruction result in the image domain is calculated on the Cartesian grid ${\{x_n}\}$, as defined in \eqn{eq:x_n}. When performing the filtering operation in the image domain, as illustrated in \figref{fig:GSsystemPR_im}, ${\{\tfx\left(x_n\right)}\}$ is obtained by multiplying the IFFT result of $\mathbf{c}$ with the values of ${H_{{\rm{LSI}}}}\left( {{e^{j\omega }}} \right)$ at locations ${\omega_n = {{2\pi \Delta }}x_n}$.
Since all ${\{x_n}\}$ as defined in \eqn{eq:x_n} are within the FOV reduces \eqn{eq:filterHort} to
 \begin{equation*}
{H_{{\rm{LSI}}}}\left( {{e^{j 2 \pi \Delta x_n }}} \right) = {\rm{sin}}{{\rm{c}}^{p + 1}}\left( {\frac{n}{N}} \right),\,\,n \in \left[ { - {N}/{2},{N}/{2}} \right) \cap \mathbb{Z}.
 \end{equation*}

The sparse matrix $\mathbf{\Phi}$ defined in \eqn{eq:SparseSysMat3}, is given by
${\left\{ \mathbf{{\Phi}} \right\}_{m,n}} = {\beta ^p}\left( {{\kappa _m} - {k_n}} \right)$,
for a given set of sampling and reconstruction coordinates. The number of non-zero elements in the matrix $\mathbf{{\Phi}}$ is a function of the support of the kernel function and the number of samples $M$. For the 2D case this amounts to
\begin{equation}\label{eq:SparseSysMatPhiNNZ}
{\rm{NNZ}}\left( \mathbf{{\Phi}}\right) \simeq \mathnormal{M} \pi \left(\mathnormal{p}+1\right)^2.
\end{equation}

\subsection{Dense grid interpolation}\label{par:DenseGridInt}
We have seen above that the intermediate subspace $\mathcal{Q}$ introduces an error into the reconstruction process.
The approximation error can be reduced by resampling onto a denser uniform grid in k-space. This is done by scaling ${{\Delta}}$ in \eqn{eq:q_n} by an oversampling factor $\oversamplingfactor > 1$, i.e., ${\Delta} \rightarrow {\Delta}/\oversamplingfactor$. The oversampling increases $N$, the total number of cartesian reconstruction points in k-space, as well as the field of view reconstructed in the image domain, by a factor of $\oversamplingfactor$ for each dimension of the problem, i.e., $N \rightarrow N \oversamplingfactor^{\text{dim}}$, where $\text{{dim}}$ is the problem dimension.
Increasing $\oversamplingfactor$ reduces the approximation error with a penalty of increasing the computational load.
From a geometric viewpoint increasing the density in the reconstruction subspace $\mathcal{Q}$, spanned by $\{{q}\left( k / {{\Delta}} - n\right)\}$, causes the subspace to become larger and consequently closer to $\mathcal{A}$ and to $\mathcal{S}$ thereby decreasing the approximation error \eqn{eq:bounded_err3}.

For $\oversamplingfactor > 1$,
the reconstruction filter \eqn{eq:filter2} needs to be modified accordingly:
\begin{equation*}\label{eq:filter2_old}
H_{\rm{LSI}}\left( {{e^{j \omega}}} \right) = \left\{ \begin{array}{l}
\frac{{{R_{{{{\mathcal{A}}}}\mathcal{Q}}}\left( {{e^{j \omega}}} \right)}}{{{R_{{{{\mathcal{A}}}}A}}\left( {{e^{j \omega}}} \right)}},\,{{{R_{{{{\mathcal{A}}}}A}}\left( {{e^{j \omega}}} \right)}} \ne 0\\
0,\,\,\,\,\,\,\,\,\,\,\,\,\,\,\,\,\,\,\,\,\,\,{{{R_{{{{\mathcal{A}}}}A}}\left( {{e^{j \omega}}} \right)}} = 0
\end{array} \right.,
 \end{equation*}
with
\begin{equation*}
\frac{{{R_{{\cal A}{\cal Q}}}\left( {{e^{j\omega }}} \right)}}{{{R_{{\cal A}{\cal A}}}\left( {{e^{j\omega }}} \right)}} =
\frac{{\sum\limits_{n \in \mathbb{Z}} {{\rm{rect}}\left( {\frac{{\oversamplingfactor \omega }}{{2\pi }} - \oversamplingfactor n} \right){{\rm{sinc}}^{p + 1}}\left( {\frac{{\oversamplingfactor \omega }}{{2\pi }} - \oversamplingfactor n} \right)} }}{{\sum\limits_{n \in \mathbb{Z}} {{\rm{rect}}\left( {\frac{{\oversamplingfactor \omega }}{{2\pi }} - \oversamplingfactor n} \right)} }},
 \end{equation*}
where the image domain region beyond the original FOV is set to $0$.

It should be noted that both convolutional gridding and NUFFT employ an oversampling factor $\oversamplingfactor$  to improve performance at the expense of increased computational complexity\cite{beatty2005rapid}. In most cases it was found sufficient to use an oversampling factor of $\oversamplingfactor=2$.

\subsection{Iterating SPURS}
Another way to improve the reconstruction results is to use a simple iterative scheme.
In a single iteration of SPURS we obtain $\mathbf{d}$ which is a vector of coefficients from which we can reconstruct the continuous function $\tfk\left(k\right) = A\mathbf{d}$. %
By operating with the sampling operator ${\SN ^*}$ to resample the reconstructed function $\tfk\left(k\right)$ on the nonuniform grid we obtain $\tilde{\mathbf{b}} = {\SN ^*}{\tfk}$, which approximates the original set of samples $\mathbf{b} = {\SN ^*}{\fk}$.
We define an error vector $\pmb{\mathbf{\varepsilon}} = \tilde{\mathbf{b}} - \mathbf{b}$. %
Achieving $\pmb{\mathbf{\varepsilon}}=\mathbf{0}$ means that a function $\tfk \in \mathcal{A}$ has been found which is consistent with the given vector of samples $\mathbf{b}$.

In \cite{aldroubi1998exact} it was proven that for a function $\fk$, known to belong to a class of spline-like spaces\footnote{Bandlimited functions are a limiting case for this class.} $\mathcal{A}$, the exact reconstruction of $\fk$ from its samples $b\left[m\right] = {{\fk}}\left( {{\kappa _m}} \right)$ can be achieved, provided that the sampling set ${\left\{ {{\kappa _m}} \right\}}$ is ``sufficiently dense"\cite{aldroubi2000beurling,aldroubi2001nonuniform}.
A reconstruction process was proposed and proven to converge to $\fk$ by iteratively operating with an interpolator and a bounded projector onto the spline-like space $\mathcal{A}$. It was noted that the interpolator can be generalized to any set $\{q_n = {q}\left( k - n\right)\}$ which forms a bounded uniform partition of unity, i.e. $\sum\nolimits_n {q\left( {k - n} \right)}  = 1$. %
The convergence of the result to $\fk$ was proven among others in the $L^p$-norm and in the sup-norm which implies uniform convergence.
In this section we utilize SPURS to employ a fast iterative algorithm which fits into the framework proposed in \cite{aldroubi1998exact}.
By iteratively operating with ${{E_{\mathcal{Q}{\mSN ^ \bot }}}}$ and ${P_{\mathcal{A}}}$ from \eqn{eq:GSsystem2}, and as long as the sampling set is dense enough, $\tfk$ converges to $\fk$.

The first step of the algorithm, operates on the vector of samples $\mathbf{b}$ with the operator ${{{{G}}}} = {\left( {A^*A} \right)^\dag }A^*Q{\left( {\SNs Q} \right)^\dag }$ to produce the vector of coefficients $\mathbf{d}$. This first iteration is designated ${\mathbf{d_0}} = {{{{G}}}}{\mathbf{b_0}} = {{{{G}}}}{\mathbf{b}}$ (i.e. ${\mathbf{b_0}} = {\mathbf{b}}$) which is performed by \eqn{eq:cSQb} and \eqn{eq:dSASQc}.
Using ${\mathbf{d_0}}$ we evaluate the function $\tfk_0 = A\mathbf{d_0}$ which is the first approximation of $\fk$. Let us define the continuous error function
\begin{equation}\label{eq:error_func}
\varepsilon_p = \fk - \tfk_p,\,\,\,\,\varepsilon_p \in \mathcal{A}
 \end{equation}
which can be evaluated on the sampling points for each iteration $p$
\begin{equation*}\label{eq:It_epsilon}
\varepsilon_p\left[m\right] = \varepsilon_p\left( {{\kappa _m}} \right) = f\left( {{\kappa _m}} \right) - \tfk_p\left( {{\kappa _m}} \right) = \left\{\mathbf{b} - {\SNs }A\mathbf{d_p}\right\}_m.
 \end{equation*}

\begin{figure}
 \centering
  \includegraphics[width=0.45\textwidth]{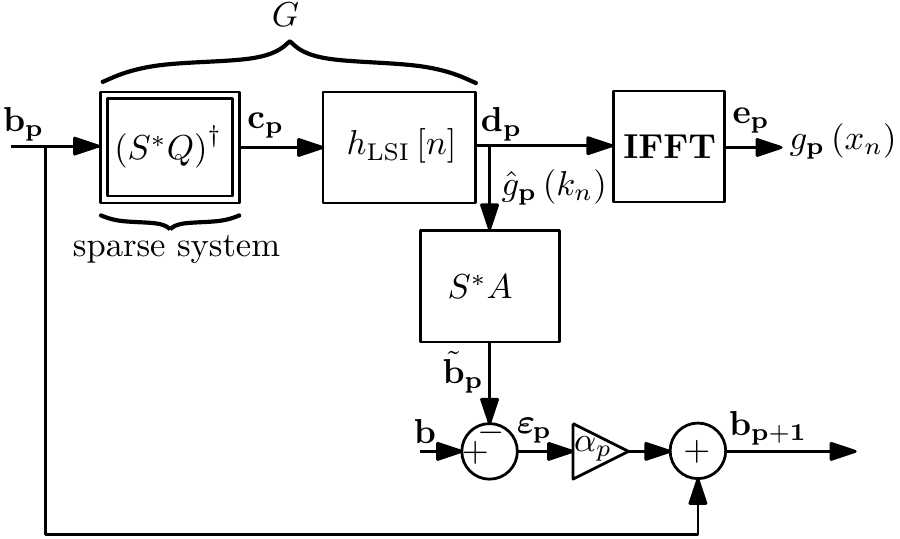} \\
      \caption{Iterative SPURS algorithm block diagram.}
    \label{fig:GSsystemPRIt}
\end{figure}

We now proceed to the second iteration. Using the error vector ${\pmb{\mathbf{\varepsilon}}\mathbf{_0}}$, the new measurement vector $\mathbf{b_1} = \mathbf{b_0} + \alpha{\pmb{\mathbf{\varepsilon}}\mathbf{_0}}$ is calculated, where $\alpha$ controls the iteration step size, and $\mathbf{d_1} = {{G}}\mathbf{b_1}$. Continuing the iterations leads to
\begin{equation*}\label{eq:It_SPURS}
\left. \begin{array}{l}
{\mathbf{b_0}} = {\mathbf{b}}\\
\mathbf{d_{p}} = {{{{G}}}}{\mathbf{b_{p}}} \\
{\mathbf{b_{p+1}}} = \mathbf{b_p} + {\alpha_{p}}\left( \underbrace{\mathbf{b} - \overbrace{\SNs A{\mathbf{d_p}}}^{\tilde{\mathbf{b}}\mathbf{_p}}}_{{\pmb{\mathbf{\varepsilon}}\mathbf{_p}}} \right) .
\end{array} \right.
 \end{equation*}
The complete iterative process is depicted in \figref{fig:GSsystemPRIt}.

According to \cite{aldroubi1998exact}, a sufficient condition for convergence of $\tfk _p$ to $\fk$ is that the sampling set ${\left\{ {{\kappa _m}} \right\}}$ is ${\gamma _0}$-dense\footnote{
A set $\left\{ {{\kappa _m}} \right\}$ is ${\gamma _0}$-dense in ${\mathbb{R}^d}$ if ${\mathbb{R}^d} = \mathop  \cup \limits_m {B_\gamma }\left( {{\kappa _m}} \right),\,\,\forall \gamma  > {\gamma _0}$, where ${B_\gamma }$ is a ball of radius $\gamma$ with center ${{\kappa _m}}$}, which implies that the maximal distance between a sampling point and its nearest neighbor is $2{\gamma _0}$.
Moreover, for $\gamma_0$ sufficiently small, $\left\| {{{\varepsilon}}{_{p+1}}} \right\| \le \eta \left\| {{{\varepsilon}}{_p}} \right\|$ where $\eta<1$, therefore $\left\| {{{\varepsilon}}{_p}} \right\| \rightarrow 0$.
From \eqn{eq:error_func}, $\left\|{{{{\varepsilon}}_p}}\right\| \rightarrow 0$ is equivalent to $\tfk_p\left( {{k}} \right) \rightarrow \fk\left( {{k}} \right)$ for all $k$ as $p \rightarrow \infty$.
The contraction factor $\eta$ is a decreasing function of the density, which means that the algorithm converges more rapidly for denser sets.

The scalar ${\alpha_{p}}$ controls the iteration step size. For a constant $\alpha _p$ the convergence rate might be slow. In order to improve the convergence rate $\alpha _p$ may be chosen at each iteration such that the norm of the error $\left\|{\pmb{\mathbf{\varepsilon}}\mathbf{_p}}\right\|$ is minimized, where ${\pmb{\mathbf{\varepsilon}}\mathbf{_p}} = {\mathbf{b} - \SNs A{\mathbf{d_p}}}$. The error progression between iterations is ${\pmb{\mathbf{\varepsilon}}\mathbf{_{p+1}}}  = \left( {I - {\alpha _p}\SNs A{{{{G}}}}} \right){\pmb{\mathbf{\varepsilon}}\mathbf{_p}}$, leading to an optimal step
  \begin{equation*}\label{eq:calc_alpha}
{\alpha _p} = \arg \mathop {\min }\limits_\alpha  {\left\| {\left( {I - \alpha \SNs A{{{{G}}}}} \right){\pmb{\mathbf{\varepsilon}}\mathbf{_p}}} \right\|^2} = \frac{{{{\left( {\pmb{\mathbf{\varepsilon}}\mathbf{_p}} \right)}^ * }\SNs A{{{{G}}}}{\pmb{\mathbf{\varepsilon}}\mathbf{_p}}}}{{{{\left\| {\SNs A{{{{G}}}}{{\pmb{\mathbf{\varepsilon}}\mathbf{_p}}}} \right\|}^2}}}.
 \end{equation*}

In our simulations, presented in Section \ref{sec:Simulations}, we evaluated the performance of SPURS both as a direct method and as an iterative method.

\subsection{Computational Complexity}\label{ComputationalComplexity}
 The computational complexity of SPURS is proportional to the number of non-zeros (NNZ) in the $\mathbf{LU}$ factors of $\mathbf{\Psi}$ which is constructed according to \eqn{eq:SparseTableau} from $\mathbf{{\Phi}}$, $\mathbf{{\Gamma }}$ and ${\rho}$. The number of non-zeros in the $\left(M+N\oversamplingfactor\right) \times \left(M+N\oversamplingfactor\right)$ matrix $\mathbf{\Psi}$ is twice the NNZ of $\mathbf{{\Phi}}$ as given in \eqn{eq:SparseSysMatPhiNNZ} for the case of 2D imaging and a B-spline kernel, plus an additional $M + N \oversamplingfactor^{\text{dim}}$ non-zeros on the main diagonal. This amounts to
 \begin{equation*}
  \text{NNZ}\left( \mathbf{\Psi}_{\rm{2D}} \right) \simeq M \pi {\left( { {\left(p+1\right)}\oversamplingfactor/2 } \right)^{\text{2}}} + M + N\oversamplingfactor^{\text{2}},
 \end{equation*}
 where for the more general case,
 \begin{equation}\label{eq:NNZ_Psy}
  \text{NNZ}\left( \mathbf{\Psi} \right) \simeq M \pi_{\rm{dim}} {\left(  {\supp{\left(q\right)}\oversamplingfactor/2 } \right)^{\text{dim}}} + M + N\oversamplingfactor^{\text{dim}},
 \end{equation}
where $M$ and ${N\oversamplingfactor^{\text{dim}}}$ are the number of non-Cartesian and Cartesian grid points, respectively, $\text{{dim}}$ is the problem dimension, $\pi_{\rm{dim}}$ is a constant that depends on the dimension (for ${\rm{dim}}=2$, $\pi_{\rm{dim}} = \pi$, for ${\rm{dim}}=3$, $\pi_{\rm{dim}} = 4\pi /3$ etc.), $\oversamplingfactor$ is the oversampling factor and ${\supp{\left(q\right)}}$ is the support of the kernel function $q$, e.g., for B-splines of degree $p$, ${\supp{\left(\beta^p\right)}} = p+1$.

In practice, when $\mathbf{\Psi}$ is sparse, its $\mathbf{L}$ and $\mathbf{U}$ factors which are used to recover $\mathbf{c}$ preserve a similar degree of sparsity, with a certain increase in $\text{NNZ}$ termed ``fill-in''.
The computational complexity of the forward and backward substitution stage is $O\left( {\text{NNZ}\left( \mathbf{L} + \mathbf{U} \right)} \right)$ which, despite the fill-in, is of the same order of magnitude as $\text{NNZ}\left( \mathbf{\Psi}\right)$ given in \eqn{eq:NNZ_Psy}.
Assuming that the filtering stage is performed in the image domain, it adds a complexity of $O\left( {{N}} \right)$, whereas the IFFT stage adds a complexity of $O\left( {{{N}\oversamplingfactor^{{\rm{dim}}}}\log \left( {N^{1/{\rm{dim}}}\oversamplingfactor} \right) }\right)$ for an image with $N^{1/{\rm{dim}}}$ pixels in each dimension. Thus, the online solution phase of SPURS has computational complexity
 \begin{equation}\label{eq:SPURS_CC}
      O\left( {M{{\left( {\supp{\left(q\right)}}\oversamplingfactor \right)}^{\text{dim}}}} + {{{N}\oversamplingfactor^{{{\rm{dim}}}}}\log \left( {N^{1/{\rm{dim}}}\oversamplingfactor} \right) } \right).
 \end{equation}
It can be shown that \eqn{eq:SPURS_CC} is comparable to that of convolutional gridding or of a single iteration of the NUFFT.

In the iterative scheme, an additional stage of calculating ${\tilde{\mathbf{b}}\mathbf{_p}} = \SNs A{\mathbf{d_p}}$ is performed. This adds $O\left(M \times N\oversamplingfactor^{\text{dim}}\right)$ operations to each iteration. Therefore, iterative SPURS has computational complexity
 \begin{equation}\label{eq:SPURS_CC_It}
      O\left( {M{{\left( {\supp{\left(q\right)}}\oversamplingfactor \right)}^{\text{dim}}}} + {{{N}\oversamplingfactor^{{\rm{dim}}}}\left(M+\log \left( {N^{1/{\rm{dim}}}\oversamplingfactor} \right) \right)} \right)
 \end{equation}
per iteration.
In practical situations, $N$ and $M$ are of a similar order of magnitude, therefore, the leading term in \eqn{eq:SPURS_CC_It} is $O\left(MN\oversamplingfactor^{\text{dim}}\right)$. It is noteworthy to compare this to the leading term of \eqn{eq:SPURS_CC}, $O\left( {{{N}\oversamplingfactor^{{\rm{dim}}}}\log \left( {N^{1/{\rm{dim}}}\oversamplingfactor} \right) }\right)$, which is considerably smaller. The latter is also the leading term in the complexity of convolutional gridding, rBURS or a single iteration of NUFFT. Therefore, the improved performance exhibited by additional iterations of SPURS comes with a certain penalty in terms of the computational burden as compared with a similar number of iterations of NUFFT.

\section{Numerical simulation}\label{sec:Simulations}
In this section we perform image reconstruction from numerically generated k-space samples of analytical phantoms and compare the performance of SPURS to that of other methods.
The numerical experiments are implemented in Matlab (The code is available online at \cite{SPURS2016code}). %
Computer simulations are used to compare the performance of SPURS with that of convolutional gridding \cite{jackson1991selection}, rBURS\cite{rosenfeld1998optimal,rosenfeld2002new}, and the (inverse) NUFFT method \cite{dutt1993fast} as implemented by the NFFT package \cite{keiner2009using,guerquin2012realistic} specifically using the application provided for MRI reconstruction \cite{knopp2007note}.

The NUFFT uses a Kaiser-Bessel window with cut-off parameter $m=6$ (i.e., ${\rm{support}} = 12$), Voronoi weights for density compensation, and oversampling factor of $\oversamplingfactor = 2$. Convolutional gridding uses the same parameters and is simply implemented as a single iteration of NUFFT.
In rBURS $\delta \kappa = 1.2$, $\Delta k = 3$ are used, with two values of oversampling ($\oversamplingfactor = 1,2$, denoted rBURS and rBURSx2 respectively).
Unless specified otherwise the SPURS kernel used is a B-spline of degree $3$ (i.e., ${\rm{support}} = 4$) with an oversampling factor of $\oversamplingfactor = 2$. %
The results for SPURS are presented for a single iteration and for the iterative scheme.

The simulation employs a %
 realistic analytical MRI brain phantom \cite{guerquin2012realistic}, %
 of dimensions ${256}\times{256}$, i.e., $N=65536$.

Two types of sampling trajectories are demonstrated: radial and spiral.
  The k-space coordinates for the radial trajectory are given by:
 \begin{equation}\label{eq:Radial_traj}
{\left( {{\kappa _x},{\kappa _y}} \right)_{r,s}} = N\left( {\frac{r}{N_{\rm{bins}}} - 0.5} \right)\left( {\cos {\omega _s},\sin {\omega _s}} \right),
  \end{equation}
  with
   \begin{equation*}
{\omega _s} = \frac{{\pi s}}{{{N_{{\rm{spokes}}}}}}.
  \end{equation*}
Here, ${N_{{\rm{spokes}}}}$ denotes the number of radial spokes, with $s = 0, \ldots ,{N_{{\rm{spokes}}}} - 1$;
  $N_{\rm{bins}}$ is the number of sampling points along each spoke, with $r =  {0, \ldots ,N_{\rm{bins}} - 1}$.
Thus,  $M = {N_{{\rm{spokes}}}} \times {N_{\rm{bins}}}$.

The spiral trajectory comprises a single arm Archimedean constant-velocity spiral with $M$ sampling points along the trajectory, and k-space coordinates given by:
 \begin{equation}\label{eq:Spiral_traj}
{\left( {{\kappa _x},{\kappa _y}} \right)_j} = {\frac{N}{2}} \sqrt {\frac{{j}}{{M}}} \left( {\cos {\omega _j},\sin {\omega _j}} \right),\,j = 0, \ldots ,M - 1,
  \end{equation}
where, %
${\omega _j} = 2\pi\sqrt {{{j}}/{{\pi}}}$\
ensures that the k-space sampling density is approximately uniform.

White Gaussian noise (WGN) is added to the samples to achieve a desired input signal to noise ratio (ISNR).
For each experiment the SNR of the reconstructed image %
is calculated with respect to the true phantom image. The SNR measure assesses the pixel difference between the true and the reconstructed phantom image,
and is defined by
 \begin{equation*}
{\rm{SNR}}\left( {\tfx\left( {{x_n}} \right),\fx\left( {{x_n}} \right)} \right) = 10\log \frac{{\frac{1}{N}\sum\limits_{n = 1}^N {{{\left( {\fx\left( {{x_n}} \right)} \right)}^2}} }}{{\frac{1}{N}\sum\limits_{n = 1}^N {{{\left( {\tfx\left( {{x_n}} \right) - \fx\left( {{x_n}} \right)} \right)}^2}} }}\left[ {dB} \right],
  \end{equation*}
where $\fx\left( x_n \right)$ are the pixel values of the original image and $\tfx\left( x_n \right)$ of the reconstructed image.
The SNR measure does not take into account structure in the image, and along with other traditional methods such as PSNR and mean squared error (MSE) have proven to be inconsistent with the human visual system (HVS). The Structural Similarity (SSIM) index\cite{wang2004image} was designed to improve on those metrics. SSIM provides a measure of the structural similarity between the ground truth and the estimated images by assessing the visual impact of three characteristics of an image: luminance, contrast and structure.
For each pixel in the image, the SSIM index is calculated using surrounding pixels enclosed in a Gaussian window with standard deviation $1.5$:
 \begin{equation*}
{\rm{SSIM}}(\fx\left( x_n \right),\tfx\left( x_n \right)) = \frac{{(2{\mu _\fx}{\mu _\tfx} + {c_1})(2{\sigma _{\fx\tfx}} + {c_2})}}{{(\mu _\fx^2 + \mu _\tfx^2 + {c_1})(\sigma _\fx^2 + \sigma _\tfx^2 + {c_2})}},
  \end{equation*}
where ${\mu _\fx}$ is the average of $\fx\left( x_n \right)$ in the Gaussian window,
${\sigma _\fx^2}$ is the variance of $\fx\left( x_n \right)$ in the Gaussian window,
${\sigma _{\fx\tfx}}$ is the covariance betwenn $\fx\left( x_n \right)$ and $\tfx\left( x_n \right)$ in the Gaussian window, and ${c_1}$ and ${c_2}$ are two variables to stabilize the division with weak denominator.
In our results we present the mean of the SSIM value over the whole image, denoted MSSIM.

\begin{figure}[!t] %
    \centering
    \includegraphics[trim={0 0 0cm 0cm},clip,width=0.47\textwidth]{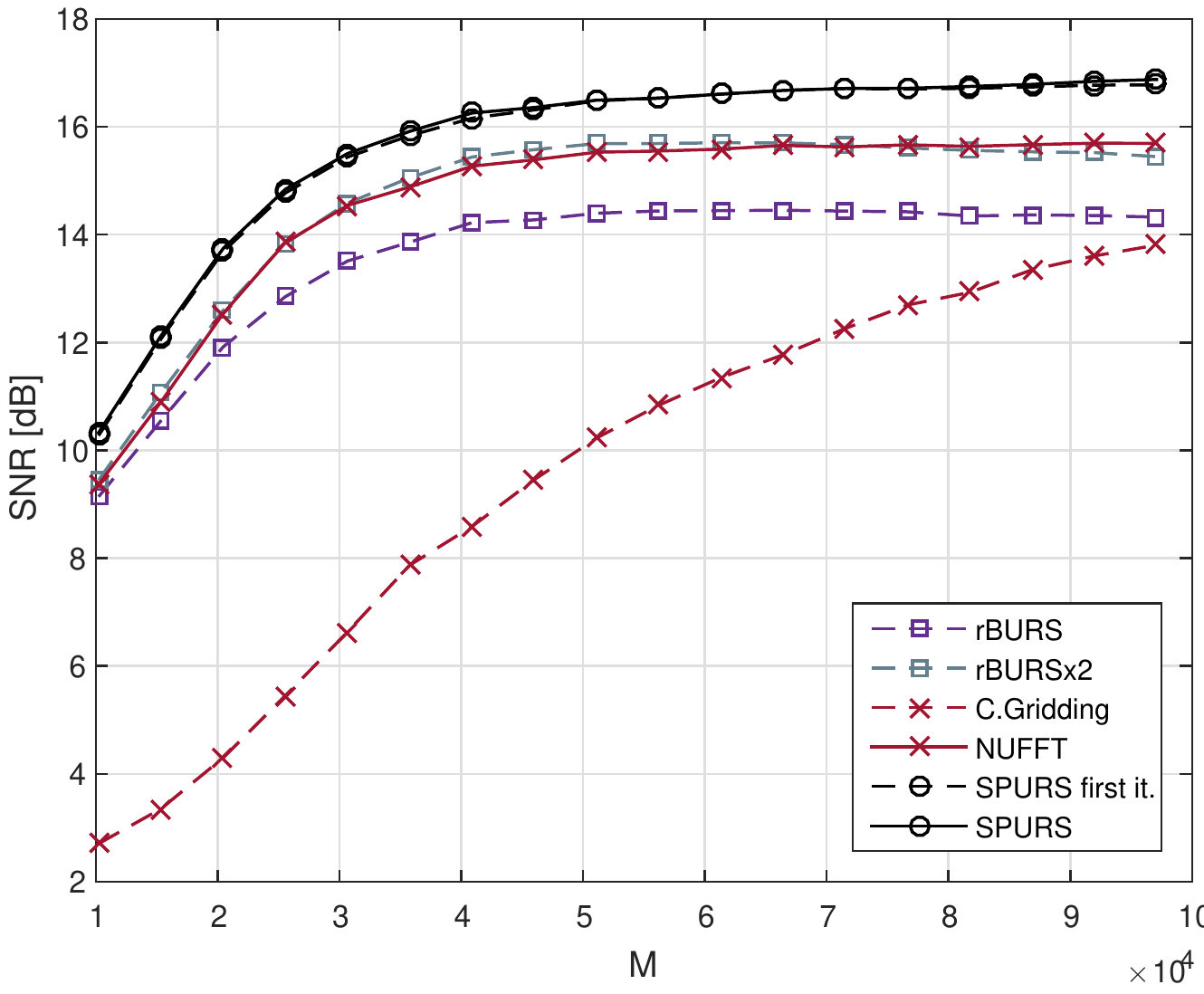}
    \caption{SNR as a function of $M$ for an analytical brain phantom sampled on a radial trajectory with ISNR = $30$ dB.}
    \label{fig:EXP_M_iSNR_30_Radial_SNR}
\end{figure}

\begin{figure}[!t] %
    \centering
    \includegraphics[trim={0 0 0cm 0cm},clip,width=0.47\textwidth]{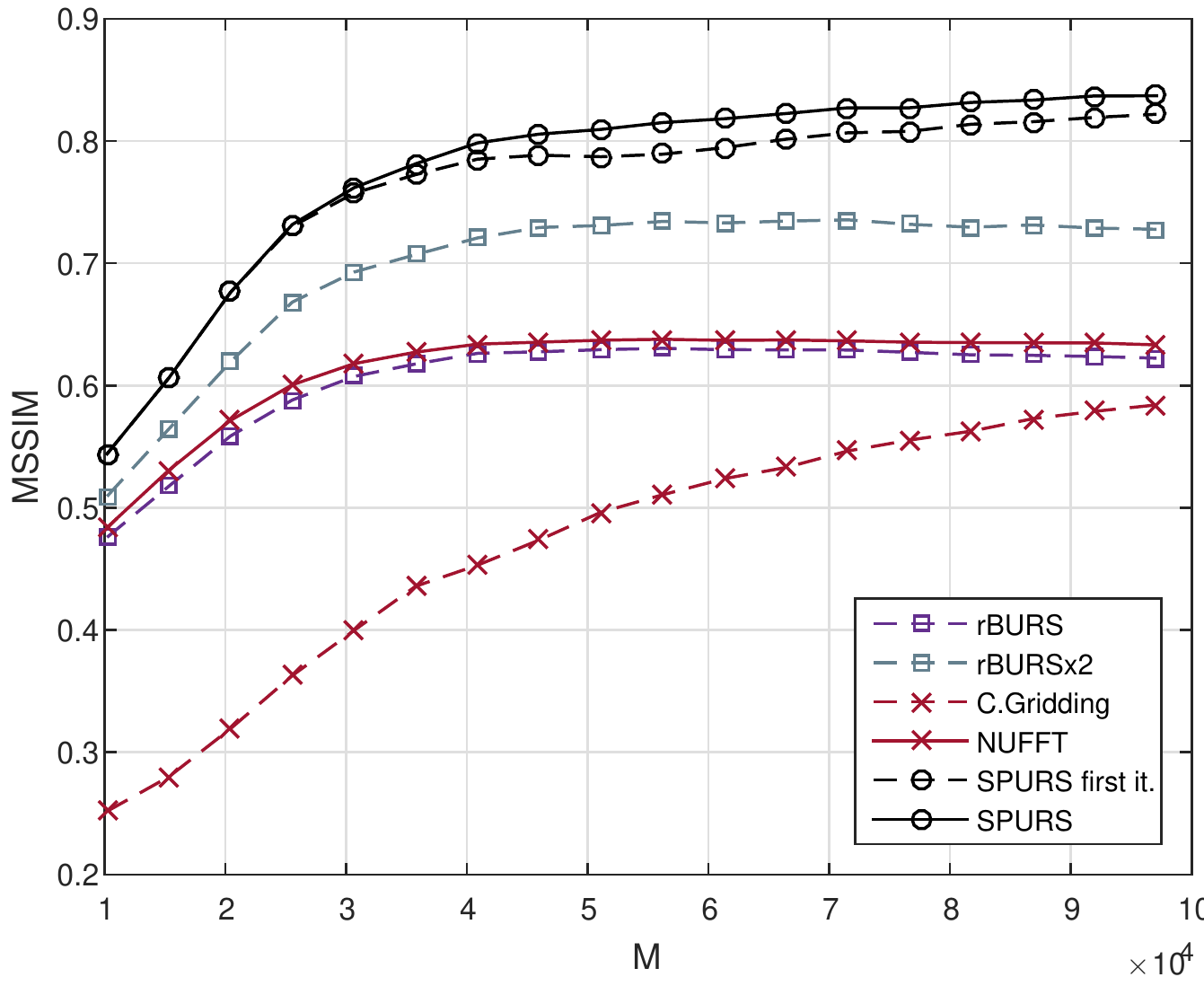}
    \caption{MSSIM as a function of $M$ for an analytical brain phantom sampled on a radial trajectory with ISNR = $30$ dB.}
    \label{fig:EXP_M_iSNR_30_Radial_MSSIM}
\end{figure}

\begin{figure}[!t] %
    \centering
    \includegraphics[trim={0 0 0cm 0cm},clip,width=0.47\textwidth]{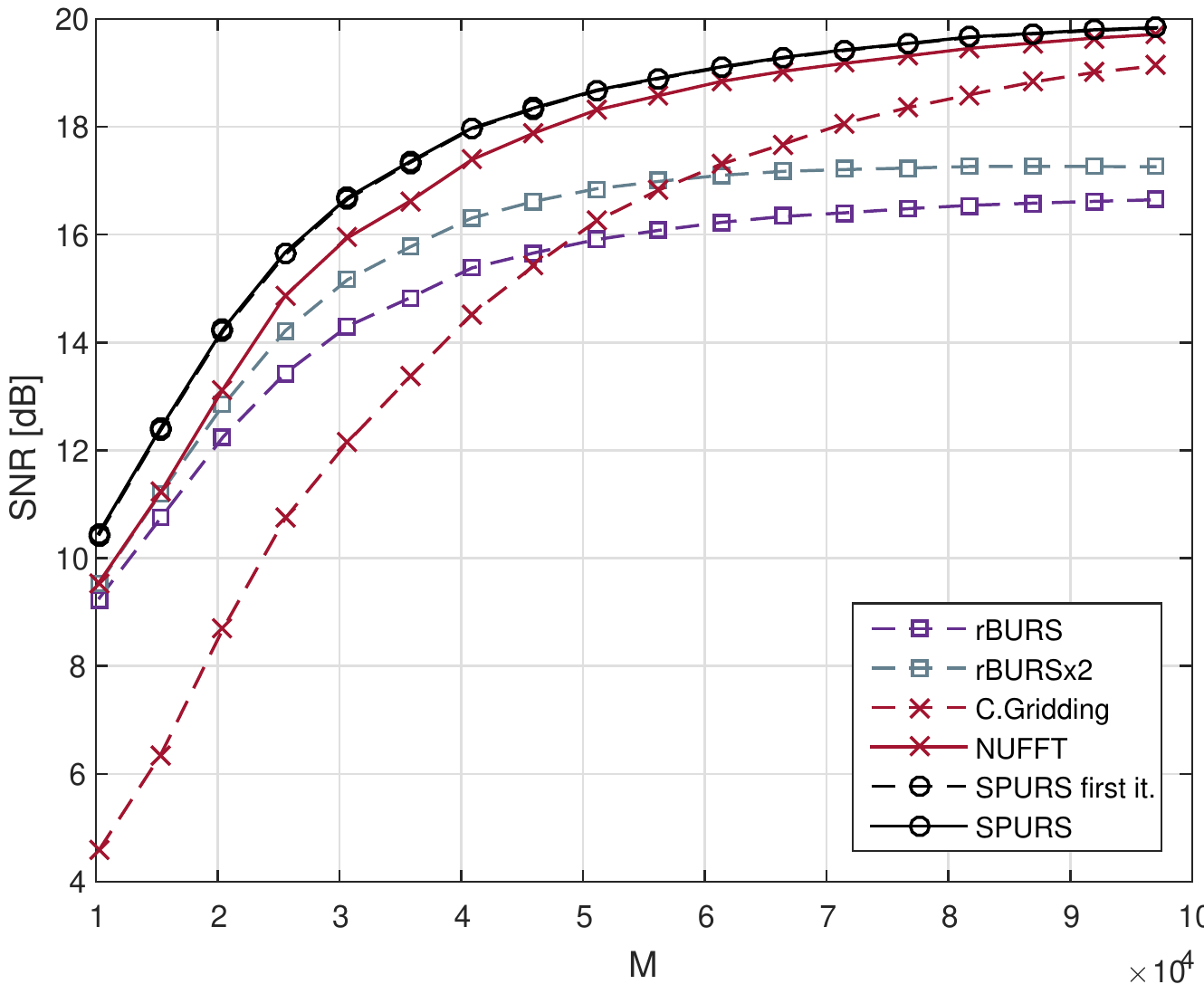}
    \caption{SNR as a function of $M$ for an analytical brain phantom sampled on a radial trajectory with no sampling noise added i.e. ISNR = $\infty$.}
    \label{fig:EXP_M_iSNR_inf_Radial_SNR}
\end{figure}

\begin{figure}[!t] %
    \centering
    \includegraphics[trim={0 0 0cm 0cm},clip,width=0.47\textwidth]{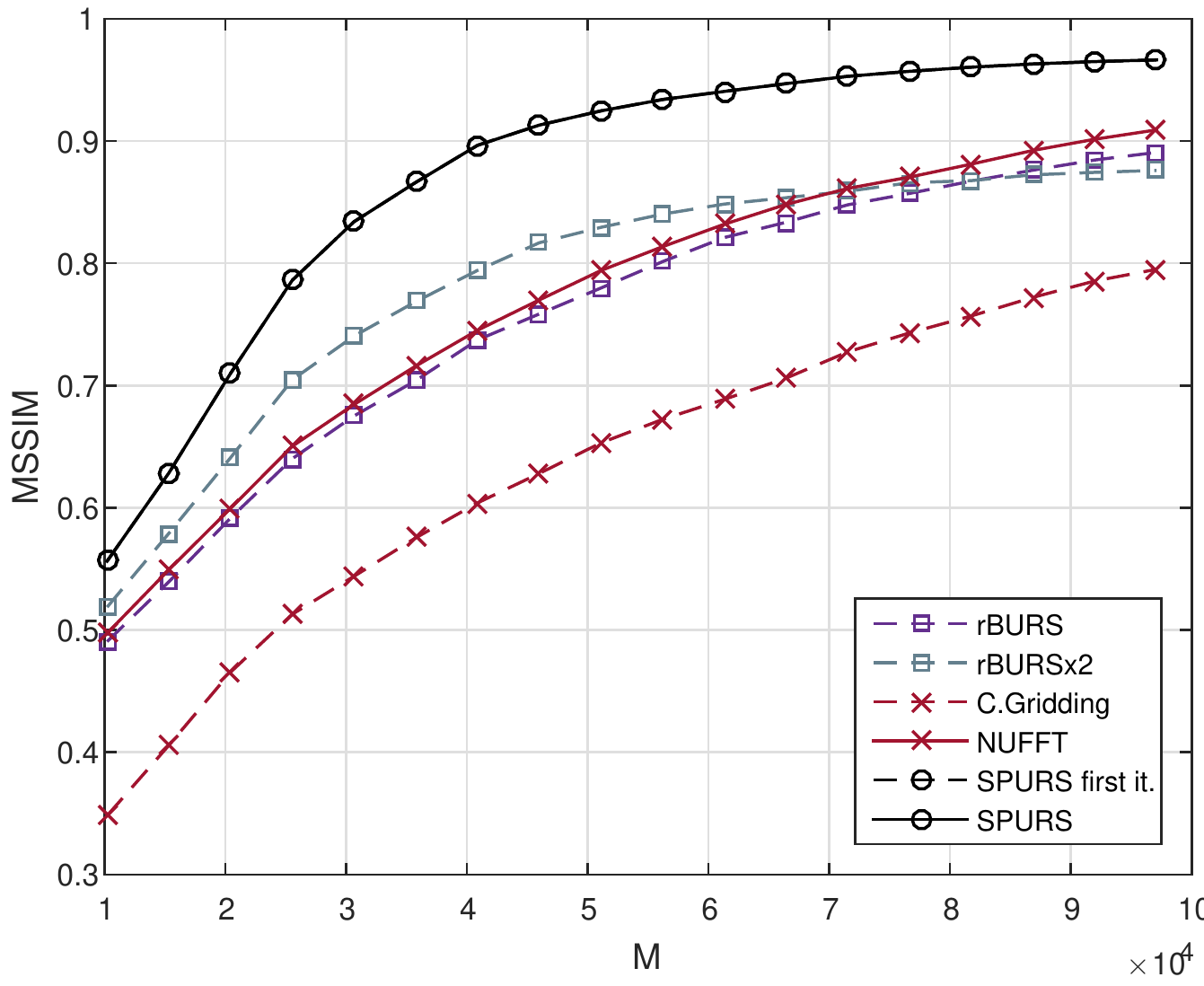}
    \caption{MSSIM as a function of $M$ for an analytical brain phantom sampled on a radial trajectory with no sampling noise added i.e. ISNR = $\infty$.}
    \label{fig:EXP_M_iSNR_inf_Radial_MSSIM}
\end{figure}

In the first experiment a radial trajectory is used \eqn{eq:Radial_traj}, where the value of ${N_{\rm{bins}}} = 512$ and ${N_{{\rm{spokes}}}}$ is varied between $20$ and $190$, which results in $M$ values between $10240$ and $97280$.
Noise was added to the samples to achieve an ISNR of $30$ dB. Figures \ref{fig:EXP_M_iSNR_30_Radial_SNR} and \ref{fig:EXP_M_iSNR_30_Radial_MSSIM} present the SNR and MSSIM of the reconstructed image as a function of $M$, the number of sampling points in k-space. For reconstruction methods which can be iterated, results are shown for both a single iteration (dashed line) and the final result after the algorithm has converged (solid line).
The same experiment was repeated with noiseless samples and is presented in Figures \ref{fig:EXP_M_iSNR_inf_Radial_SNR} and \ref{fig:EXP_M_iSNR_inf_Radial_MSSIM}.

\begin{figure}[!htbp] %
    \centering
    \includegraphics[trim={0cm 0 0cm 0cm},clip,width=0.47\textwidth]{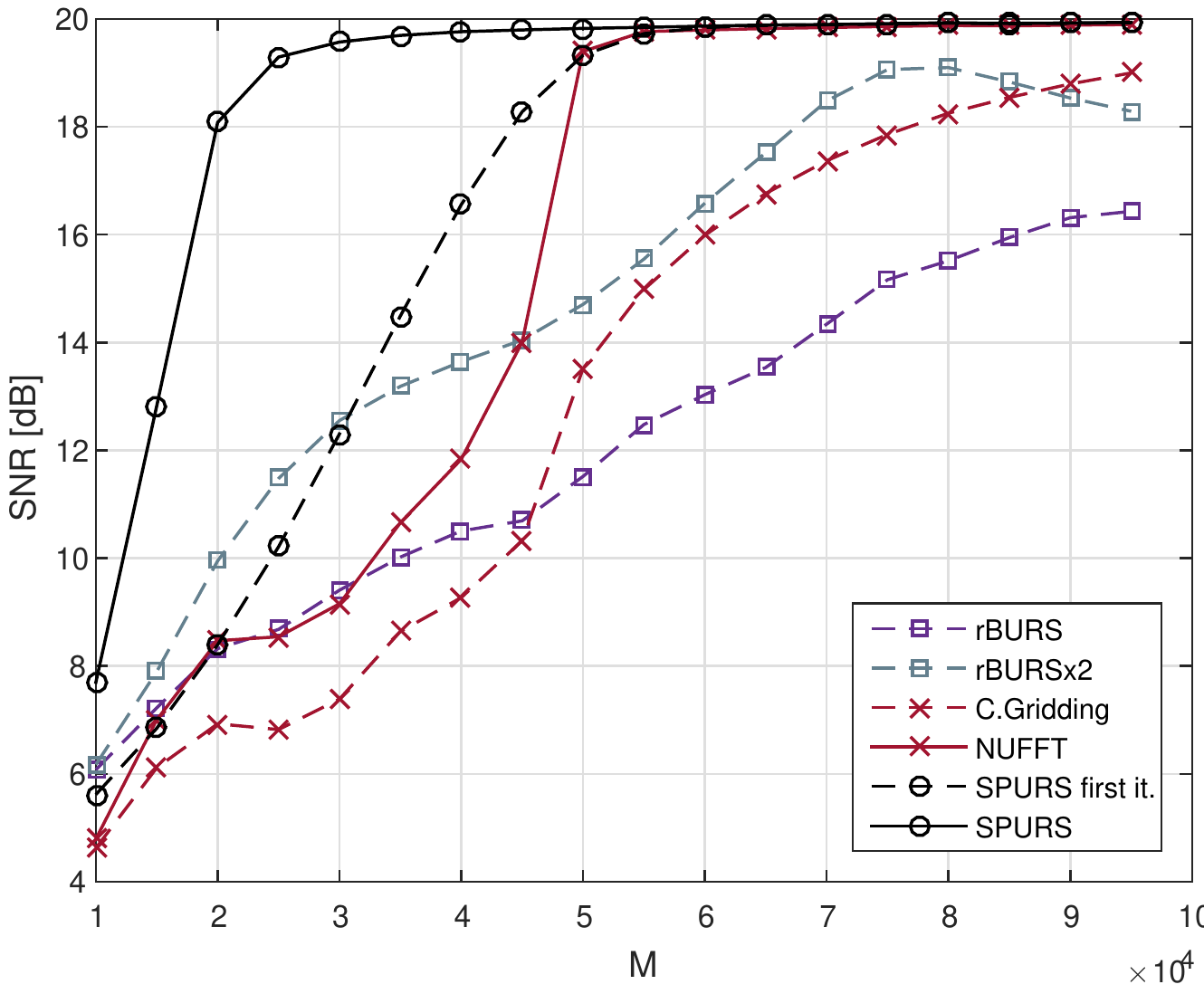}
    \caption{SNR as a function of $M$ for an analytical brain phantom sampled on a spiral trajectory with ISNR = $30$ dB.}
    \label{fig:EXP_M_iSNR_30_Spiral_SNR}
\end{figure}

\begin{figure}[!htbp] %
    \centering
    \includegraphics[trim={0cm 0 0cm 0cm},clip,width=0.47\textwidth]{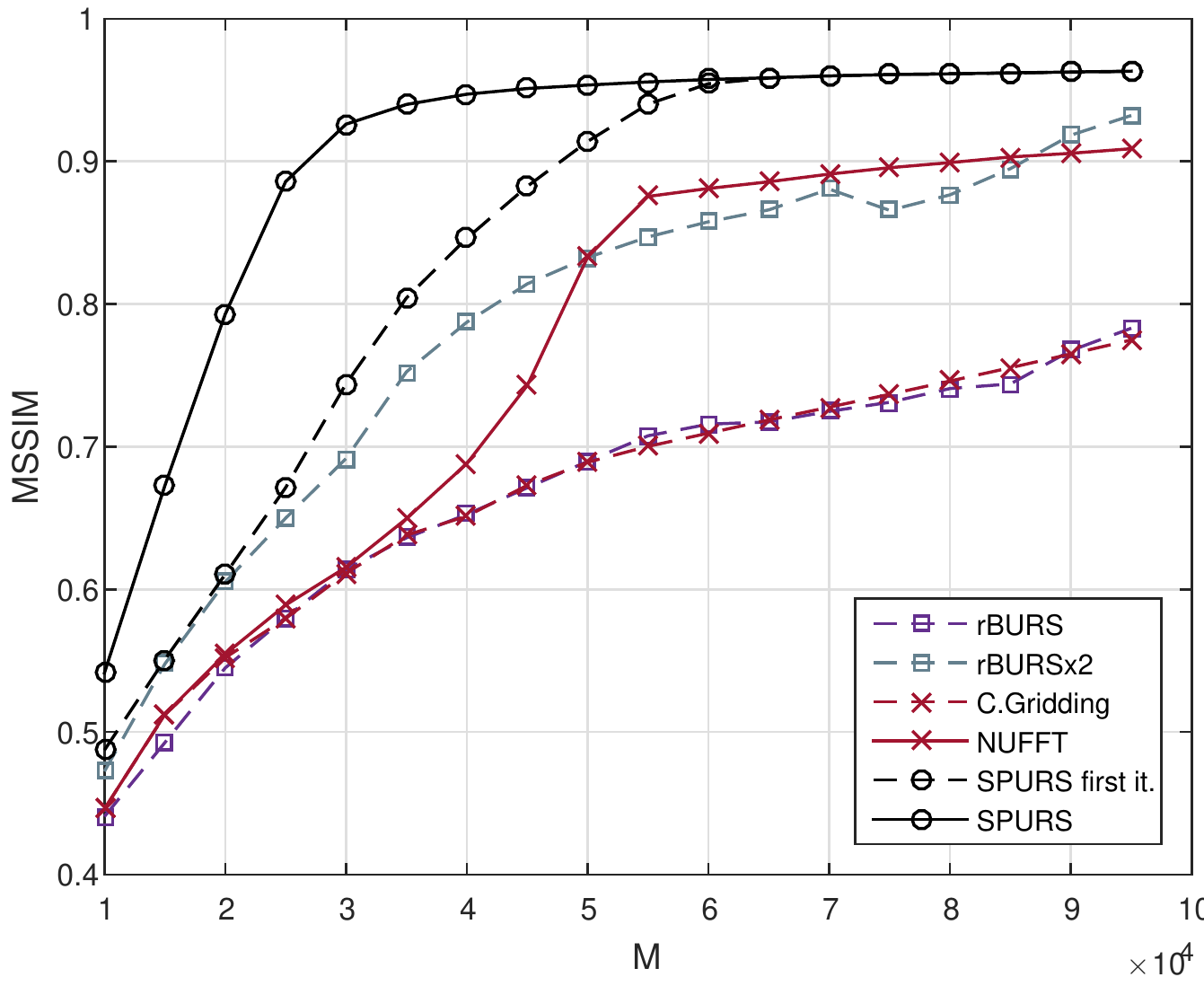}
    \caption{MSSIM as a function of $M$ for an analytical brain phantom sampled on a spiral trajectory with ISNR = $30$ dB.}
    \label{fig:EXP_M_iSNR_30_Spiral_MSSIM}
\end{figure}

\begin{figure}[!htbp] %
    \centering
    \includegraphics[trim={0cm 0 0cm 0cm},clip,width=0.47\textwidth]{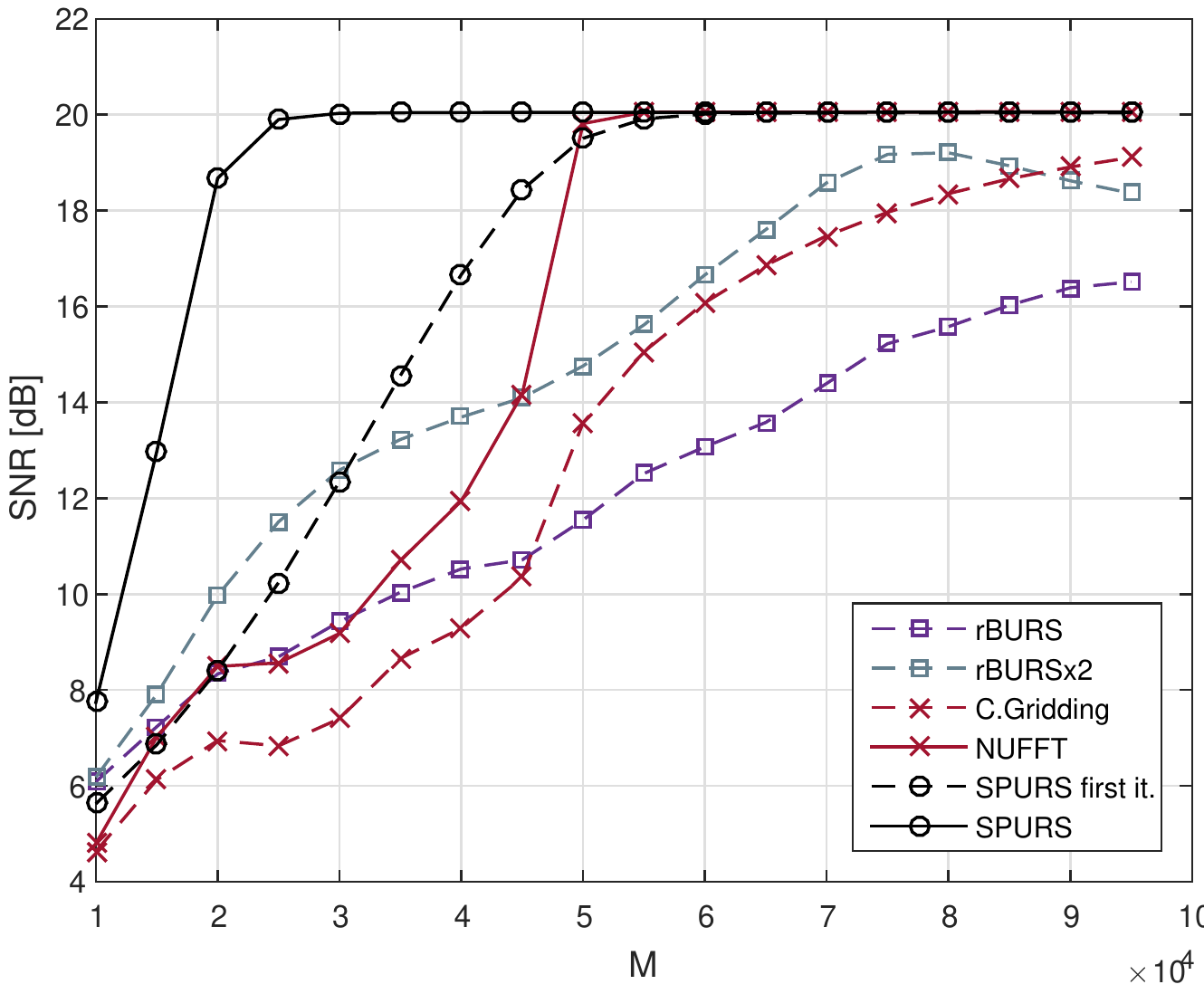}
    \caption{SNR as a function of $M$ for an analytical brain phantom sampled on a spiral trajectory with no sampling noise added i.e. ISNR = $\infty$.}
    \label{fig:EXP_M_iSNR_inf_Spiral_SNR}
\end{figure}

\begin{figure}[!htbp] %
    \centering
    \includegraphics[trim={0cm 0 0cm 0cm},clip,width=0.47\textwidth]{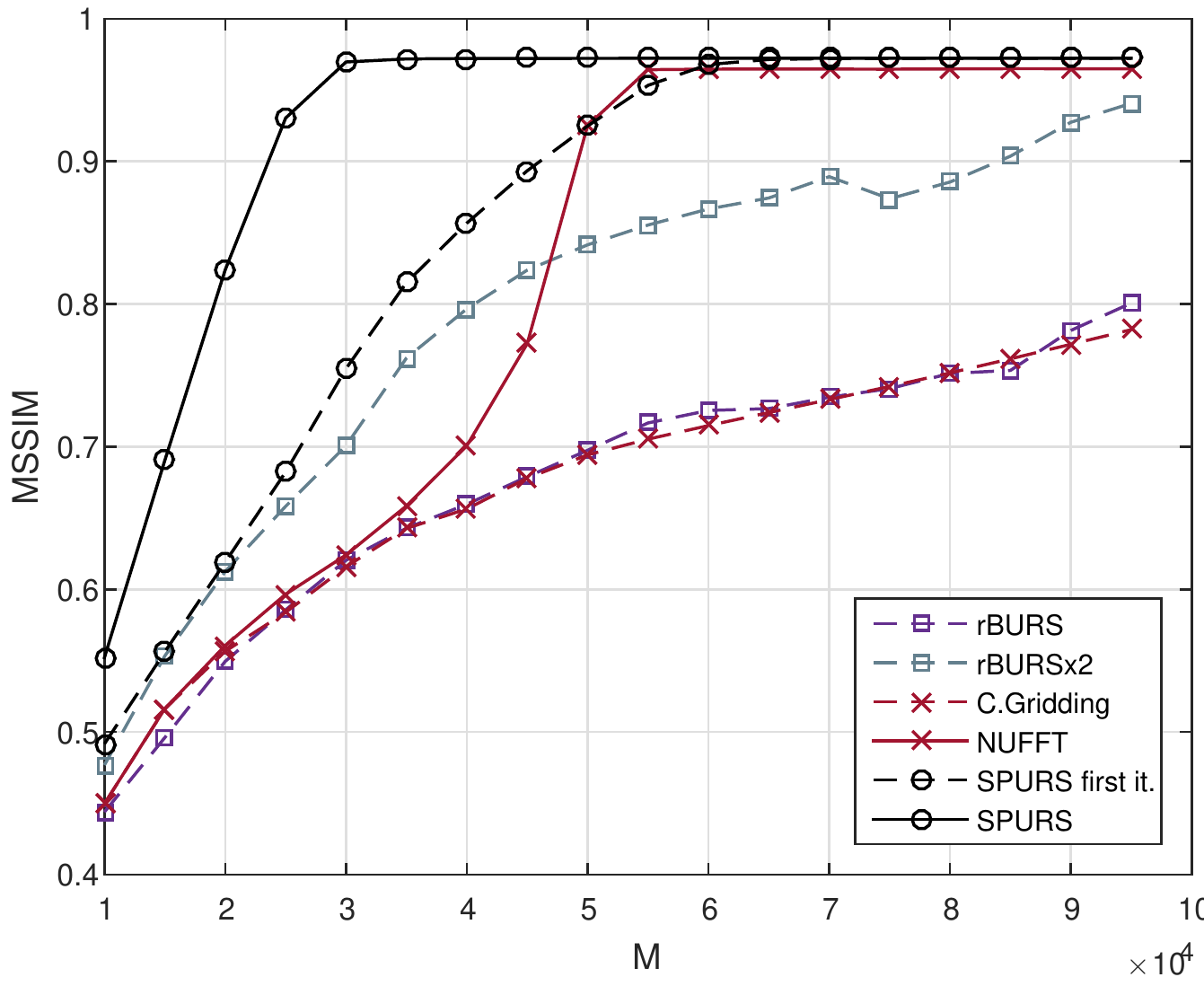}
    \caption{MSSIM as a function of $M$ for an analytical brain phantom sampled on a spiral trajectory with no sampling noise added i.e. ISNR = $\infty$.}
    \label{fig:EXP_M_iSNR_inf_Spiral_MSSIM}
\end{figure}

The second experiment is similar to the first with a spiral trajectory as described by \eqn{eq:Spiral_traj}. The number of samples $M$ is varied by increments of $5000$ between $10000$ and $85000$ with ${\rm{ISNR}} = 30 {\rm{dB}}$. The results are presented in Figures \ref{fig:EXP_M_iSNR_30_Spiral_SNR} and \ref{fig:EXP_M_iSNR_30_Spiral_MSSIM}. Here too, the experiment was repeated with noiseless samples and the results are presented in Figures \ref{fig:EXP_M_iSNR_inf_Spiral_SNR} and \ref{fig:EXP_M_iSNR_inf_Spiral_MSSIM}.

\begin{figure}[!h]
\centering
\begin{tabular}{cc}
  \includegraphics[trim={1.5cm 0cm 1.5cm 1cm},clip,width=0.22\textwidth]{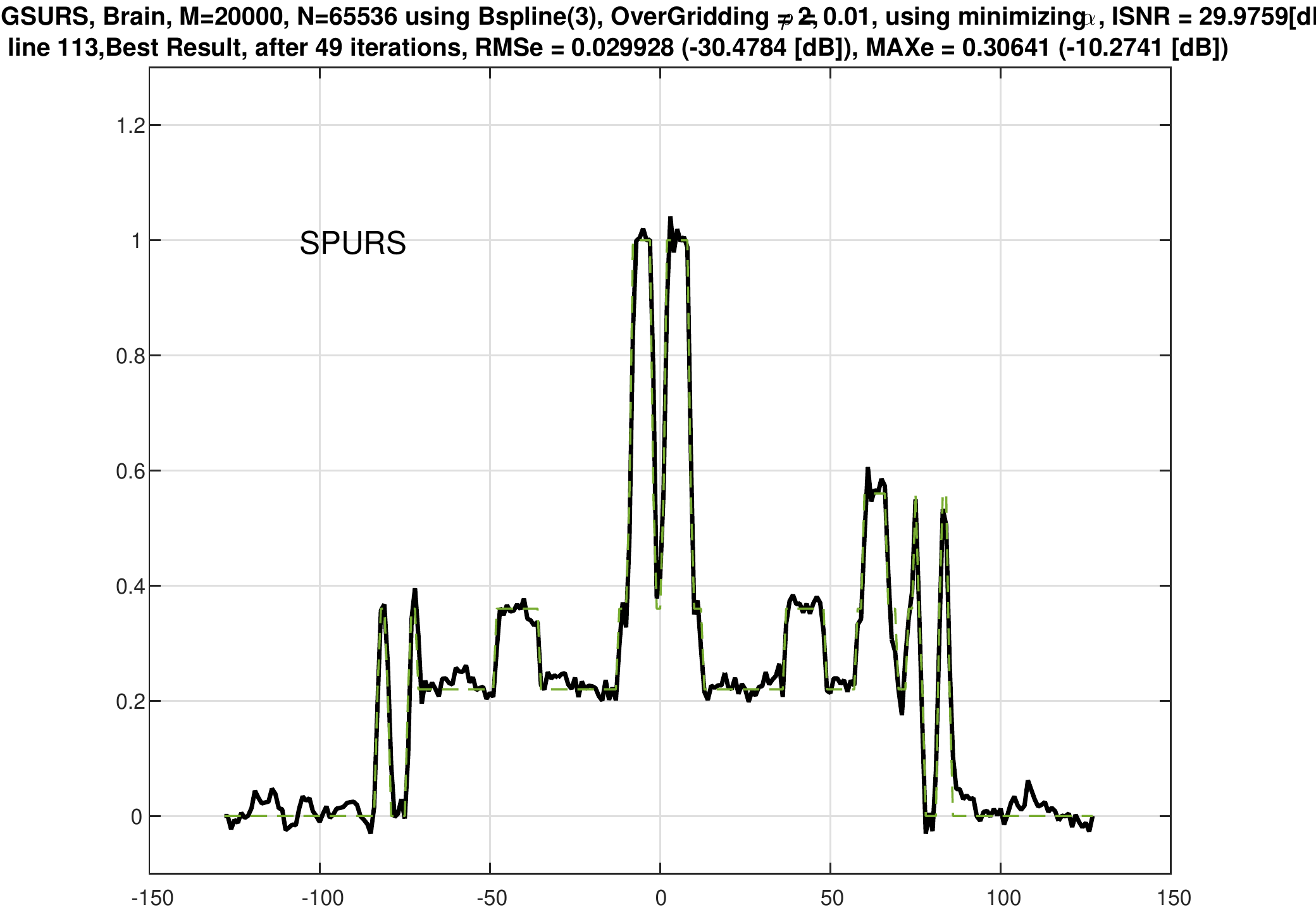} &
  \includegraphics[trim={5.5cm 2.5cm 5.5cm 2.5cm},clip,width=0.22\textwidth]{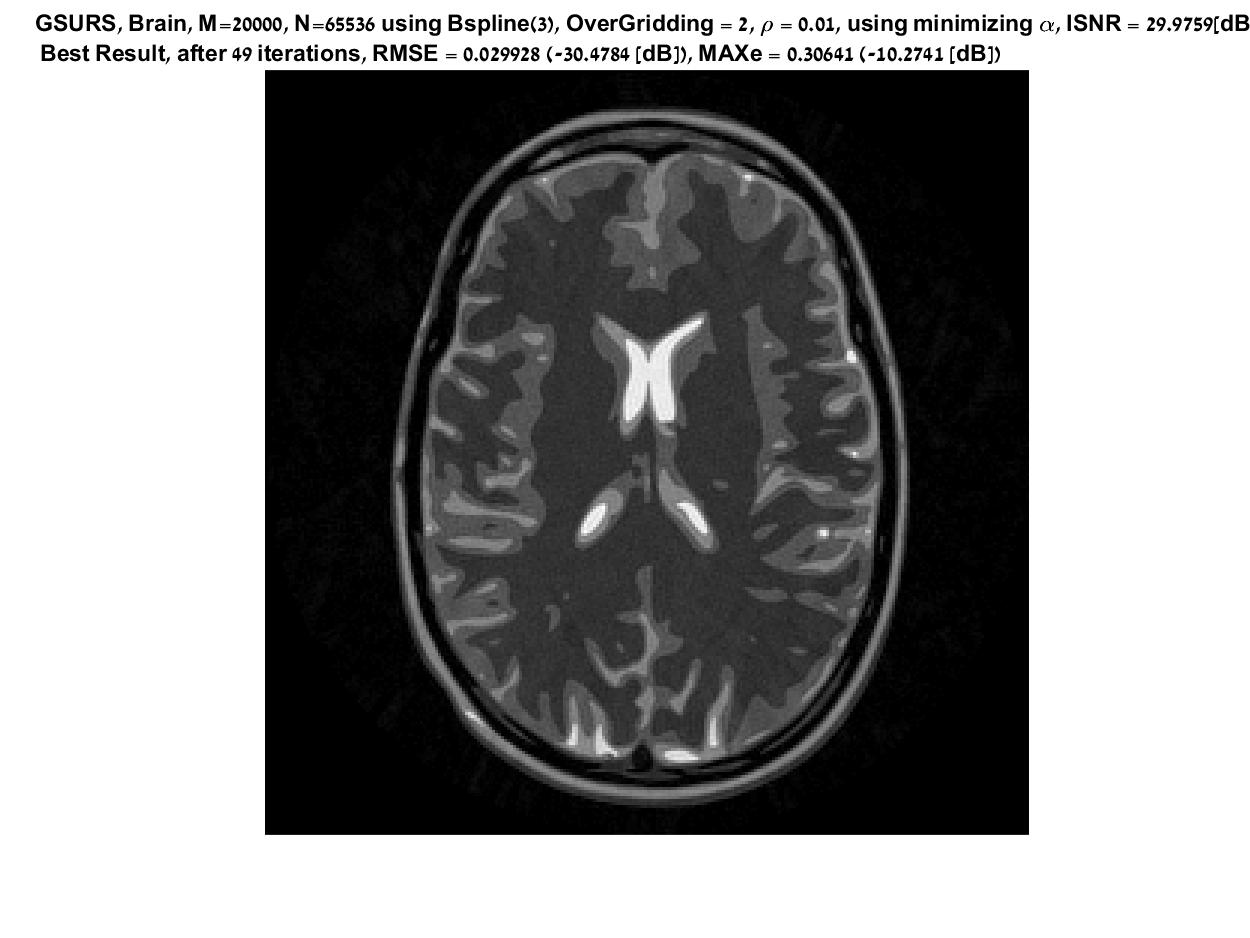}\\

  \includegraphics[trim={0cm 0cm 0cm 1cm},clip,width=0.22\textwidth]{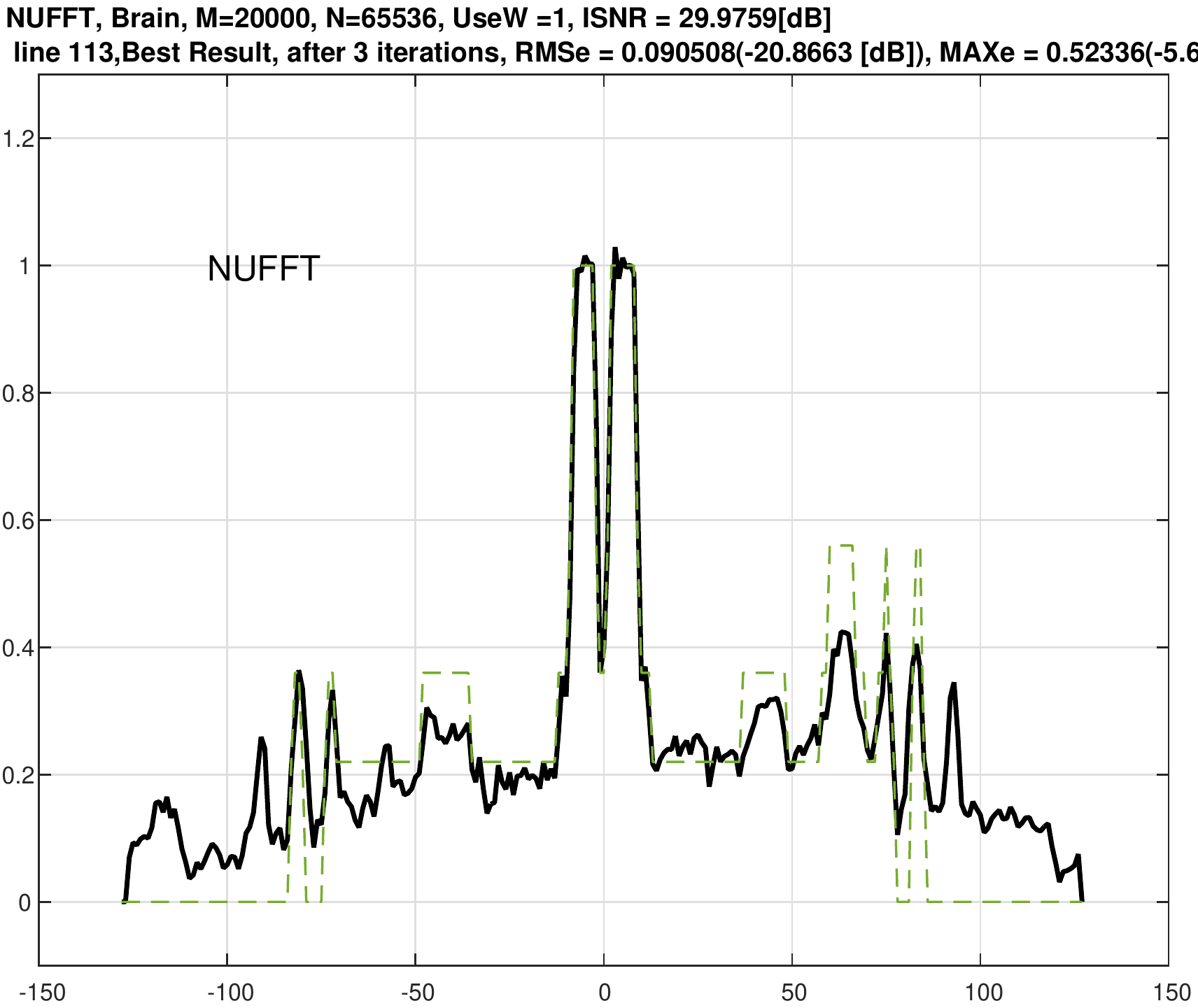} &
  \includegraphics[trim={5.5cm 2.5cm 5.5cm 2.5cm},clip,width=0.22\textwidth]{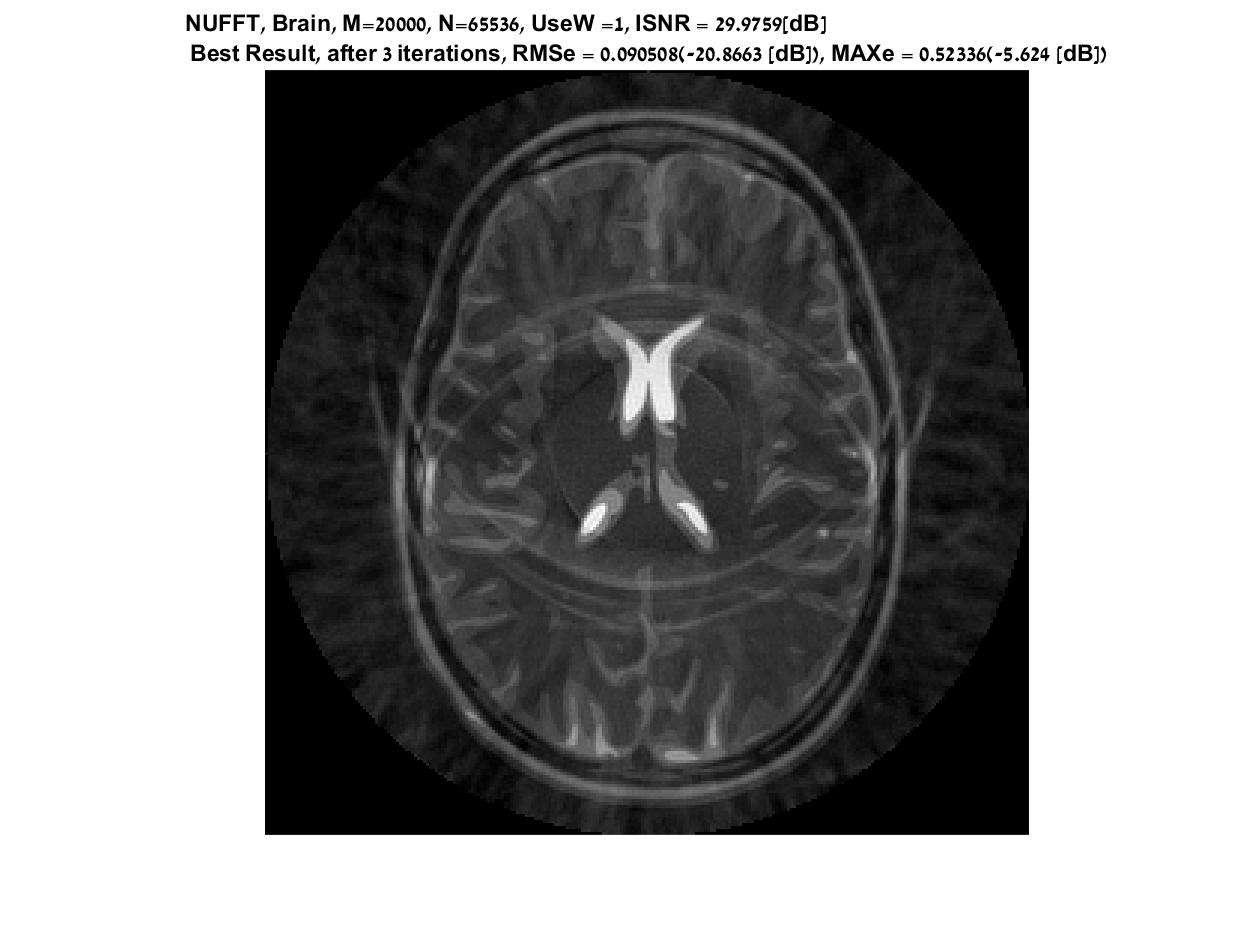}\\

  \includegraphics[trim={3.55cm 0cm 3.55cm 1cm},clip,width=0.22\textwidth]{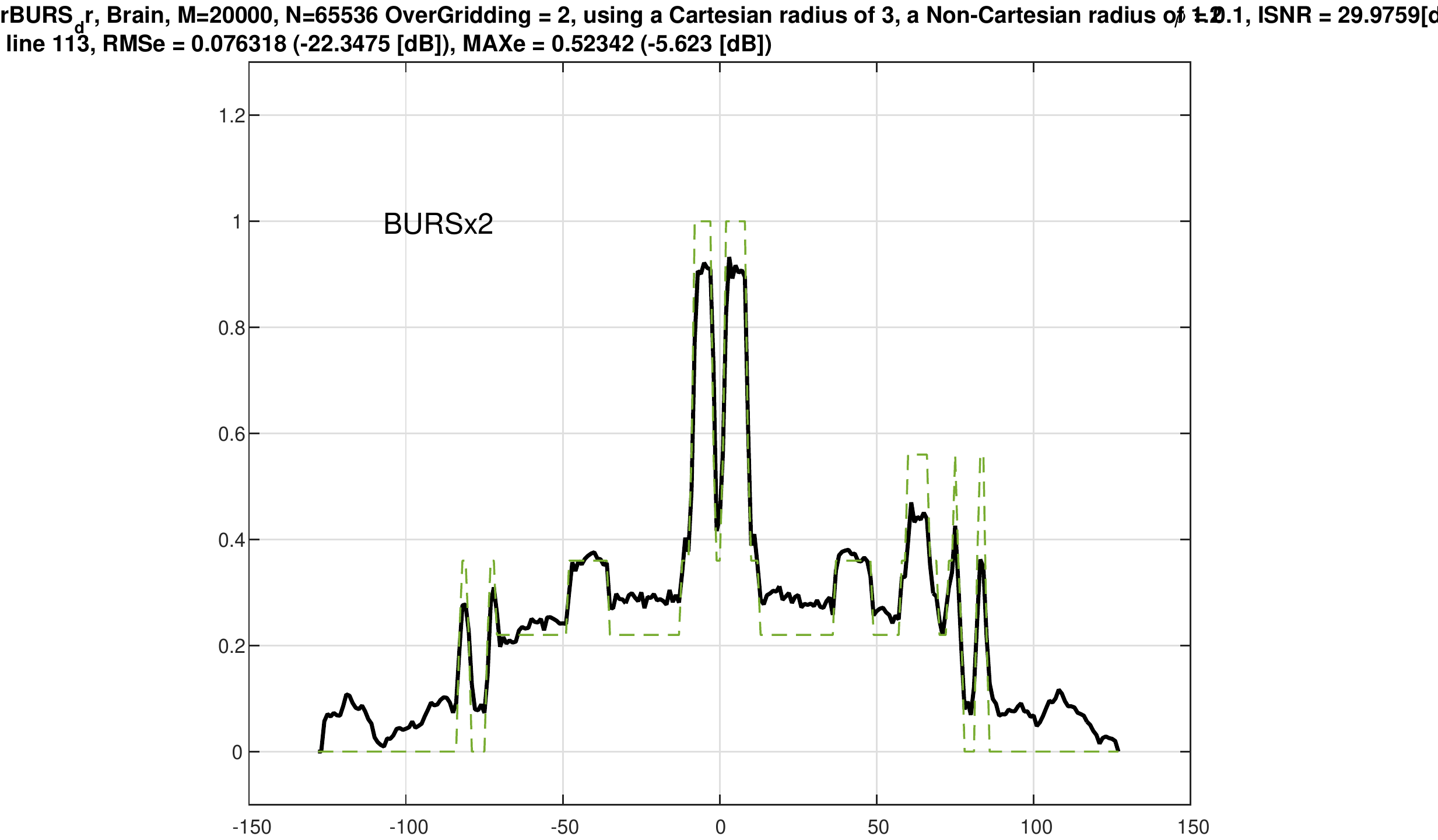} &
  \includegraphics[trim={5.5cm 2.5cm 5.5cm 2.5cm},clip,width=0.22\textwidth]{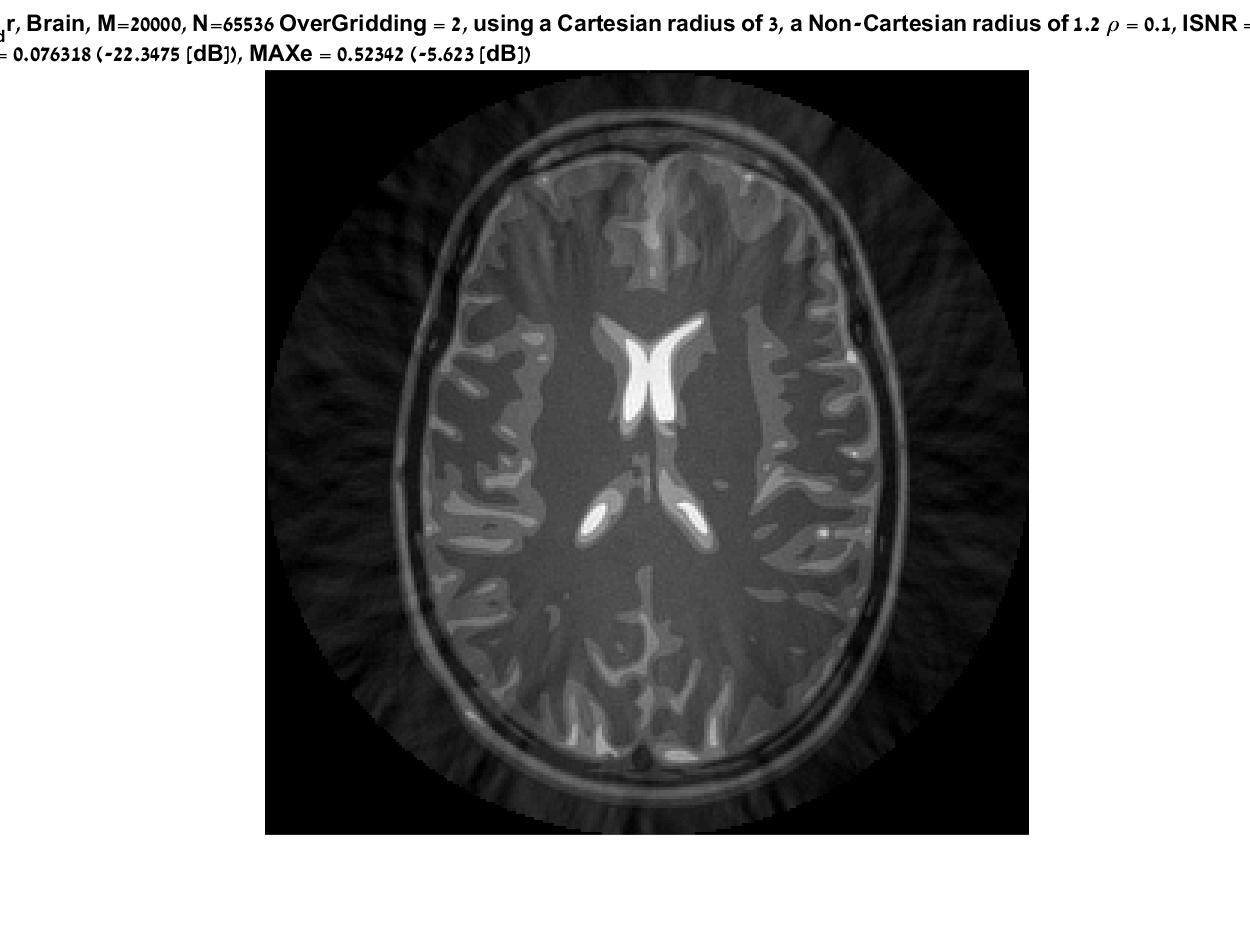}\\

  \includegraphics[trim={0cm 0cm 0cm 1cm},clip,width=0.22\textwidth]{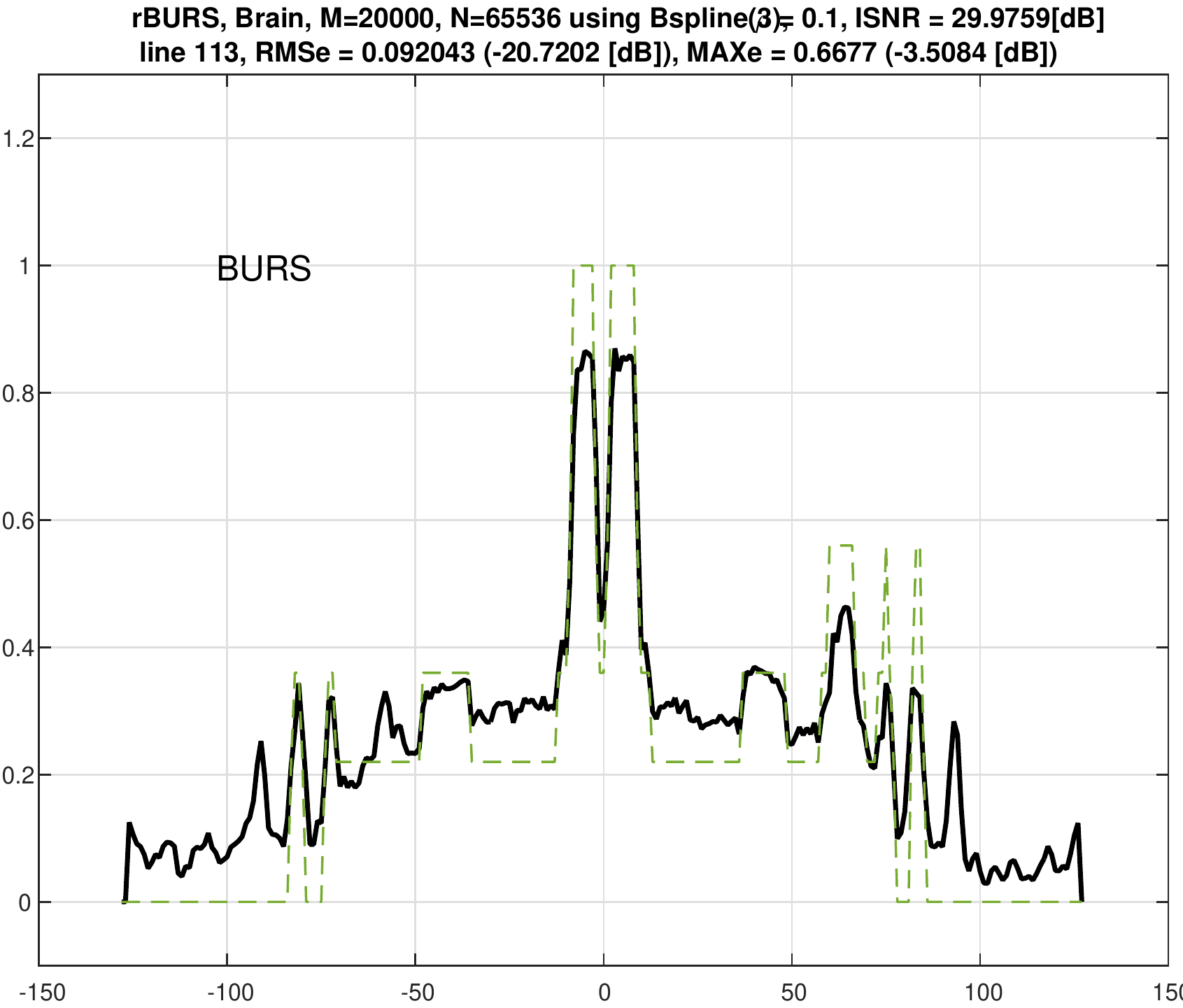} &
  \includegraphics[trim={5.5cm 2.5cm 5.5cm 2.5cm},clip,width=0.22\textwidth]{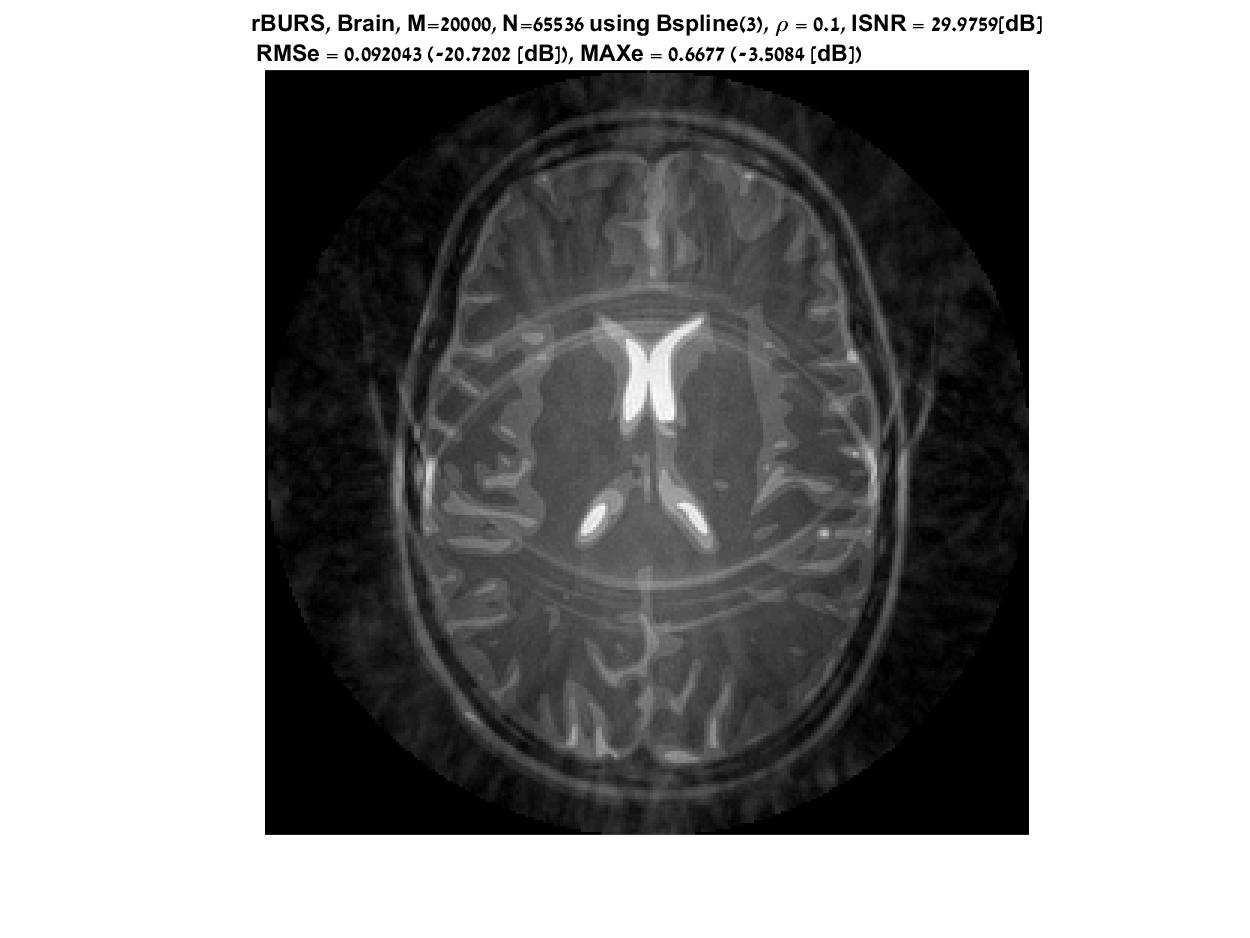}\\

  \includegraphics[trim={0cm 0cm 0cm 1cm},clip,width=0.22\textwidth]{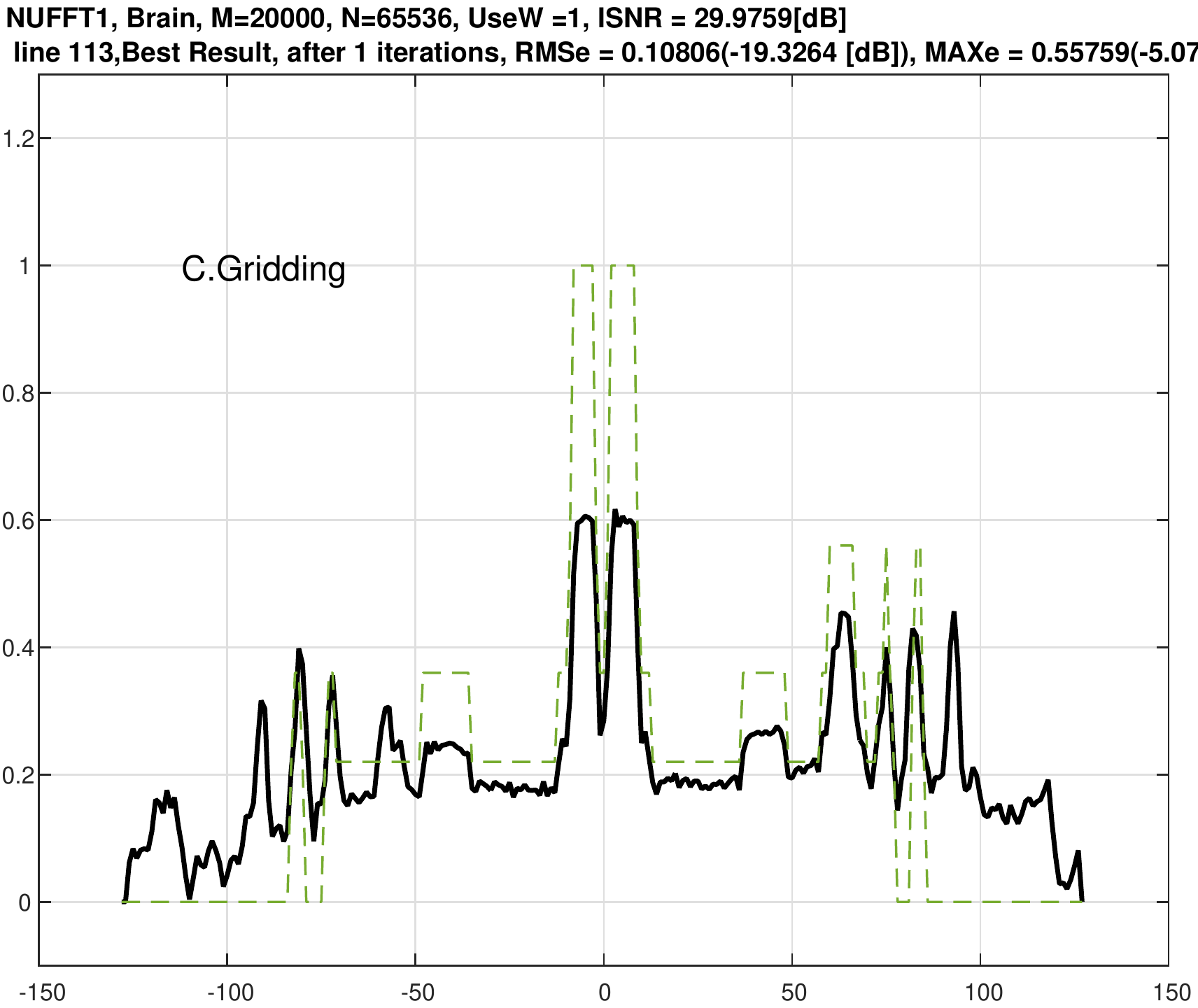} &
  \includegraphics[trim={5.5cm 2.5cm 5.5cm 2.5cm},clip,width=0.22\textwidth]{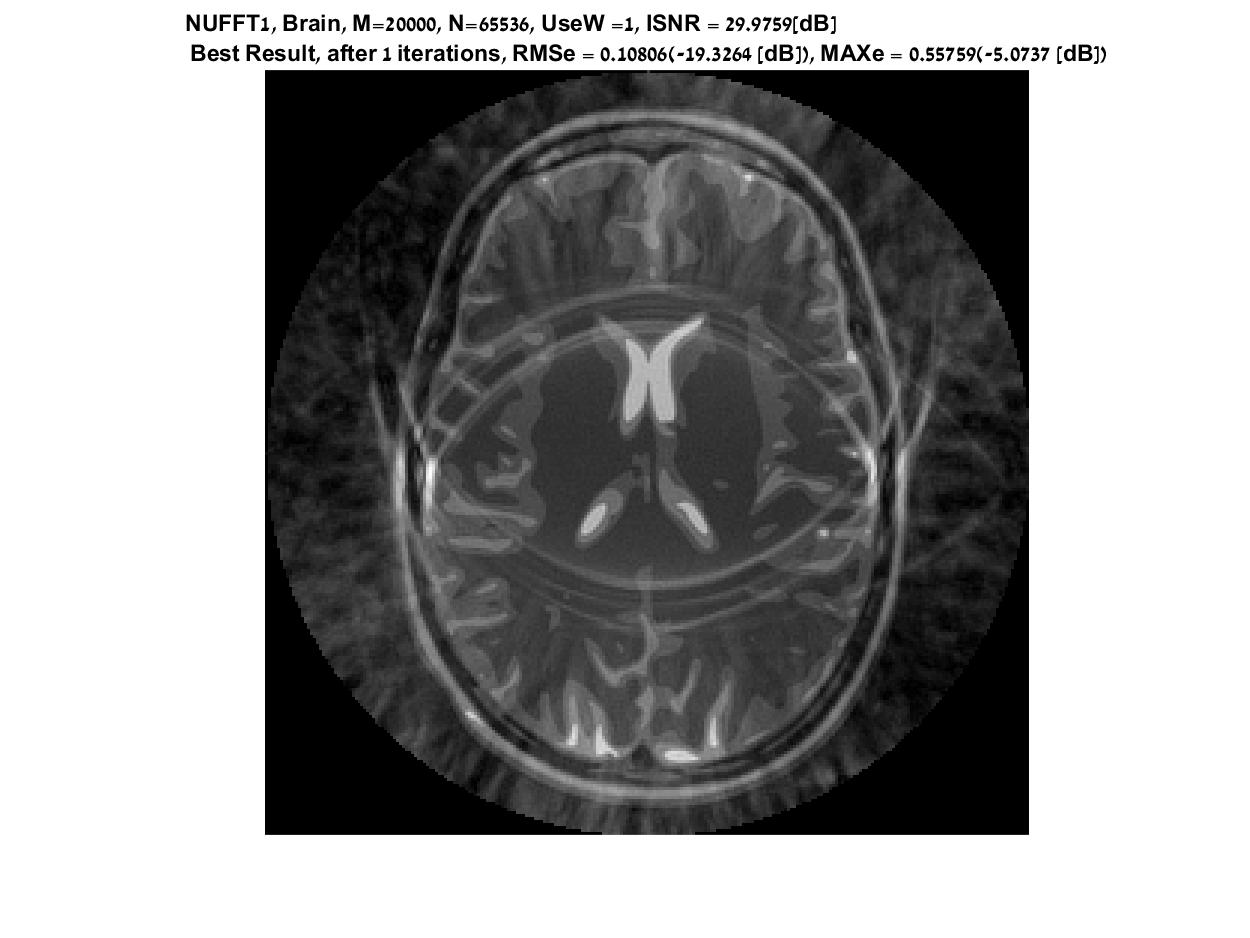}
\end{tabular}
    \caption{Results for an analytic brain phantom sampled on a spiral trajectory with $M=20000$, ISNR = $30$ dB. Right column: reconstructed images; left column: profile plots of row 113. Rows, from top to bottom, SNR and MSSIM values in parenthesis: SPURS using $\beta^3$ (18.09 dB, 0.79), NUFFT (8.47 dB, 0.55), BURS with $\oversamplingfactor = 2$(8.32 dB, 0.54), BURS with $\oversamplingfactor = 1$(9.95 dB, 0.61), convolutional gridding (6.93 dB, 0.55).}
    \label{fig:EXP_M_iSNR_Spiral_img20k}
\end{figure}

\begin{figure}[!h]
\centering
\begin{tabular}{cc}
  \includegraphics[trim={1.5cm 0cm 1.5cm 1cm},clip,width=0.22\textwidth]{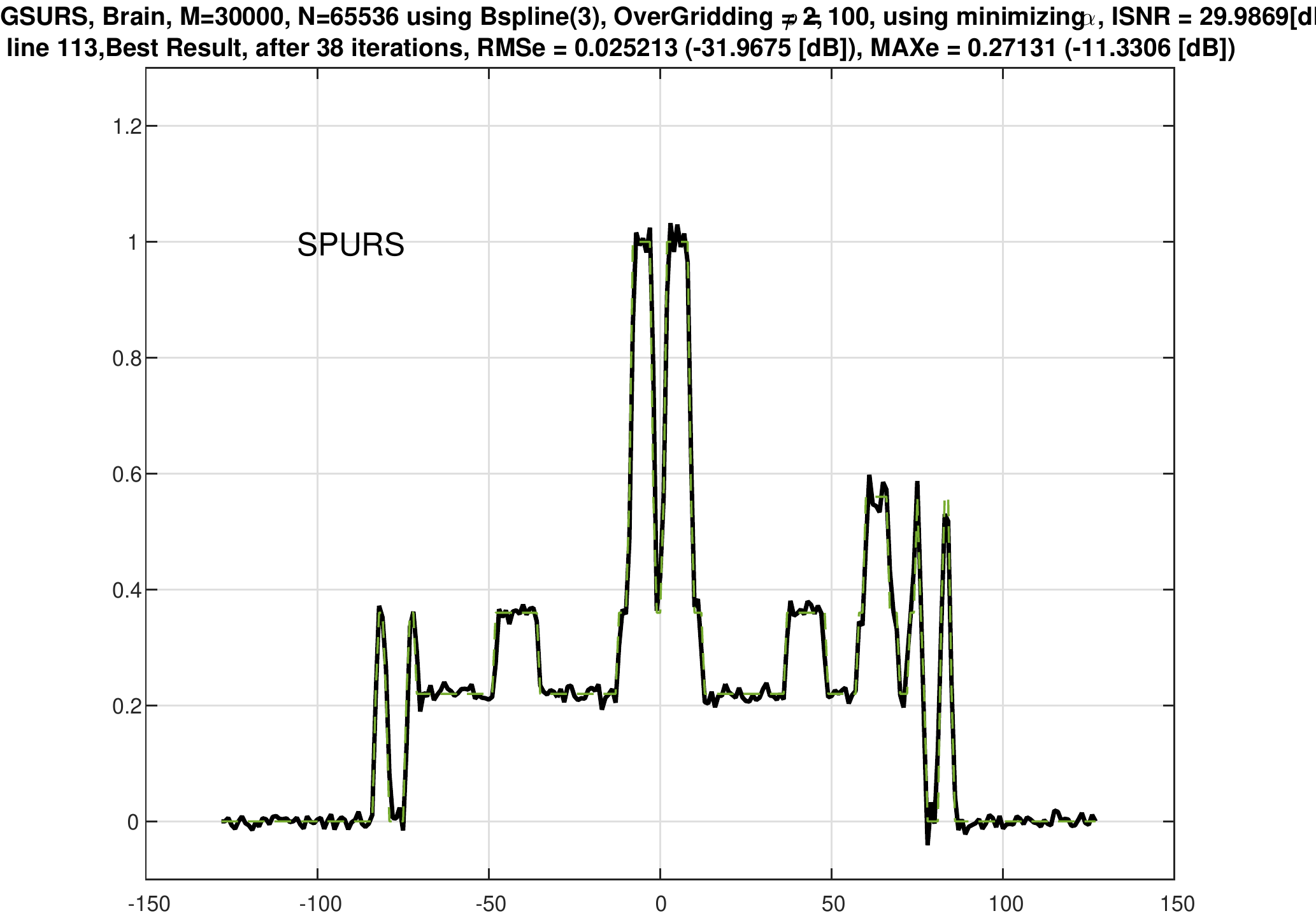} &
  \includegraphics[trim={5.5cm 2.5cm 5.5cm 2.5cm},clip,width=0.22\textwidth]{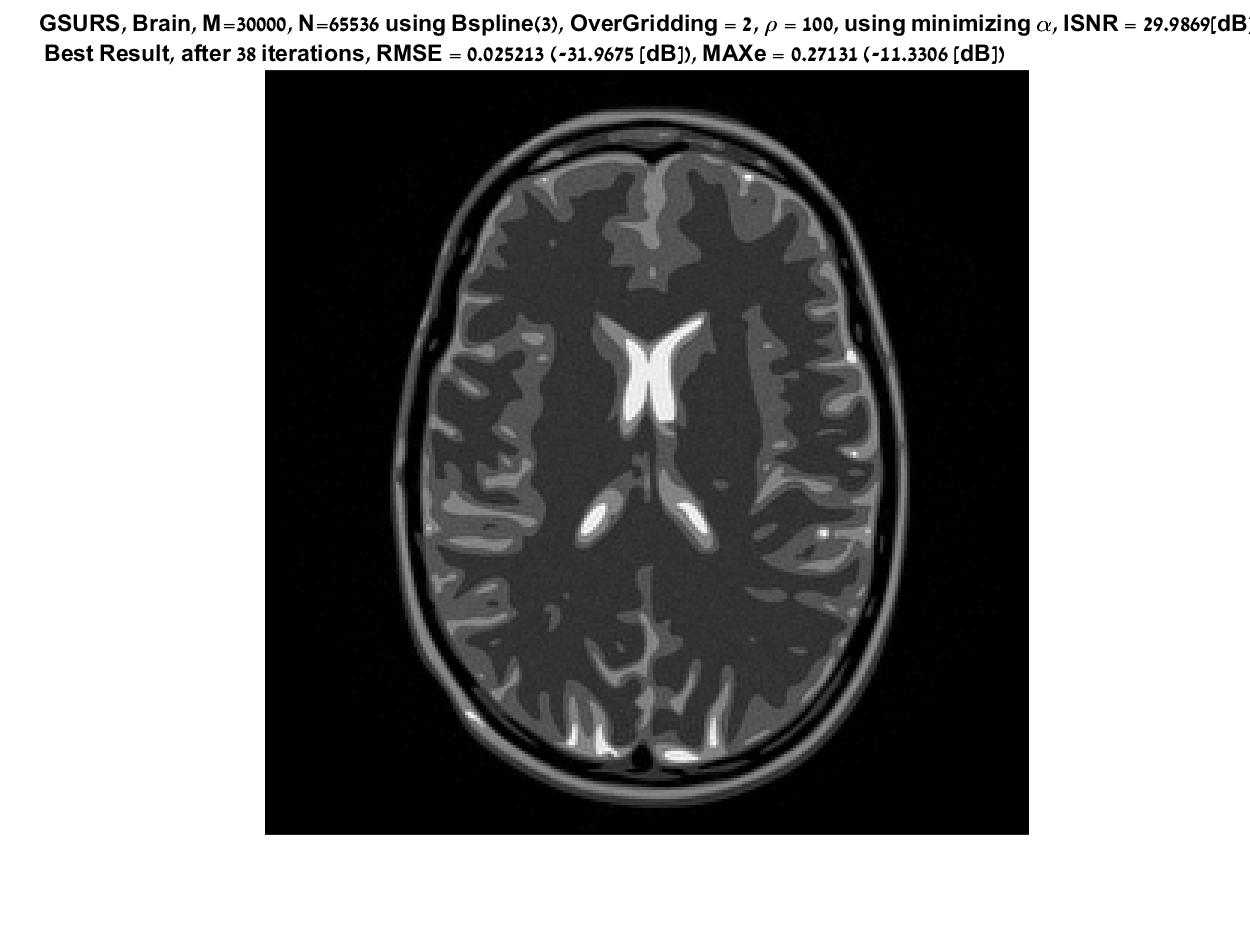}\\

  \includegraphics[trim={0cm 0cm 0cm 1cm},clip,width=0.22\textwidth]{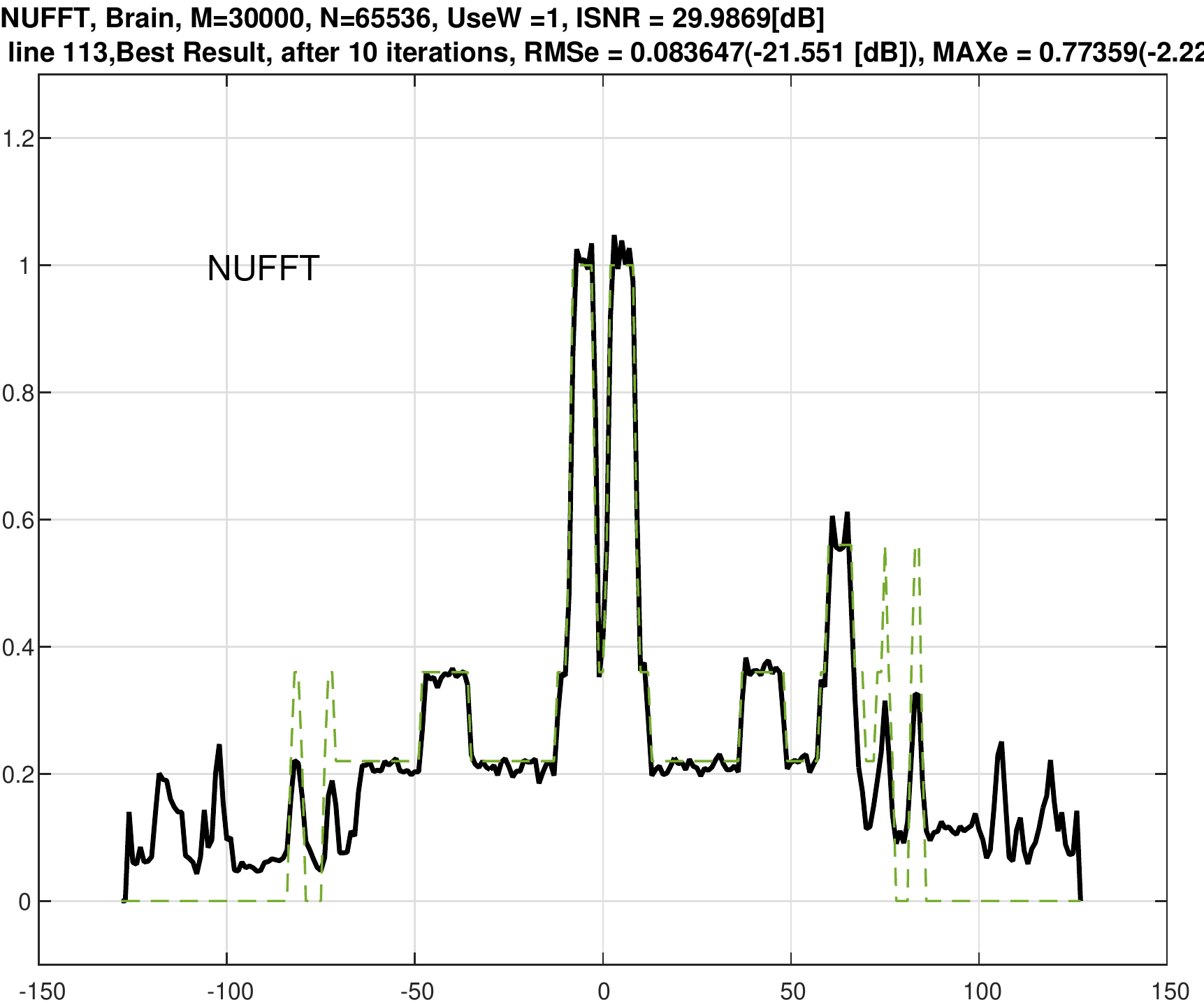} &
  \includegraphics[trim={5.5cm 2.5cm 5.5cm 2.5cm},clip,width=0.22\textwidth]{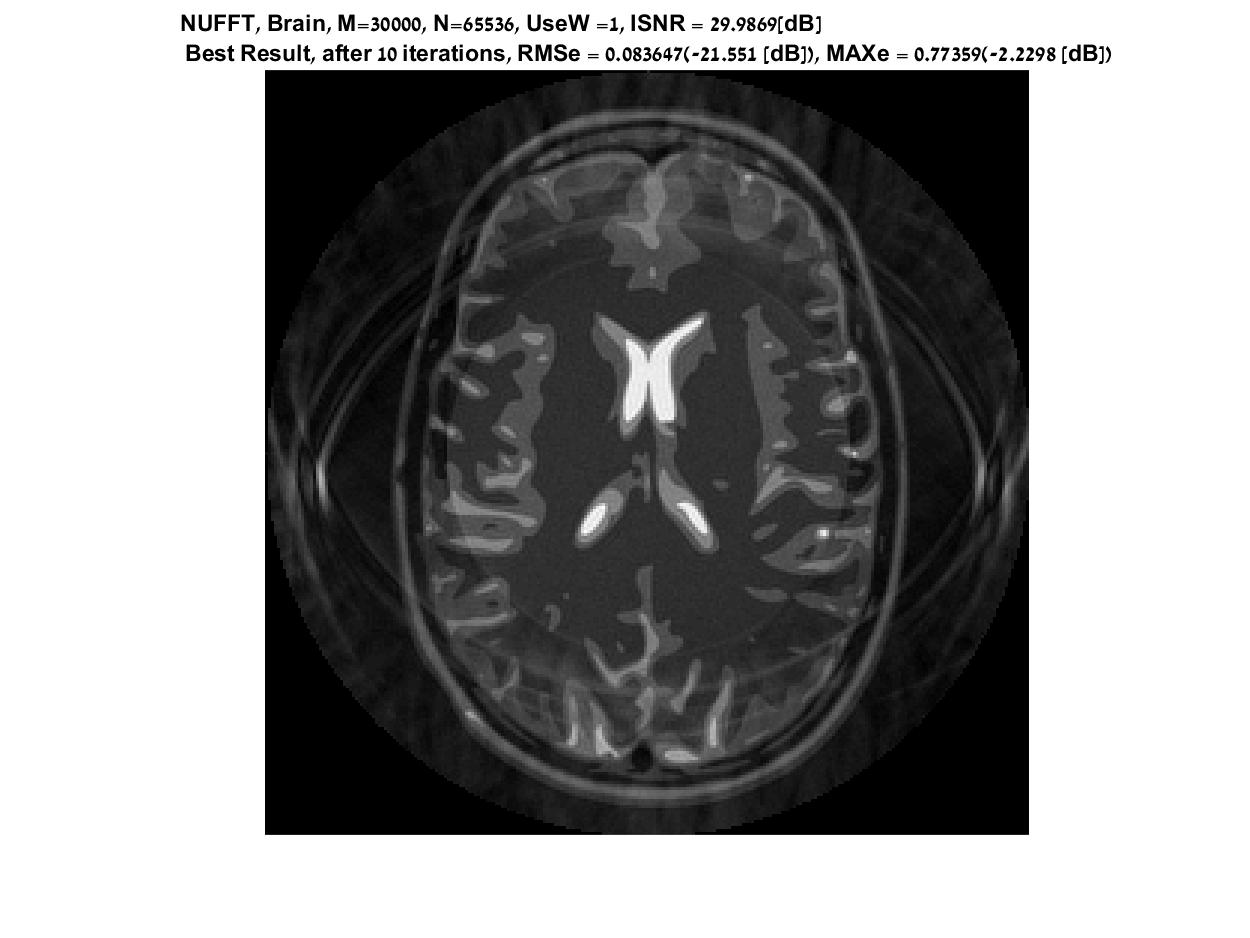}\\

  \includegraphics[trim={3.55cm 0cm 3.55cm 1cm},clip,width=0.22\textwidth]{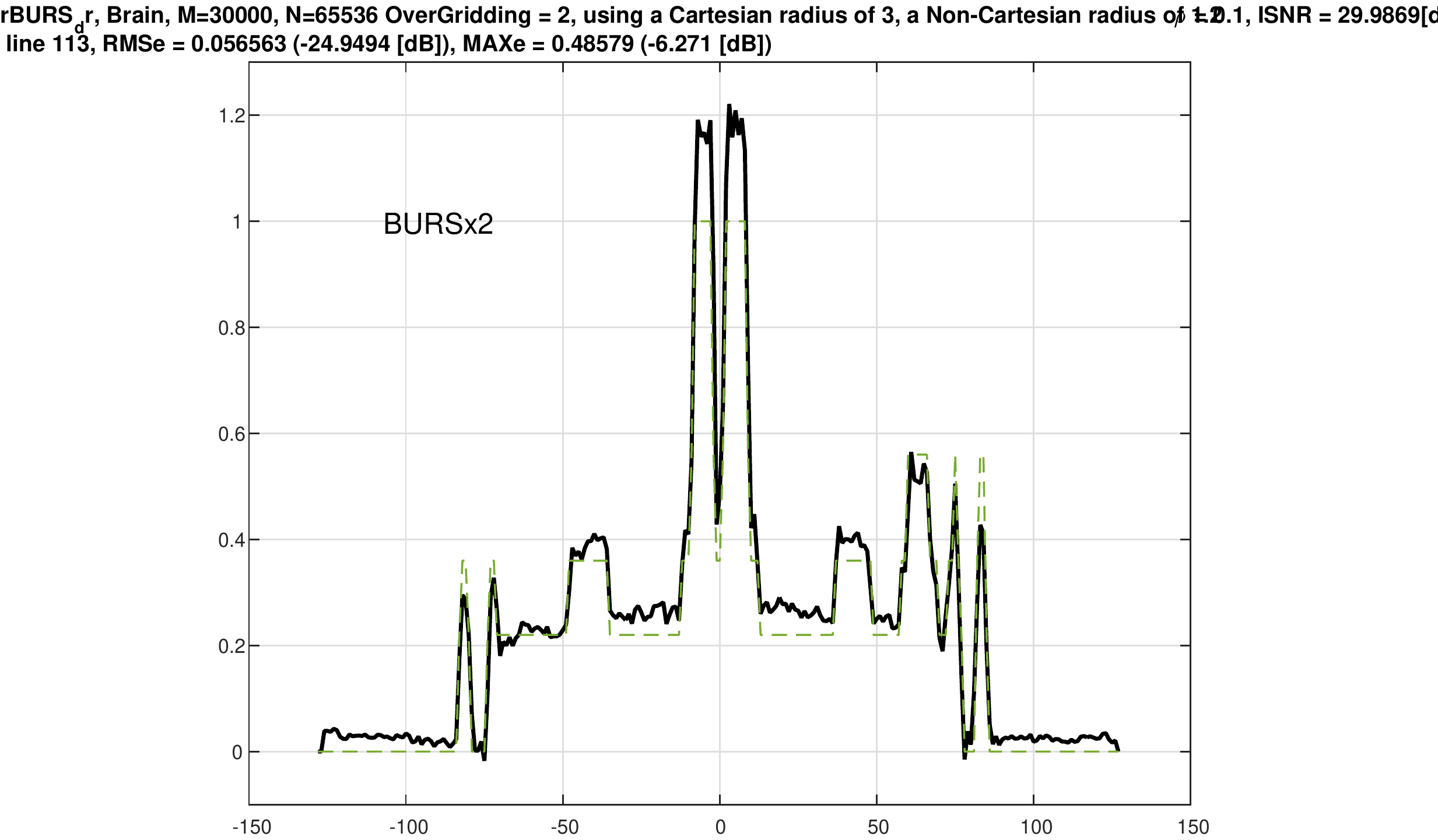} &
  \includegraphics[trim={5.5cm 2.5cm 5.5cm 2.5cm},clip,width=0.22\textwidth]{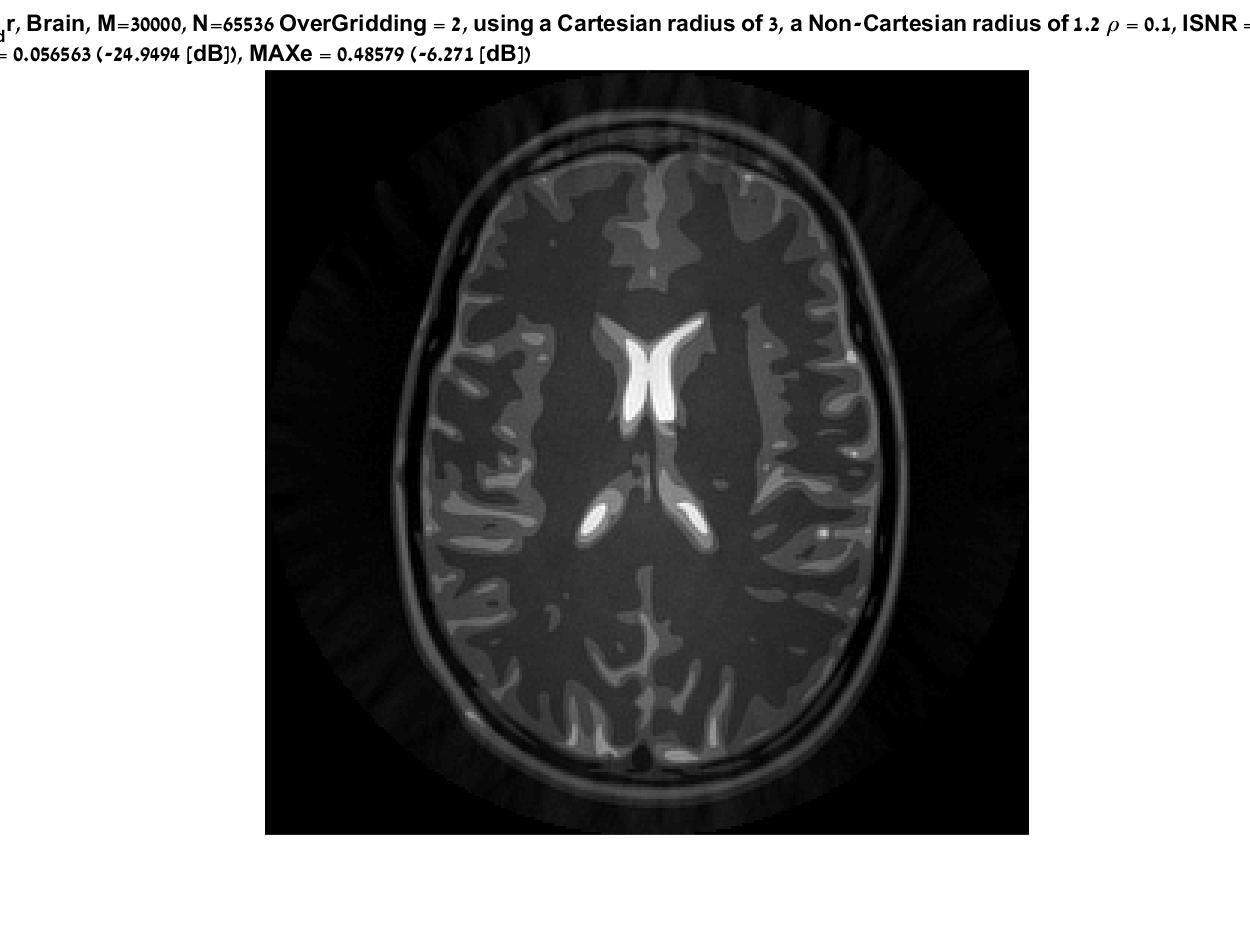}\\

  \includegraphics[trim={0cm 0cm 0cm 1cm},clip,width=0.22\textwidth]{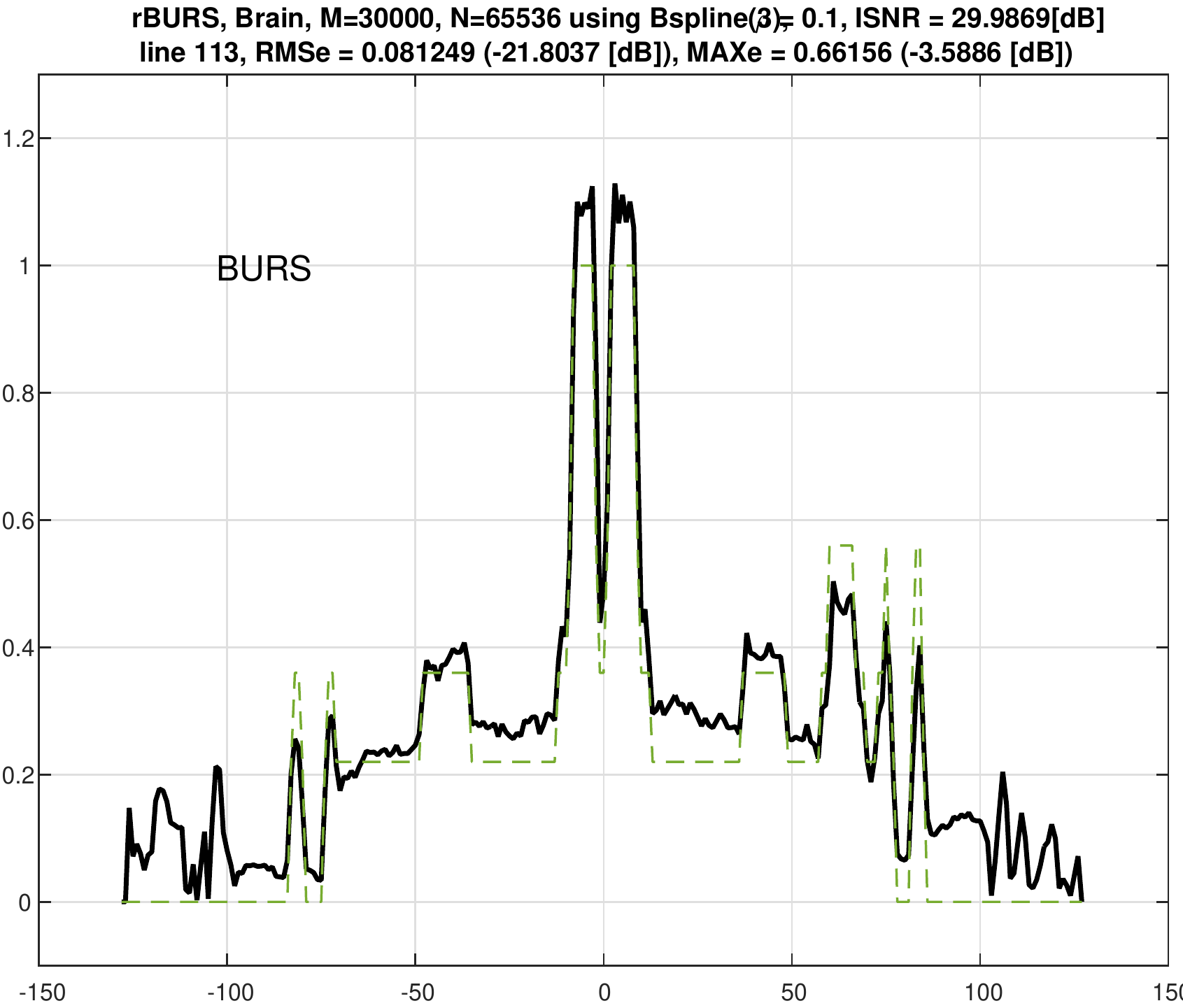} &
  \includegraphics[trim={5.5cm 2.5cm 5.5cm 2.5cm},clip,width=0.22\textwidth]{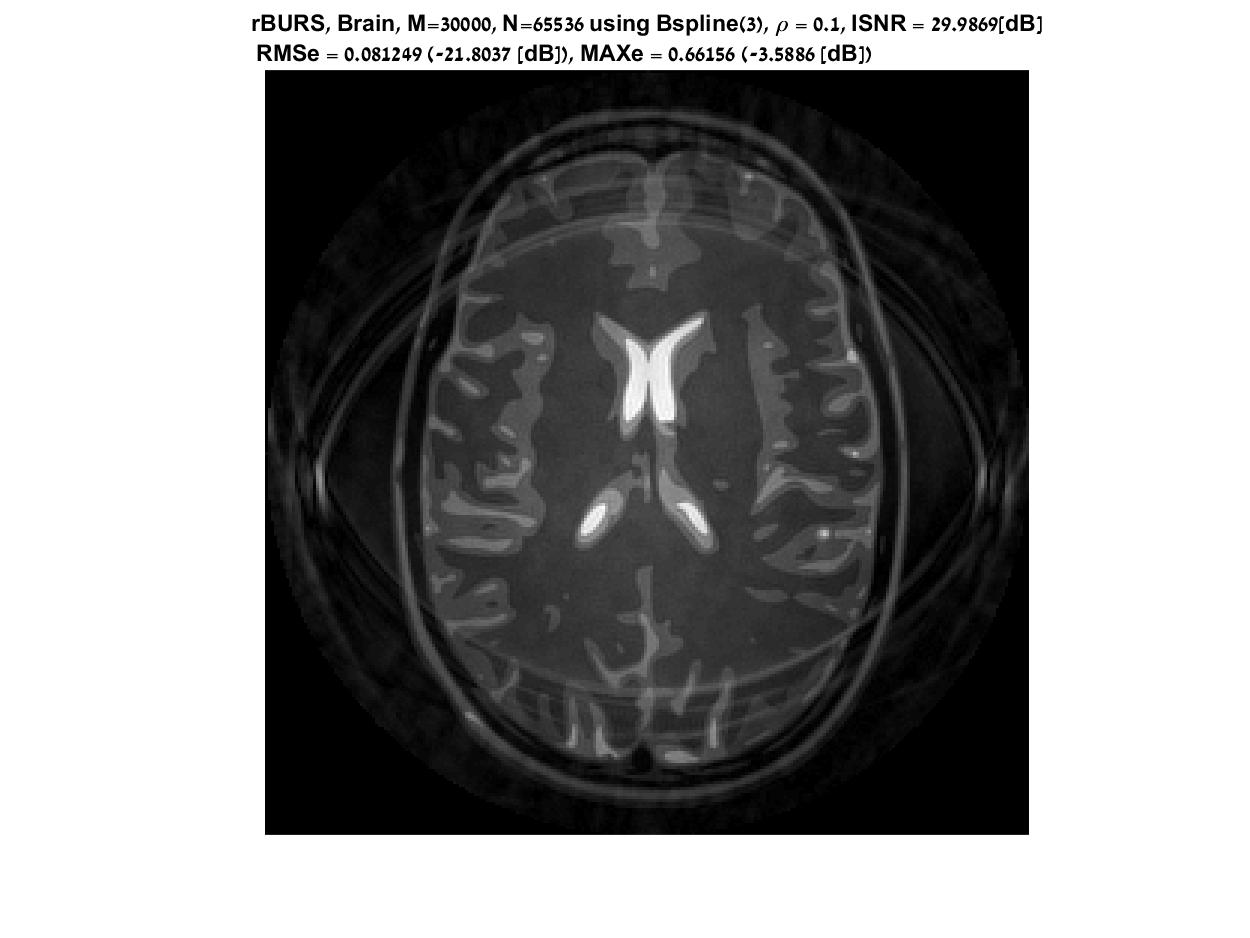}\\

  \includegraphics[trim={0cm 0cm 0cm 1cm},clip,width=0.22\textwidth]{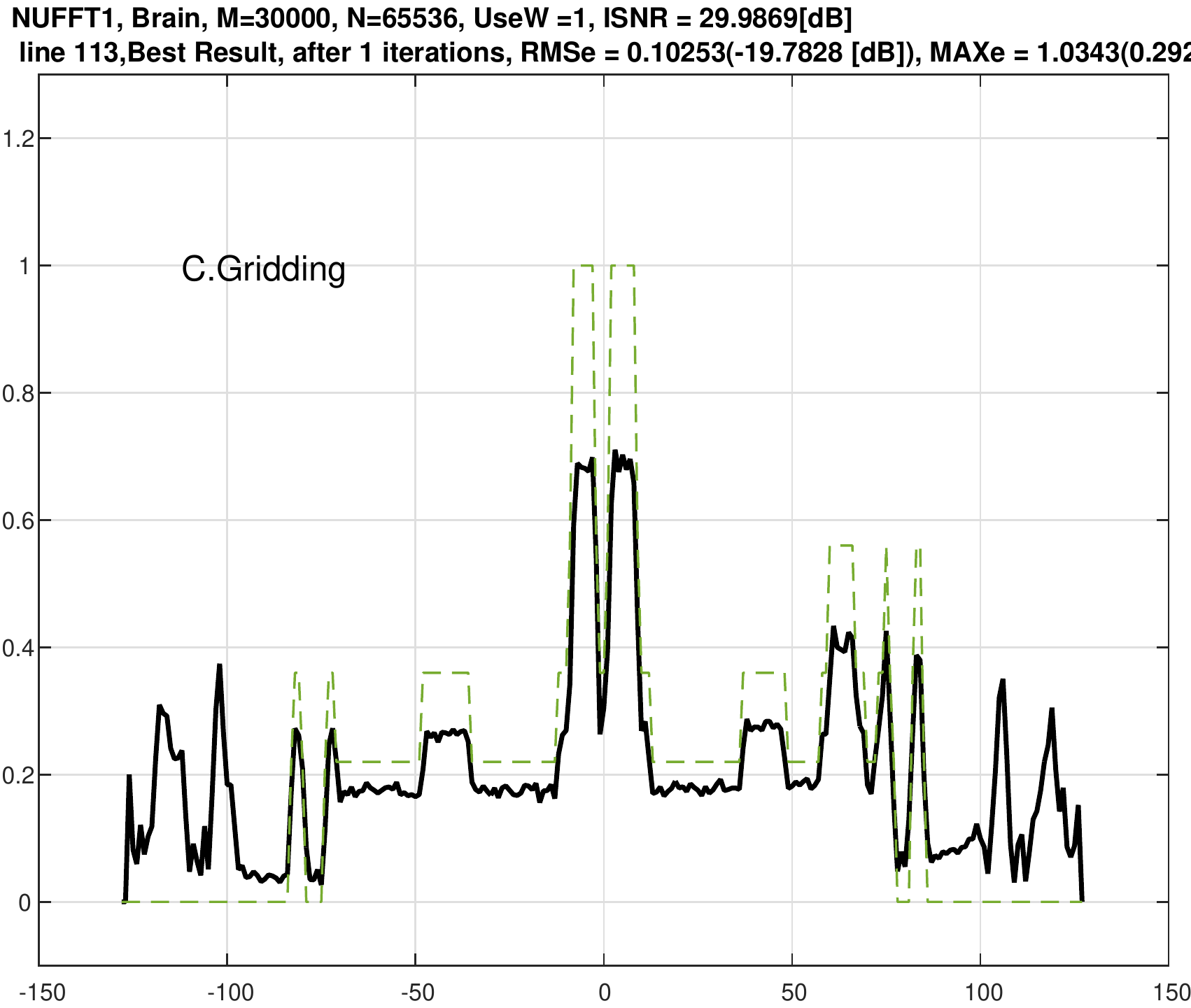} &
  \includegraphics[trim={5.5cm 2.5cm 5.5cm 2.5cm},clip,width=0.22\textwidth]{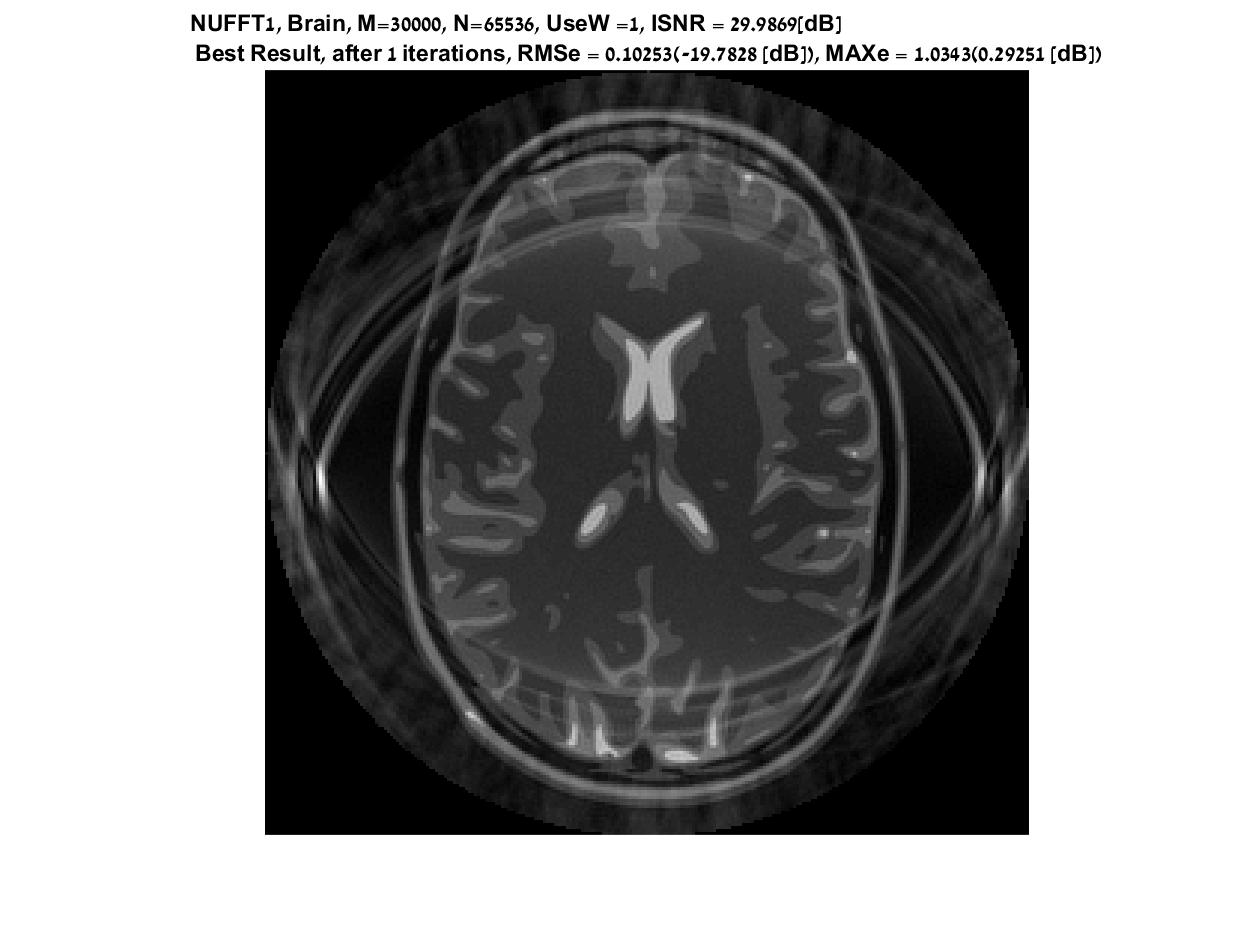}
\end{tabular}
    \caption{Results for an analytic brain phantom sampled on a spiral trajectory with $M=30000$, ISNR = $30$ dB. Right column: reconstructed images; left column: profile plots of row 113. Rows, from top to bottom, SNR and MSSIM values in parenthesis: SPURS using $\beta^3$ (19.57 dB, 0.93), NUFFT (9.15 dB, 0.61), BURS with $\oversamplingfactor = 2$(9.40 dB, 0.61), BURS with $\oversamplingfactor = 1$(12.55 dB, 0.69), convolutional gridding (7.38 dB, 0.61).}
    \label{fig:EXP_M_iSNR_Spiral_img30k}
\end{figure}

\figref{fig:EXP_M_iSNR_Spiral_img30k} exhibits the reconstruction results with the spiral trajectory with ${\rm{ISNR}} = 30 {\rm{dB}}$ for $M=30000$. The reconstructed images are displayed alongside profile plots of row $113$. The same is also presented in Figures \ref{fig:EXP_M_iSNR_Spiral_img20k} for $M=20000$.

\begin{figure}[!htbp]
    \centering %
    \includegraphics[trim={0 0 0 0cm},clip,width=0.47\textwidth]{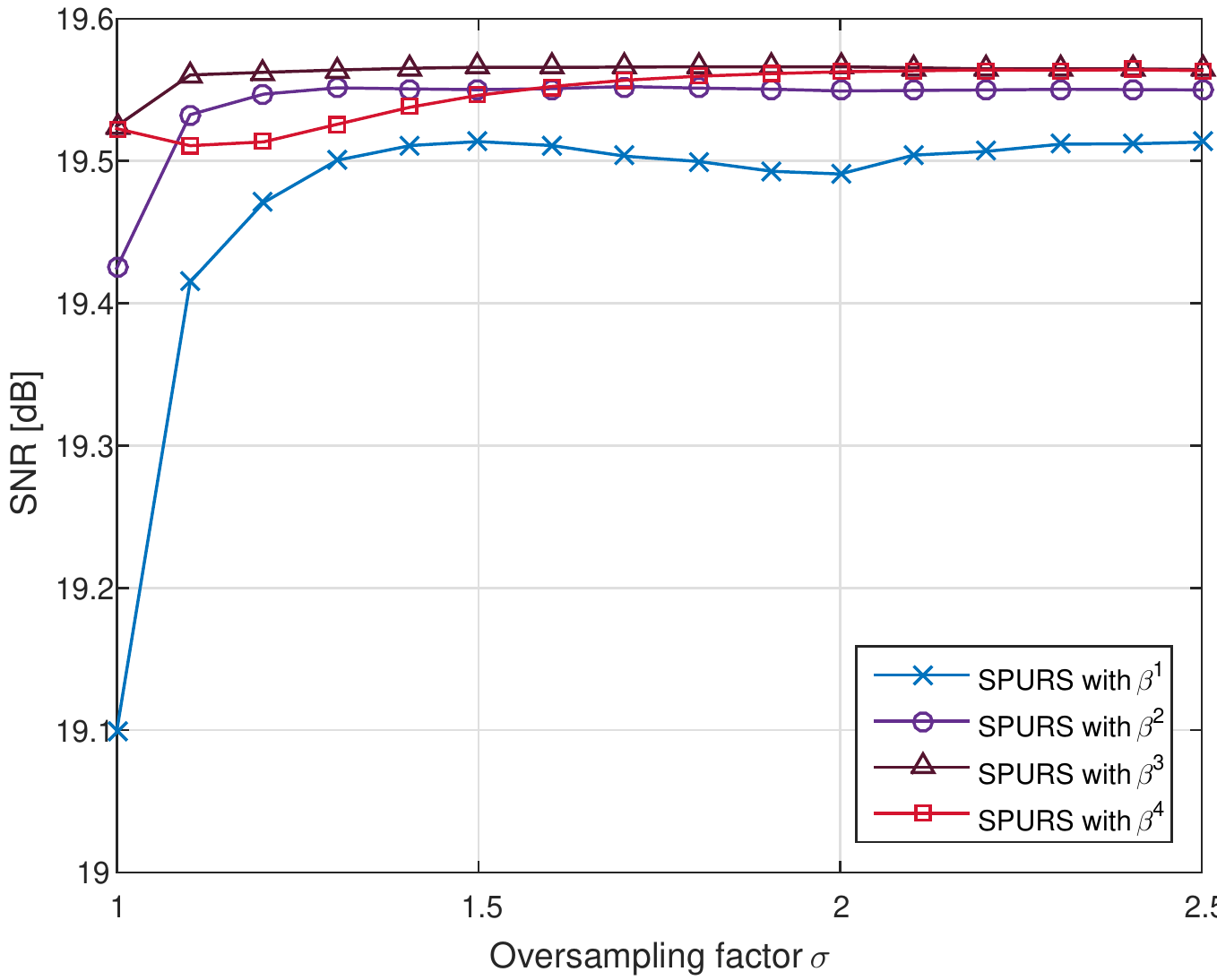}
    \caption{SNR as a function of the oversampling factor $\oversamplingfactor$ and the spline degree for an analytical brain phantom sampled on a spiral trajectory with $M=30000$ and ISNR = $30$ dB.}
    \label{fig:EXP_OS_vs_SD_SNR}
\end{figure}

Figure \ref{fig:EXP_OS_vs_SD_SNR} demonstrates the influence of the oversampling factor $\oversamplingfactor$ and the degree of the B-spline kernel function on the approximation error for the analytical brain phantom sampled on a spiral trajectory with $M=30000$ and ISNR of $30$ dB.

\begin{figure}[!htbp] %
    \centering
    \includegraphics[trim={0 0 0 0cm},clip,width=0.47\textwidth]{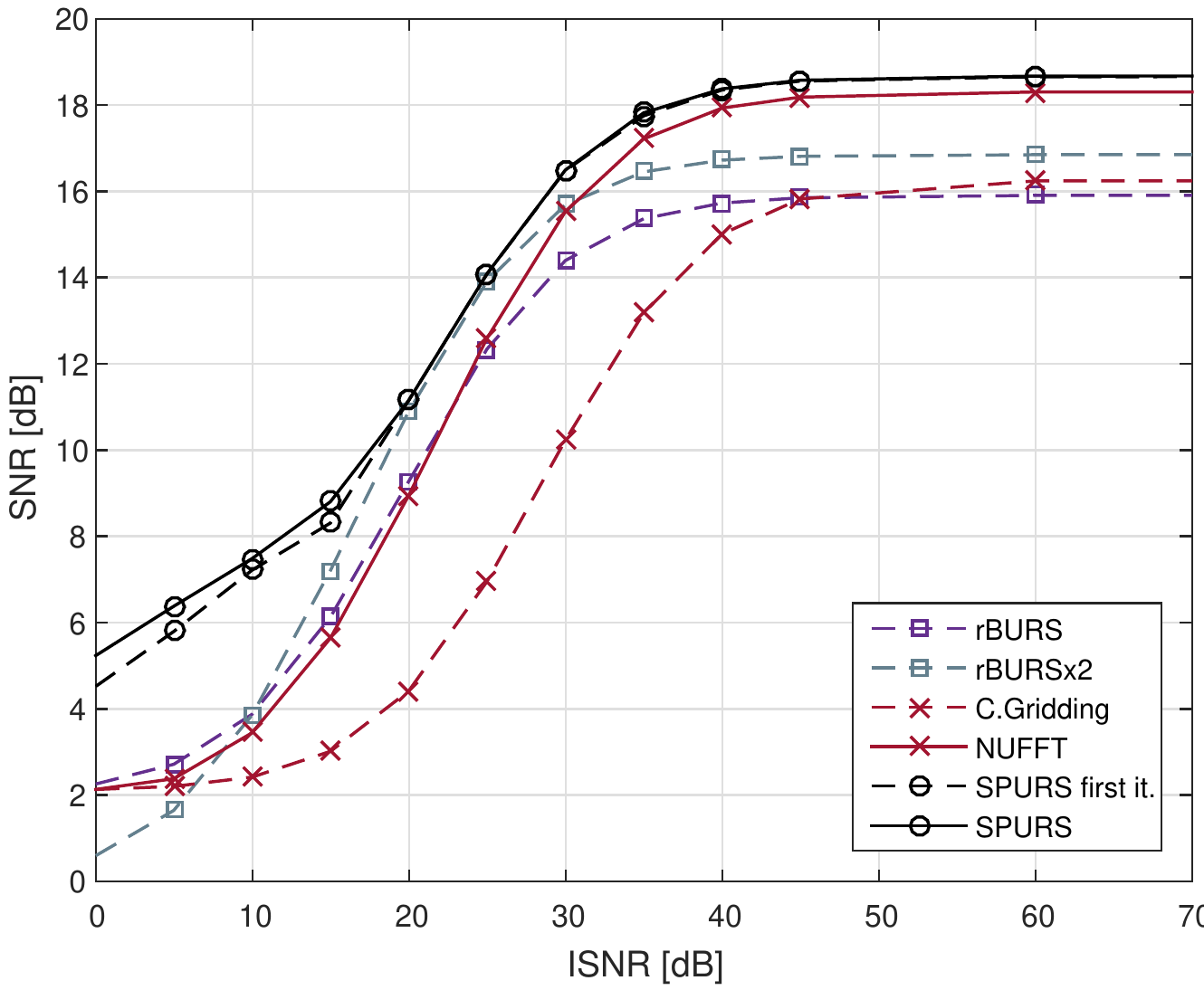}
    \caption{SNR as a function of ISNR for an analytical brain phantom sampled on a Radial trajectory with M = $51200$ samples.}
    \label{fig:EXP_iSNR_SNR_Radial51k}
\end{figure}

\begin{figure}[!htbp] %
    \centering
    \includegraphics[trim={0 0 0 0cm},clip,width=0.47\textwidth]{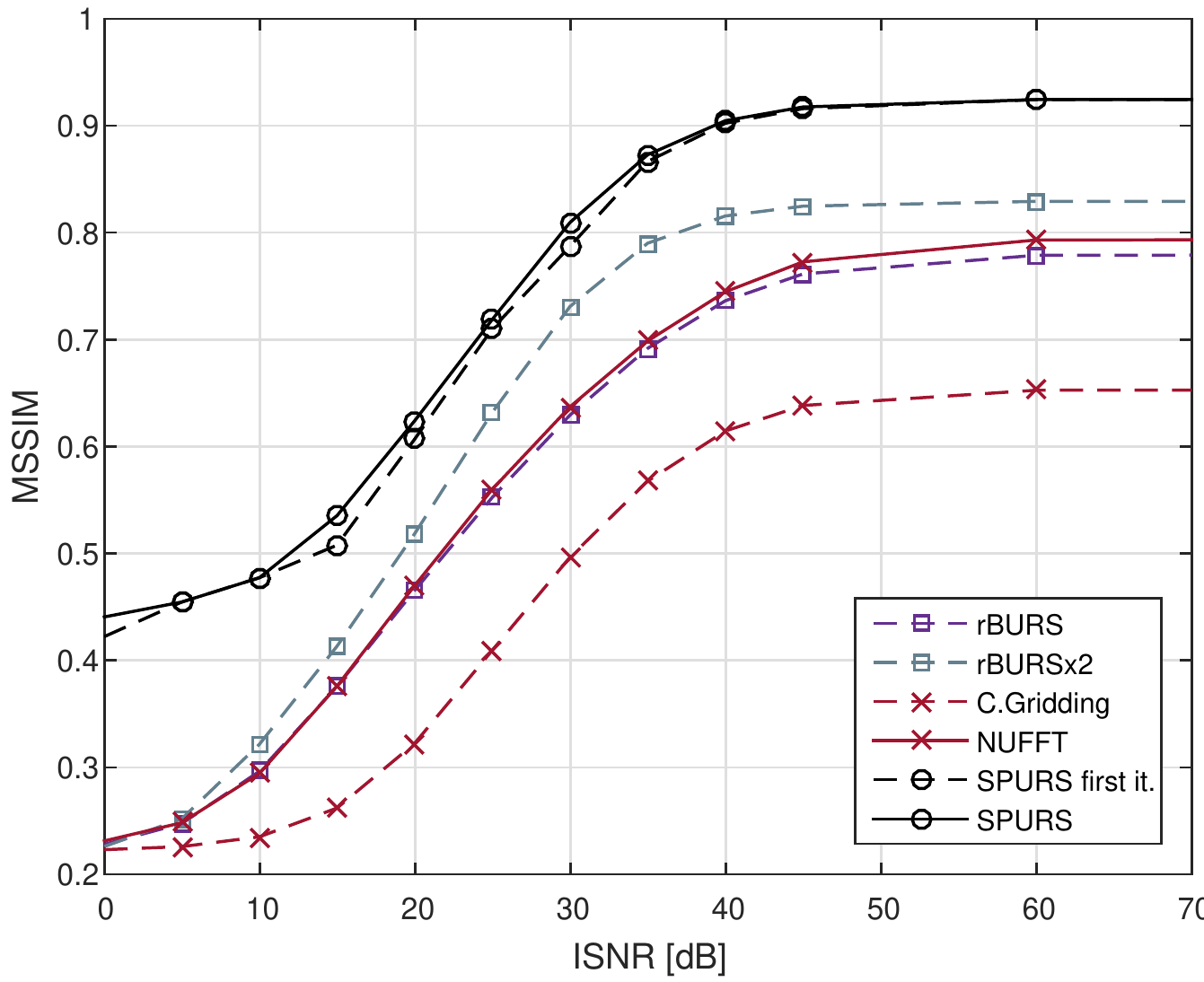}
    \caption{MSSIM as a function of ISNR for an analytical brain phantom sampled on a Radial trajectory with M = $51200$ samples.}
    \label{fig:EXP_M_iSNR_MSSIM_Radial51k}
\end{figure}

\begin{figure}[!htbp] %
    \centering
    \includegraphics[trim={0 0 0 0cm},clip,width=0.47\textwidth]{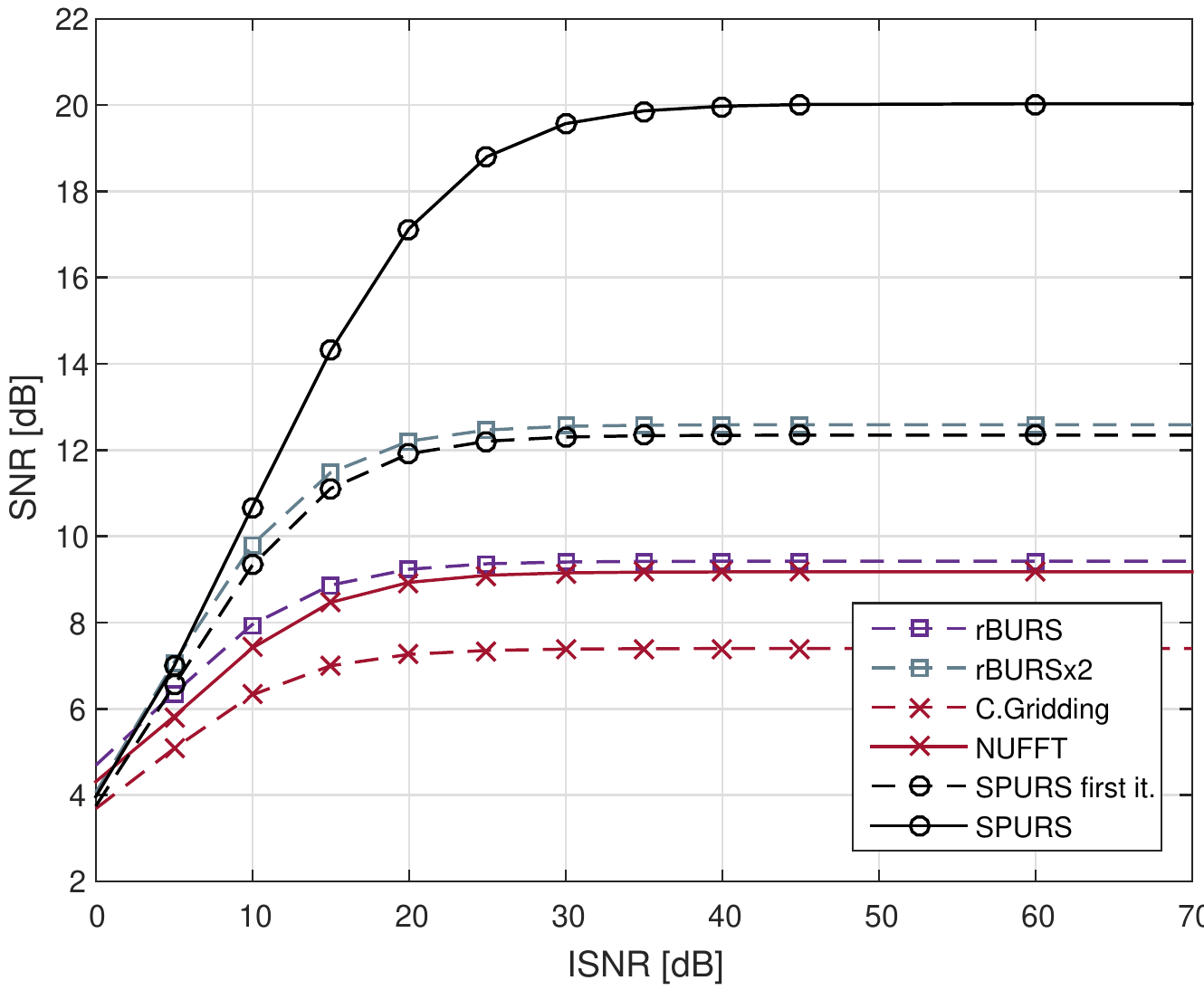}
    \caption{SNR as a function of ISNR for an analytical brain phantom sampled on a spiral trajectory with M = $30000$ samples.}
    \label{fig:EXP_iSNR_SNR_Spiral30k}
\end{figure}

\begin{figure}[!htbp] %
    \centering
    \includegraphics[trim={0 0 0 0cm},clip,width=0.47\textwidth]{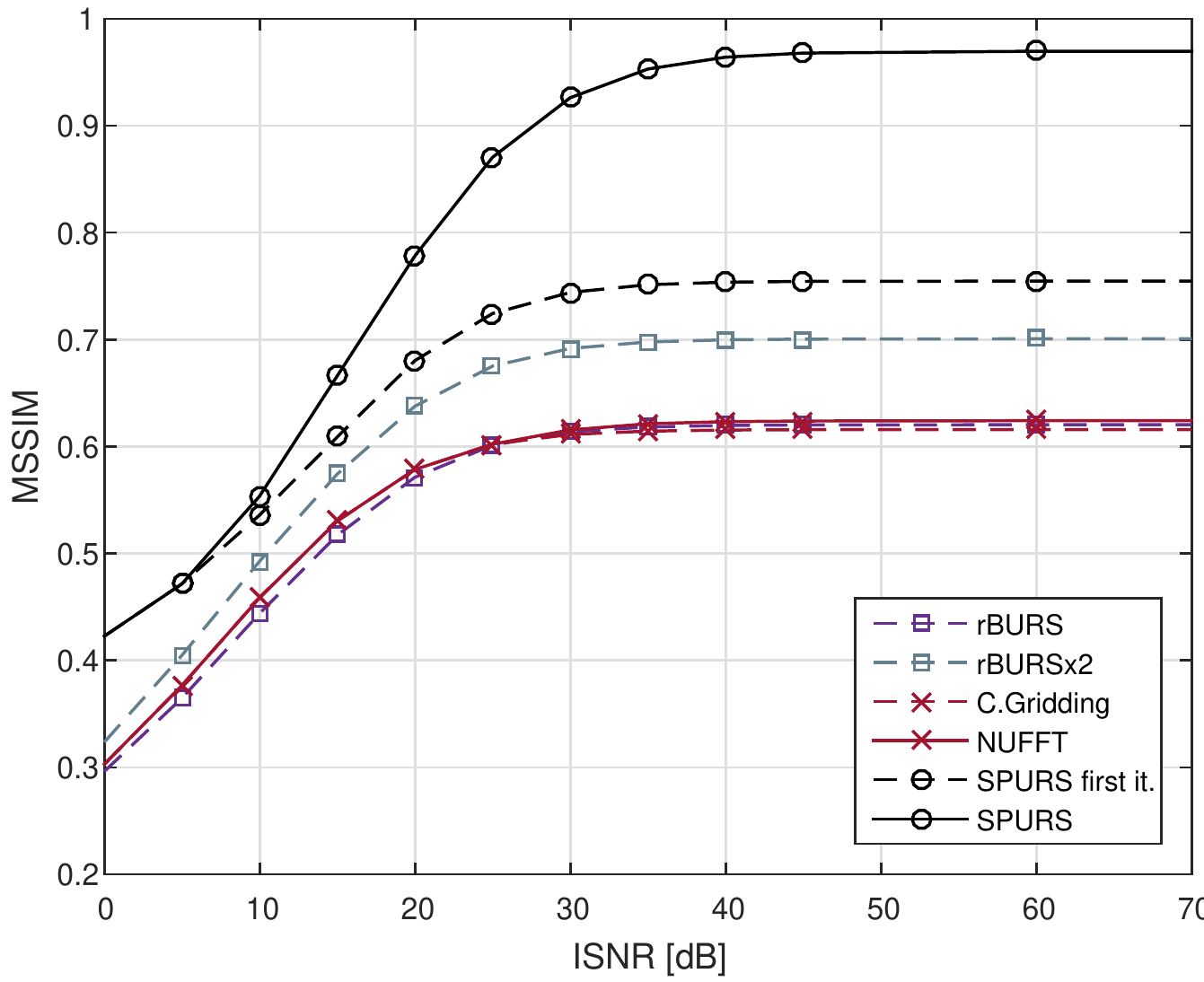}
    \caption{MSSIM as a function of ISNR for an analytical brain phantom sampled on a spiral trajectory with M = $30000$ samples.}
    \label{fig:EXP_M_iSNR_MSSIM_Spiral30k}
\end{figure}

\begin{figure}[!htbp] %
    \centering
    \includegraphics[trim={0 0 0 0cm},clip,width=0.47\textwidth]{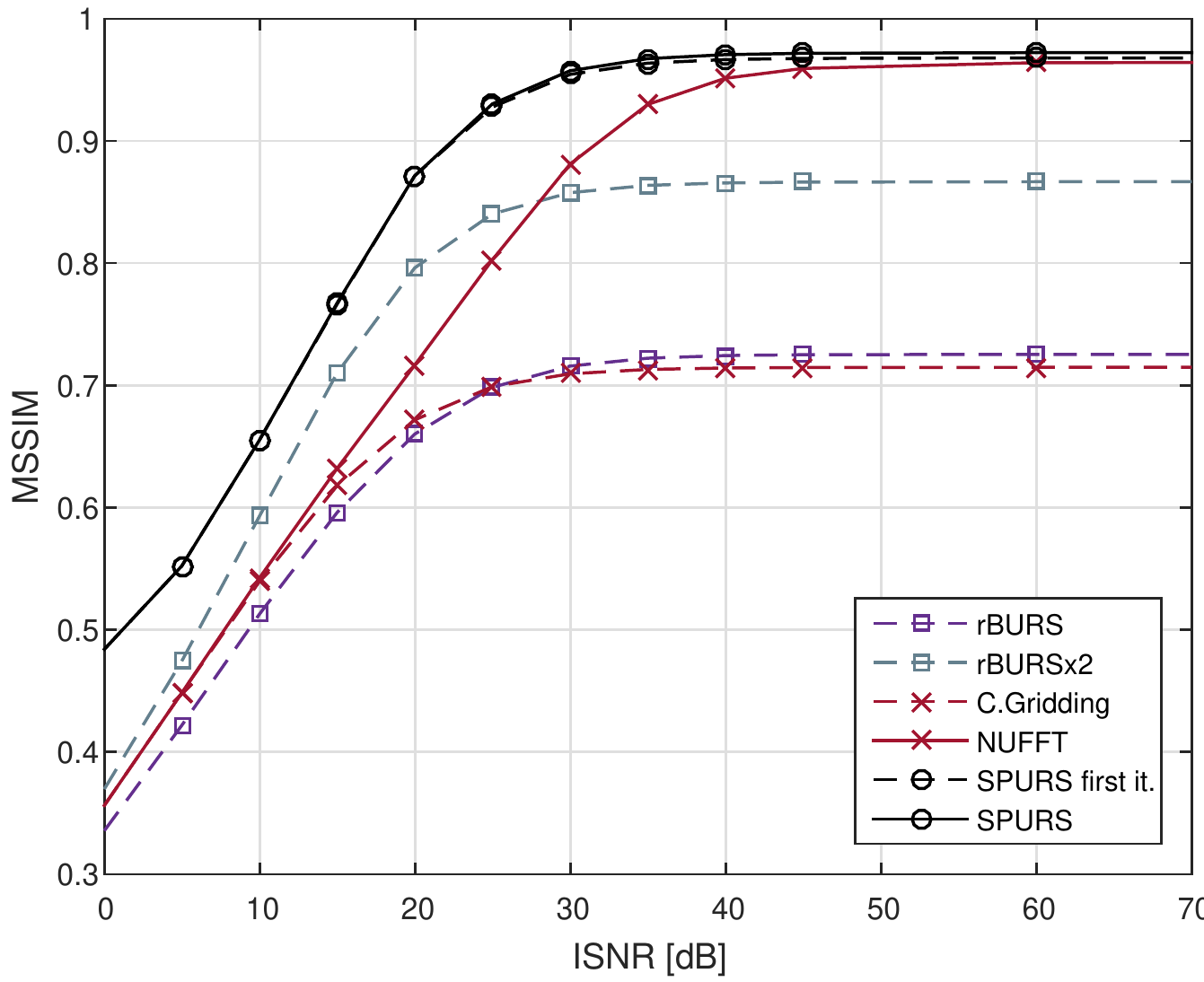}
    \caption{MSSIM as a function of ISNR for an analytical brain phantom sampled on a spiral trajectory with M = $60000$ samples.}
    \label{fig:EXP_M_iSNR_MSSIM_Spiral60k}
\end{figure}

Figures \ref{fig:EXP_iSNR_SNR_Radial51k} and \ref{fig:EXP_M_iSNR_MSSIM_Radial51k} demonstrate the influence of the input SNR on the reconstruction result for the analytical brain phantom sampled on a Radial trajectory with $M=51200$. The same is presented for sampling done on a spiral trajectory with $M=30000$ in Figures \ref{fig:EXP_iSNR_SNR_Spiral30k} and \ref{fig:EXP_M_iSNR_MSSIM_Spiral30k}, and with $M=60000$ in \figref{fig:EXP_M_iSNR_MSSIM_Spiral60k}.
The ISNR value is varied between $0$ dB and a noiseless input (on the right hand side of the plot).

\section{Discussion}\label{sec:Discussion}
The first two experiments compare the performance of the different reconstruction algorithms as a function of the number of measurement points $M$. As expected, the performance of all the methods deteriorates as the number of samples is reduced. In general, better results were achieved for the case of spiral sampling compared to radial sampling, as a function both of the ISNR and of $M$. This is most likely due to the fact that the spiral trajectory has a much more uniform density than that of the radial trajectory which is over-sampled in the proximity of the $k$ space origin, with sampling density decreasing as the distance from the $k$ space origin increases.%

Comparing the MSSIM and SNR values %
for high values of $M$, it is easy to see that a single iteration of SPURS performs better than all the other methods over the entire range of $M$ and ISNR values. It is important to notice how the performance gap between SPURS and the other methods increases as the ISNR decreases. %
For example, the gap between SPURS and the other methods is larger in \figref{fig:EXP_M_iSNR_30_Radial_MSSIM} (or \figref{fig:EXP_M_iSNR_30_Spiral_MSSIM}) than in \figref{fig:EXP_M_iSNR_inf_Radial_MSSIM} (or \figref{fig:EXP_M_iSNR_inf_Spiral_MSSIM}).
These results show that SPURS has better noise performance then the other reconstruction method tested.

When sampling on a radial trajectory, there is a small performance gain in iterating SPURS for the noisy case (\figref{fig:EXP_M_iSNR_30_Radial_MSSIM}) and no gain for the noiseless case (\figref{fig:EXP_M_iSNR_inf_Radial_MSSIM}). This means that SPURS reaches its maximal or near-maximal performance within a single iteration, as opposed to NUFFT which requires $6$ to $10$ iterations to reach its best reconstruction result, which is still inferior to that achieved by a single iteration of SPURS.
In the noiseless setting, the quality measures for all the reconstruction methods improve as $M$ is increased, as is naturally expected. On the other hand, when sampling noise is introduced, only SPURS is able to benefit from the extra samples and continues to improve the MSSIM results for $M>45000$ (\figref{fig:EXP_M_iSNR_30_Radial_MSSIM}).

When sampling on a spiral trajectory, SPURS further demonstrates its superior performance over the other methods. For $M$ values high enough, SPURS, NUFFT and rBURS with $\oversamplingfactor = 2$ achieve very good results, but the performance curve for each method levels off for different values of $M$ (Figures \ref{fig:EXP_M_iSNR_30_Spiral_SNR}, \ref{fig:EXP_M_iSNR_30_Spiral_MSSIM}, \ref{fig:EXP_M_iSNR_inf_Spiral_SNR} and \ref{fig:EXP_M_iSNR_inf_Spiral_MSSIM}).
Iterative SPURS levels off for values as low as $M = 20000$, requiring about $10$ iterations to converge to its best result. For these low $M$ values, significant artifacts appear in the reconstructed image produced by all methods excluding SPURS as presented in \figref{fig:EXP_M_iSNR_Spiral_img20k} for $M=20000$ and \figref{fig:EXP_M_iSNR_Spiral_img30k} for $M=30000$. The performance curve of the NUFFT method and of a single iteration of SPURS level off at around $M=50000$. For $M=50000$ and higher, a single iteration of SPURS produces marginally better results than those produced by NUFFT, which requires about $10$ iterations to converge. Among the other non-iterative methods, both rBURS with $\oversamplingfactor = 2$ and convolutional gridding perform similarly well for $M>50000$, however the results are still inferior to those of a single iteration of SPURS, all of which have similar computational complexity.

Similar trends are exhibited in Figures \ref{fig:EXP_iSNR_SNR_Radial51k},\ref{fig:EXP_M_iSNR_MSSIM_Radial51k},\ref{fig:EXP_iSNR_SNR_Spiral30k} and \ref{fig:EXP_M_iSNR_MSSIM_Spiral30k}, which demonstrate the noise performance at a given number of sampling points $M$.
It is shown once again that a single iteration of SPURS outperforms the other methods and that in some cases the results can be further improved by iterating SPURS.
Figure \ref{fig:EXP_M_iSNR_MSSIM_Spiral60k} shows the noise performance for a large number of sampling points $M=60000$. For high values of ISNR, the performance of a single iteration of SPURS is similar to that of NUFFT, however, at low values of ISNR the advantage of SPURS over the other methods is apparent.

\begin{figure}
    \centering %
    \includegraphics[trim={0 0 0 0cm},clip,width=0.47\textwidth]{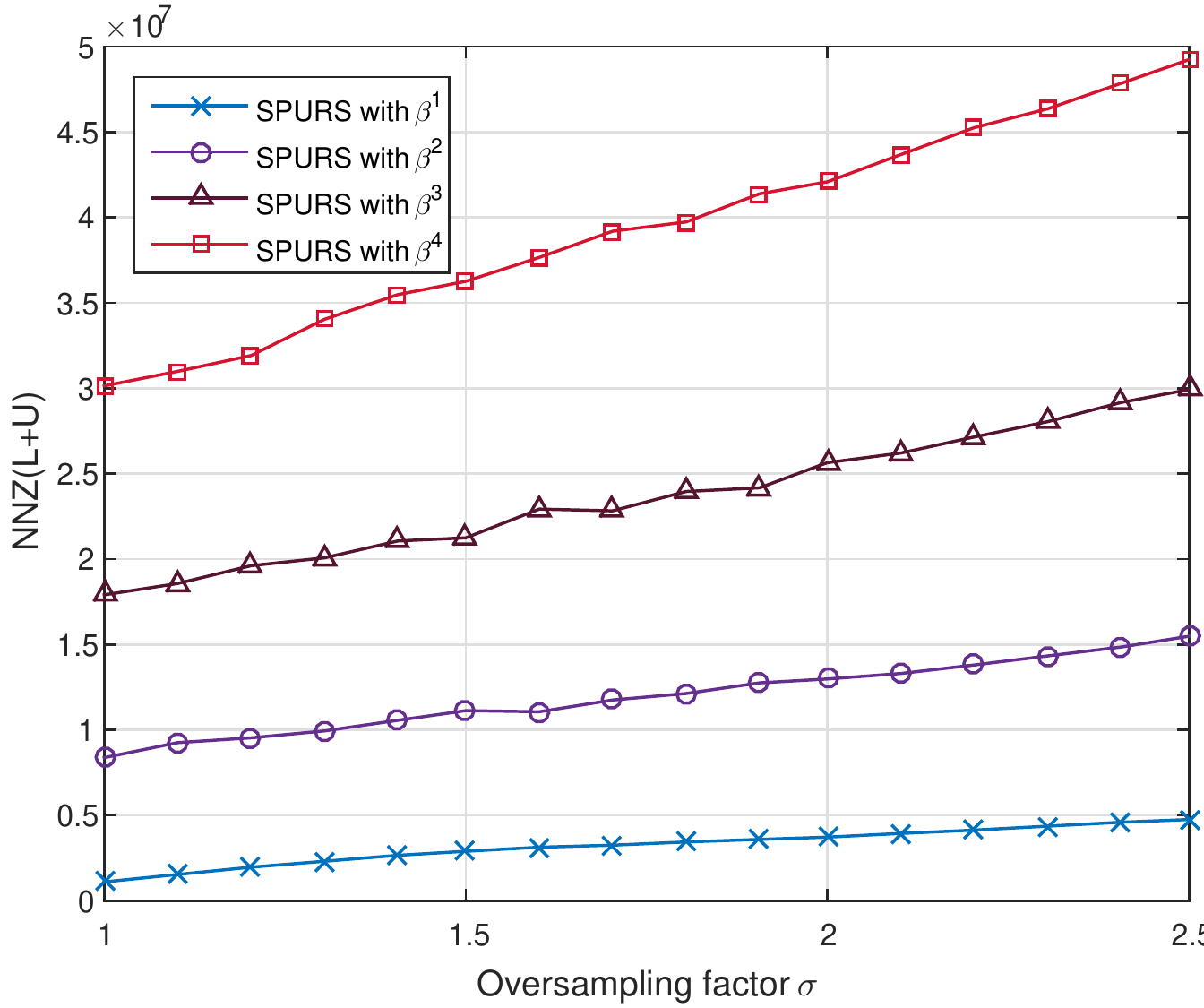}
    \caption{The number of non zeros in the $L+U$ matrices which factor $\mathbf{\Psi}$ as a function of the oversampling factor $\oversamplingfactor$ and the spline degree for a spiral sampling trajectory with $M=30000$.}
    \label{fig:EXP_OS_vs_SD_NNZ}
\end{figure}

Figure \ref{fig:EXP_OS_vs_SD_SNR} shows the influence of the oversampling factor $\oversamplingfactor$ and the support of the kernel function $q$ on the performance of SPURS. It can be seen that the SNR levels off at $\oversamplingfactor=1.2$. Moreover, the degree of the B-spline has a relatively small impact on the performance; in particular, even using a B-spline of degree $1$ (which has a support of $2$ k-space samples, and is equivalent to linear interpolation) incurs merely a $0.1$ dB penalty in SNR with respect to higher degree splines. As presented in Section \ref{ComputationalComplexity}, both $\oversamplingfactor$ and the spline support affect the computational complexity in a way that it is advantageous to keep them at a minimum.
For example, in \figref{fig:EXP_M_iSNR_Spiral_img30k} a spiral trajectory with $M=30000$ and ${\rm{ISNR}} = 30 {\rm{dB}}$ is employed, with SPURS using $\oversamplingfactor=2$ and $\beta^3$. The reconstruction results has ${\rm{SNR}} = 19.57 {\rm{dB}}$. According to \figref{fig:EXP_OS_vs_SD_NNZ}, selecting $\oversamplingfactor=1.2$ and $\beta^1$ decreases ${\rm{NNZ}}\left(\mathbf{L+U}\right)$ and thus the storage requirements by more than tenfold and the total number of operations by a factor of about $3$. The penalty in performance is negligible, and in our experiment we obtain ${\rm{SNR}} = 19.47 {\rm{dB}}$. These results are significantly better than those of all the other methods presented in  \figref{fig:EXP_M_iSNR_Spiral_img30k}.

\section{Conclusion}\label{sec:Conclusion}
A new computationally efficient method for reconstruction of functions from a non-Cartesian sampling set is presented
which derives from modern sampling theory. In the algorithm, termed SPURS, a sequence of projections is performed, with the introduction of an interim subspace $\mathcal{Q}$ comprised of integer shifts of a compactly supported kernel. A sparse set of linear equations is constructed, which allows for the application of efficient sparse equation solvers, resulting in a considerable reduction in the computational cost.
The purposed method is used for reconstruction of images sampled nonuniformly in k-space, such as in medical imaging: MRI or CT. SPURS can also be employed for other problems which reconstruct a signal from a set of non-Cartesian samples, especially those of considerable dimension and size.
After performing the offline data preparation step, which is only performed once for a given set of sampling locations, the computational burden of the online stage of SPURS is on a par with that of convolutional gridding or of a single iteration of NUFFT. %

In terms of the quality of the reconstructed images, it is demonstrated that the performance of a single iteration of the new algorithm, for different sampling SNR ratios and for various trajectories, exceeds that of both convolutional gridding, BURS and NUFFT at no additional computational cost. Iterations can further improve the results at the cost of higher computational complexity allowing to cope with reconstruction problems in which the number of available samples and the SNR are low. These scenarios are of utmost importance in modern fast imaging techniques.

In this paper we used B-spline functions as the support-limited kernel function spanning the intermediate subspace $\mathcal{Q}$.
No attempt has been made to optimize this kernel function.
Significant research has been performed in order to optimize the kernel functions employed by other reconstruction methods such as convolutional gridding \cite{jackson1991selection,sedarat2000optimality} and NUFFT \cite{fessler2003nonuniform}. Future research could possibly improve the performance of the SPURS algorithm by optimizing the kernel used.

The sparse equation solvers used in the present research employed the default control parameters which were provided with the software package.
The factorization of the sparse system matrix can possibly be improved to run faster and produce sparser factors by tuning the control parameters of the problem or by evaluating other available solvers. %

\bibliographystyle{IEEEtran}
\bibliography{IEEEabrv,Jbib}
\end{document}